\documentclass[aps,prf,preprint,groupedaddress,longbibliography]{revtex4-1}

\usepackage{amsmath}
\usepackage{graphicx}
\usepackage{color}
\usepackage{graphicx}
\usepackage{epstopdf, epsfig}
\usepackage{amssymb}
\usepackage{tikz}
\usepackage{pgfplots}
\usepackage{soul,xcolor}
\usepackage{natbib}

\setstcolor{red}
\graphicspath{ {FIG/} }
\newcommand\omegab{\boldsymbol{\omega}}
\newcommand\taub{\boldsymbol{\tau}}

\providecommand\upi{\pi}

\newcommand{\td}[2] {\frac{d #1}{d #2}}

\newcommand{\pd}[2] {\frac{\partial #1}{\partial #2}}

\renewcommand{\vec}[1]{{\bf#1}}

\renewcommand{\div}{\vec{\nabla \cdot}}

\newcommand{\grad}{\vec \nabla }

\newcommand{\T}{\theta}

\begin{document}

\title{Inertial migration in dilute and semi-dilute suspensions of rigid particles in laminar square duct flow} 

\author{H.~Tabaei~Kazerooni}
\email{seyed.tabaeikazerooni@ruhr-uni-bochum.de}
\affiliation{SeRC and Linn\'e FLOW Centre, KTH Mechanics, SE-100 44 Stockholm, Sweden}
\affiliation{Ruhr-Universit\"at Bochum, Chair of Hydraulic Fluid Machinery, Universit\"atsstra\ss e 150, 44801
Bochum, Germany}

\author{W.~Fornari}
\email{fornari@mech.kth.se}
\affiliation{ SeRC and Linn\'e FLOW Centre, KTH Mechanics, SE-100 44 Stockholm, Sweden}

\author{J.~Hussong}
\email{jeanette.hussong@rub.de}
\affiliation{Ruhr-Universit\"at Bochum, Chair of Hydraulic Fluid Machinery, Universit\"atsstra\ss e 150, 44801
Bochum, Germany}

\author{L.~Brandt}
\email{luca@mech.kth.se}
\affiliation{SeRC and Linn\'e FLOW Centre, KTH Mechanics, SE-100 44 Stockholm, Sweden}

\date{\today}

\begin{abstract}

We study the inertial migration of finite-size neutrally buoyant spherical particles in dilute and semi-dilute 
suspensions in laminar square duct flow. We perform several direct numerical simulations using an immersed boundary 
method to investigate the effects of the bulk Reynolds number $Re_b$, particle Reynolds number $Re_p$ and duct to 
particle size ratio $h/a$ at different solid volume fractions $\phi$, from very dilute conditions to $20\%$. We show that the bulk Reynolds number 
$Re_b$ is the key parameter in inertial migration of particles in dilute suspensions. At low solid volume fraction ($\phi=0.4\%$) and 
low bulk Reynolds number $(Re_b= 144)$, particles accumulate at the center 
of the duct walls. As $Re_b$ is increased, the focusing position 
moves progressively towards the corners of the duct. At higher volume fractions, $\phi=5, 10$ and $20\%$, 
and in wider ducts with $Re_b=550$, particles are found to migrate away from the duct core towards the walls. 
In particular, for $\phi=5$ and $10\%$, particles accumulate preferentially at the corners. At the highest volume fraction considered, $\phi=20\%$, 
particles sample all the volume of the duct, with a lower concentration at the duct core. 
For all cases, we find that particles reside longer times at the corners than at the wall centers. 
In a duct with lower duct to 
particle size ratio $h/a$ (i.e. with larger particles), $Re_b=144$ and $\phi=5\%$ we find that particles preferentially accumulate around the corners. 
Hence, the volume fraction plays a key role in 
defining the final distribution of particles in semi-dilute suspensions.
The presence of particles induces secondary cross-stream motions in the duct 
cross-section, for all $\phi$. The intensity of these secondary flows depends strongly 
on particle rotation rate, on the maximum concentration of particles in focusing positions, 
and on the solid volume fraction. We find that the secondary flow intensity increases with the volume fraction up to $\phi=5\%$. 
However, beyond $\phi=5\%$ excluded volume effects lead to a strong reduction of cross-stream velocities. Inhibiting particles 
from rotating also results in a substantial reduction of the secondary flow intensity, and in variations of the exact location 
of the focusing positions.
\end{abstract}

\pacs{}

\maketitle 

\section{Introduction}

The role and the importance of fluid inertia in different microfluidic applications has been recently recognized~\cite{di2009}. Due to 
finite fluid inertia, for example, it is possible to achieve enhanced mixing and efficient particle separation and focusing. To further 
develop inertial microfluidic devices, it is therefore necessary to properly understand the behaviour of suspensions at finite Reynolds numbers and 
in confined geometries.\\
Clearly, these suspensions exhibit very interesting and peculiar rheological properties. Among these we recall 
shear-thinning or thickening, the appearance of normal stress differences and jamming at high volume fractions~\cite{stickel2005,morris2009}. 
Interesting effects due to confinement in simple shear flows in the Stokes and weakly inertial regimes have also been recently 
reported~\cite{fornari2016PRL,doyeux2016}. Another important feature observed in wall-bounded flows is particle migration. In the viscous 
regime, there is an irreversible shear-induced migration of particles away from channel walls~\cite{guazz2011}. 
However, when inertial effects become important the migration mechanisms may vary. This is typically simply referred to as \emph{inertial} migration.

The inertial migration of neutrally buoyant finite-size particles in Poiseuille flow has been the object of several 
studies since the work by Segre \& Silberberg in 1962~\cite{segre1962}. These authors studied experimentally the flow 
of a dilute suspension of randomly distributed spherical particles in a laminar pipe flow. They showed that, at very low bulk Reynolds number $Re_b=O(1)$, particles 
migrate away from the pipe core region and form a stable annulus at a distance of approximately $0.6R$, being $R$ the pipe 
radius. It was later explained that the particle equilibrium position in the pipe cross-section is determined by the balance 
between the wall repulsive lubrication force~\cite{zeng2005} and the shear induced lift force on the particle 
due to the curvature of the velocity profile~\cite{matas2004lateral}. 
More recently, Matas et al.~\cite{matas2004} studied experimentally the effects of the bulk Reynolds number $Re_b$ and 
pipe to particle size ratio on the inertial migration of spherical particles at low volume fractions, $\phi < 1\%$. 
These experiments show that particles are progressively pushed towards the wall as the bulk Reynolds number is increased. 
However, at larger $Re_b$ and depending on the pipe to particle size ratio, it was also found that particles accumulate 
on an inner annulus in the pipe cross-section. Later on, the same authors~\cite{matas2009} performed an asymptotic analysis to 
investigate the equilibrium position of a sphere in laminar pipe flow based on the point particle assumption. While this 
theoretical work confirmed the progressive shift of the particle towards the pipe wall by increasing the bulk Reynolds number $Re_b$, 
it could not predict the presence of the inner particle annulus closer to the pipe center. 
Hence, the existence of the inner equilibrium position is probably related to the finite size of the particles. Most recently, Morita et al.~\cite{morita2017} performed several experiments to clarify this discrepancy between theoretical and experimental results and suggested that the occurrence of the inner annulus is a transient phenomenon that would disappear for long enough pipes. 
Concerning dense suspensions of neutrally buoyant particles in pipe flows, Han et al.~\cite{han1999} showed experimentally that 
inertial migration is a very robust phenomenon which occurs for particle Reynolds numbers $Re_p$ larger than 0.1, regardless of 
the solid volume fraction.

In the past few years, the Segre-Silberberg phenomenon has been used as a passive method for the separation and sorting of cells 
and particles in microfluidic devices~\cite{di2009,gossett2010,karimi2013}. Details on the physics of inertial migration and its 
microfluidic applications have recently been documented in a comprehensive review article by Amini and collaborators~\cite{amini2014}. 
Due to microchip manufacturing, square channels are often utilized in such applications. Due to the loss of cylindrical 
symmetry, the particle behavior is altered with respect to pipe flows. The study of particulate duct flows has hence attracted 
various researchers over the years.

Chun and Ladd~\cite{chun2006}, for example, performed numerical simulations using a Lattice-Boltzman method to study the motion of 
single particles and dilute suspensions in square duct flows. The existence of eight equilibrium positions at the duct corners and 
wall centers was reported for single particles in a range of bulk Reynolds number $Re_b$ between $100$ and $1000$. It was shown that 
at moderately high Reynolds numbers ($Re_b\textgreater500$), the equilibrium position at the wall center is not stable and particles 
move towards the duct corners. A similar pattern was found for a low solid volume fraction ($\phi=1\%$) at $Re_b=500$. Moreover, 
the appearance of particles at the inner region of the duct was also observed in addition to four equilibrium position at the 
corners for $\phi=1\%$ and $Re_b=1000$. However, as previously discussed, it seems that the presence of the particles in the 
center region of Poiseuille flows of dilute suspensions at high bulk Reynolds number $Re_b$ is a transient feature.

Later, Di Carlo et al.~\cite{di2009PRL} carried out an experimental and numerical investigation of the motion of an individual 
particle in duct flow at low Reynolds number. In particular, they explored the lift forces acting on the particle and the influence of 
the particle to duct size ratio on the particle equilibrium position. They showed that for low bulk Reynolds number the duct corners 
are unstable equilibrium positions. The duct wall centers are the only points where the wall lubrication and shear induced lift forces 
balance each other and are hence stable equilibrium position. These results has been also confirmed recently by the theoretical work of 
Hood et al.~\cite{hood2015} in which an asymptotic model was used to predict the lateral forces on a particle and determine its stable 
equilibrium position.

Choi et al.~\cite{choi2011} investigated experimentally the spatial distribution of dilute suspended particles in a duct flow at low 
bulk Reynolds numbers ($Re_b\textless120$) and different duct to particle size ratios, $h/a$ (where $h$ and $a$ are the duct half-width and the 
particle radius). For $Re_b=12$ and relatively high duct to particle size ratio ($h/a=6.25$), they observed the formation of a 
ring of particles parallel to the duct walls at a distance of around $0.6h$ from the centerline. They showed that by increasing the bulk Reynolds 
number to $120$, the particle ring breaks and four particle focusing (equilibrium) points are observed at the duct wall centers. The same behavior 
for particle distributions across the duct cross section has also been observed experimentally by Abbas et al.~\cite{abbas2014} for $Re_b \in 
[0.07; 120]$. On the other hand, for very low Reynolds number $Re\ll1$ (i.e. when inertia is negligible), particles accumulate at the duct center 
region. More recently, Miura et al.~\cite{miura2014} carried out an experimental study on the inertial migration of particles in a macroscale square 
duct for $h/a=9.2$ and for $Re_b \in [100; 1200]$. They showed that the corner equilibrium position appears only at relatively high Reynolds number ($Re_b\textgreater250$). 
These results were later confirmed by Nakagawa et al.~\cite{nakagawa2015}, who studied numerically the migration of a rigid 
sphere in duct flow, in a range of bulk Reynols number $Re_b$ from 20 to 1000. In particular, these authors show that the equilibrium 
position at the duct corner is unstable until the bulk Reynolds number $Re_b$ exceeds a critical value ($Re_b\approx260$). At this $Re_b$, 
additional equilibrium positions are shown to appear on the heteroclinic orbits close to the corners. Finally, in a recent paper Lashgari 
et al.~\cite{lash2017} performed numerical simulations to study the inertial migration of oblate particles in squared and rectangular ducts. 

Despite a considerable amount of studies on particulate duct flows, the physical understanding of the effects observed is not complete and the range 
of parameters still unexplored is vast. For example, as said before, most experimental and numerical studies focused on dilute suspensions of particles while the flow at 
higher solid volume fractions has not yet been investigated thoroughly, a relevant aspect for high-throughput applications. Therefore, the main goal of the present study is to fill this gap by exploring 
particle and flow behavior in a square duct at relatively high particle concentrations covering the range 
of solid volume fraction $\phi=0.4-20\%$. To this end, we perform interface-resolved numerical 
simulations using an immersed boundary method with lubrication and collision models for short-range interactions. 
We report the spatial distribution of particles across the duct cross section at constant bulk Reynolds number $Re_b=550$ for $\phi=0.4-20\%$.
Overall, we observe that particles depart from the duct core region and accumulate around the duct walls, 
and preferentially in the duct corners. 
In addition, we investigate the effects of the bulk and particle Reynolds 
numbers and the duct to particle size ratio on the behavior of a dilute suspension with $\phi=0.4\%$. The particular focusing positions depend mostly on the 
bulk Reynolds number $Re_b$. Changing the duct to particle size ratio, $h/a$, particle inertia varies independently of 
fluid inertia in each case and this leads to different specific arrangements of 
particles around the equilibrium positions. Moreover, a peculiar  
concentration distribution is observed for $\phi=5\%$ in a duct with $Re_b=144$ and $h/a=9$. 
Although we expected particles to accumulate at the walls and preferentially at the wall centers,  
the concentration distribution is found to be higher around the corners. 
This indicates that the particle distribution at higher volume fractions is affected by excluded volume effects.
Finally, we show that the presence of particles alters the flow in such a way that cross-stream secondary vortices 
appear around the particle focusing positions at low solid volume fraction, $\phi=0.4\%$. 
The intensity of these secondary flows depends on the maximum concentration of particles 
at these locations. For semi-dilute suspensions ($\phi \ge 5\%$), the presence of particles induces 
a pair of cross-stream secondary vortices at the duct corners. At high solid volume fraction ($\phi=20\%$), the duct core region is never 
fully depleted of particles and the intensity of secondary flows is substantially reduced. 
Overall, the secondary flow intensity initially increases with $\phi$ and then the decreases for $\phi > 5\%$.
We will also show that particle rotation plays an important role in determining the focusing positions as well as the intensity of the secondary flows.
 
\section{Methodology}
\subsection{Numerical method}

In this study, the Immersed Boundary method (IBM) proposed by Breugem~\cite{breugem2012} 
has been used to simulate dilute and semi-dilute suspensions of neutrally buoyant spherical particles in square ducts.\\
The flow field is described on a Eulerian grid by the incompressible Navier-Stokes equations:
\begin{equation}
\label{div_f}
\div \vec u_f = 0
\end{equation}
\begin{equation}
\label{NS_f}
\pd{\vec u_f}{t} + \vec u_f \cdot \grad \vec u_f = -\frac{1}{\rho_f}\grad p + \nu \grad^2 \vec u_f + \vec f
\end{equation}

where $p$ and $\vec u_f$ are the pressure and velocity fields, while $\nu$ and $\rho_f$ are the kinematic viscosity and 
density of the fluid phase. The last term on the right hand side of equation~(\ref{NS_f}), $\vec f$, is the IBM force field imposed 
to the flow to model the boundary condition at the moving particle surface (i.e. $\vec u_f|_{\partial \mathcal{V}_p} = \vec u_p + \vec \omegab_p \times \vec r$). The dynamics of the rigid particles is governed by the Newton-Euler Lagrangian equations:
\begin{align}
\label{lin-vel}
\rho_p V_p \td{\vec u_p}{t} &= \oint_{\partial \mathcal{V}_p}^{} \vec \taub \cdot \vec n\, dS\\
\label{ang-vel}
I_p \td{\vec \omegab_p}{t} &= \oint_{\partial \mathcal{V}_p}^{} \vec r \times \vec \taub \cdot \vec n\, dS
\end{align}
where $\vec u_p$ and $\vec \omegab_p$ are the linear and angular velocities of the particle centroid. 
In equations~(\ref{lin-vel}) and (\ref{ang-vel}), $V_p = 4\upi a^3/3$ and 
$I_p=2 \rho_p V_p a^2/5$ represent the particle volume and moment of inertia, 
$\vec \taub = -p \vec I + 2\nu\rho_f \left(\grad \vec u_f + \grad \vec u_f^T \right)/2$ 
is the fluid stress tensor while $\vec r$ indicates the distance from the center of the particles ($\bf{n}$ 
is the unity vector normal to the particle surface $\partial \mathcal{V}_p$). 

The fluid phase is evolved entirely on a uniform staggered Cartesian grid using a second-order finite-difference scheme. 
An explicit third order Runge-Kutta scheme has been combined with a pressure-correction method to perform the time integration at each sub-step. 
This latter integration scheme has also been used for the 
evolution of eqs.~(\ref{lin-vel}) and (\ref{ang-vel}). 
Each particle surface is described by $N_L$ uniformly distributed Lagrangian points. 
The force exchanged by the fluid on the particles is imposed on each $l-th$ Lagrangian point. 
This force is related to the Eulerian force field
$\vec f$ by the expression $\vec f(\vec x) = \sum_{l=1}^{N_L} \vec F_l \delta_d(\vec x - \vec X_l) \Delta V_l$, where $\Delta V_l$ is the volume of the cell 
containing the $l-th$ Lagrangian point while $\delta_d$ is the Dirac delta. 
Here, $\vec F_l$ is the force (per unit mass) at each Lagrangian point, and it is computed as 
$\vec F_l=(\vec U_p(\vec X_l)-\vec U_{l}^*)/\Delta t$, where $\vec U_p=\vec u_p + \vec \omegab_p \times \vec r$ is the velocity at the 
lagrangian point $l$ at the previous time-step, while $\vec U_{l}^*$ is the interpolated first prediction velocity at the same point.

An iterative algorithm with second order global accuracy in space is employed to calculate 
this force field. To maintain accuracy, eqs.~(\ref{lin-vel}) and (\ref{ang-vel}) are rearranged in terms of the IBM force field, 
\begin{align}
\label{lin-vel-ibm}
\rho_p V_p \td{\vec u_p}{t} &= -\rho_f \sum_{l=1}^{N_l} \vec F_l \Delta V_l + \rho_f \td{}{t} \int_{\mathcal{V}_p}^{} \vec u_f\, dV \\
\label{ang-vel-ibm}
I_p \td{\vec \omegab_p}{t} &= -\rho_f \sum_{l=1}^{N_l} \vec r_l \times \vec F_l \Delta V_l + \rho_f \td{}{t} \int_{\mathcal{V}_p}^{} \vec r \times \vec u_f\, dV 
\end{align}
being $\vec r_l$ the distance between the center of a particle and the $l-th$ 
Lagrangian point on its surface. 
The second terms on the right-hand sides are corrections 
that account for the inertia of the fictitious fluid contained within the particle volume. Particle-particle and particle-wall interactions are also considered. 
Well-known models based on Brenner's asymptotic solution~\cite{brenner1961} are employed to correctly predict 
the lubrication force when the gap distance between particles 
and between particles and walls is smaller than twice the mesh size. 
A soft-sphere collision model is used to account for particle-particle and particle wall collisions. 
An almost elastic rebound is ensured with a restitution coefficient set at $0.97$. 
Friction among particles and particles and walls is also considered~\cite{costa2015}.
These lubrication and collision forces are added to the right-hand side of eq.~(\ref{lin-vel-ibm}). 
A more detailed discussion of the numerical method and of the 
mentioned models can be found in previous publications~\cite{breugem2012,picano2015,fornari2016JFM,lashg2016,forn2016}.
Periodic boundary conditions are imposed in the streamwise direction. 
In the remaining directions, the \emph{stress immersed boundary method} 
is used to impose the no-slip/no-penetration conditions at the walls. 
The stress immersed boundary method has originally been developed to simulate the flow 
around rectangular-shaped obstacles in a fully Cartisian grid~\cite{breugem2014}. 
In this work, we use this method to enforce the fluid velocity to be 
zero at the duct walls. For more details on the method, 
the reader is referred to the works of Breugem and Boersma~\cite{breugem2005} and Pourquie 
et al.~\cite{pourquie2009}.

\subsection{Flow configuration}
\begin{figure}[h!]
\centering
\includegraphics[width=0.65\textwidth]{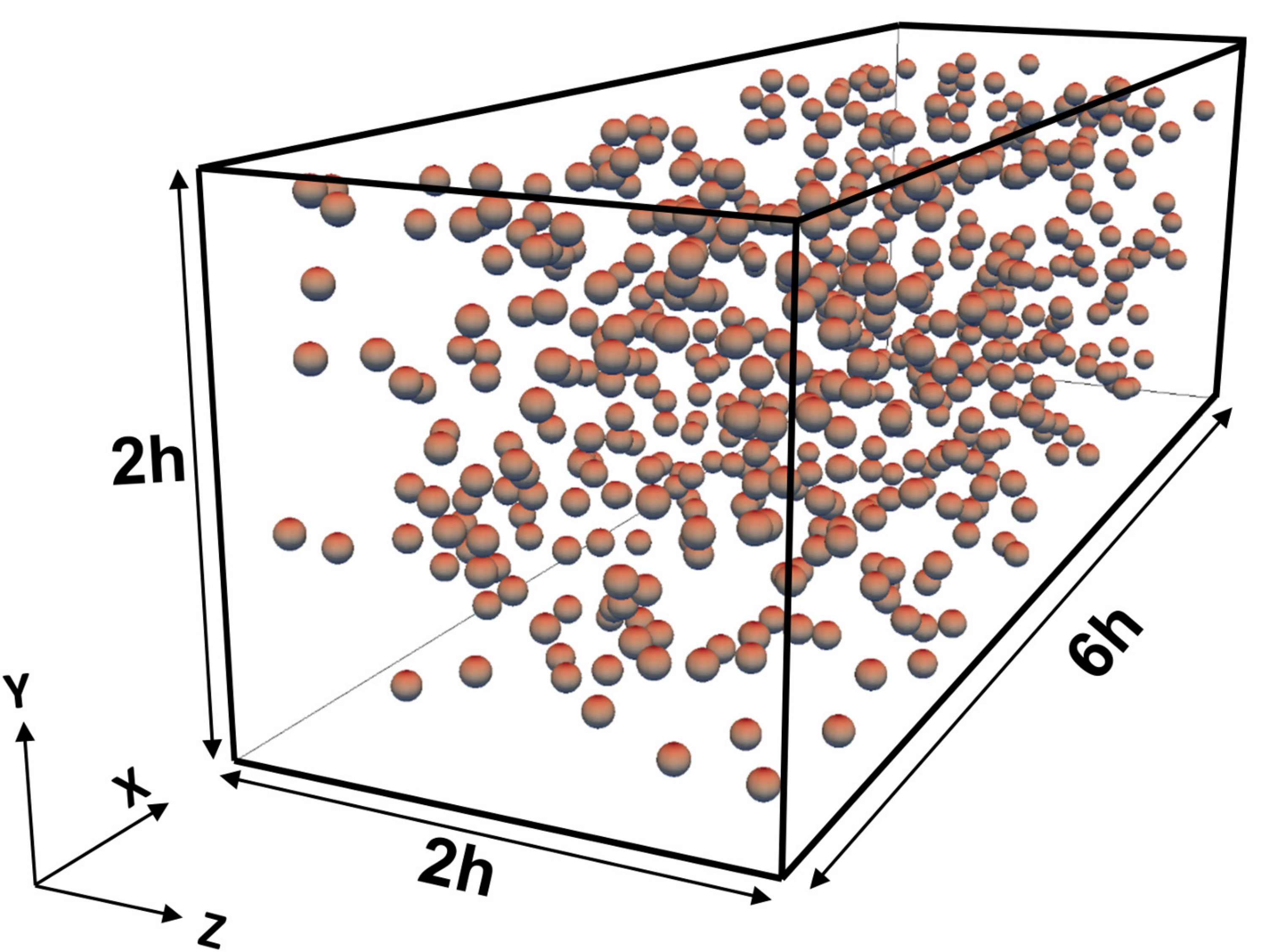}
\caption{3D view of duct geometry with few particles }
\label{fig:Snap_shot}
\end{figure}

In this work, we investigate the laminar flow of dilute and semi-dilute suspensions of 
neutrally buoyant spherical particles in straight ducts with square cross section.\\ 
Two different sets of simulations are performed. Initially we study excluded volume effects. 
To this aim we perform simulations in a Cartesian computational domain of size $L_x=6h$, $L_z=2h$ and $L_y=2h$ 
where $h$ is the duct half-width and $x$, $y$ and $z$ are the streamwise and cross-stream directions (see Fig.~\ref{fig:Snap_shot}).

The domain is discretized by a uniform ($ \Delta x = \Delta z = \Delta y$) 
cubic mesh with $1296$ $\times$ $432$ $\times$ $432$ grid points for semi-dilute cases. In terms of particle 
radii, the computational domain has a size of $108a\times36a\times36a$ with $a$ being the particle radius. A 
constant bulk velocity $U_b$ is achieved by imposing a mean pressure gradient in the streamwise direction. 
Bulk and particle Reynolds number are here defined as $Re_b=U_b2h/\nu$ and $Re_p=Re_b\left(a/h\right)^2$. 
We consider four different solid volume fractions: $\phi=0.4$, $5$, $10$, and $20\%$. In this setup these correspond 
to $134, 1670, 3340$ and $6680$ particles. In all cases, particles are initially positioned randomly in the computational 
domain with zero linear and angular velocities. Each particle is discretized with $N_l=1721$ Lagrangian control points 
while their radii are $12$ Eulerian grid points long. 
Considering $12$ grid points per particle radius ($\Delta x = 1/24$) is a good compromise in terms of computational cost and accuracy.\\
In the second part of the paper, we investigate the effects of bulk and particle Reynolds numbers 
and the duct to particle size ratio $h/a$ for dilute suspensions with a $\phi=0.4\%$. We consider different 
combinations of $Re_b$ and $h/a$ resulting in different particle 
Reynolds numbers $Re_p=Re_b\left(a/h\right)^2$.
The full set of simulations is summarized in table~\ref{tab:sim}. 
Resolution is chosen to keep 12 grid points per particle radius for the different $h/a$ ratios.

\begin{table}[h!]
  \caption{Summary of the simulations. The size of the computational domain is expressed in terms of particle radii and is denoted by $L_x, L_y, L_z$ in the 
streamwise and wall-normal directions. The number of grid points in each direction, $N_x, N_y, N_z$, is chosen to keep $12$ points per particle radius.\\}
  \label{tab:sim}
  \begin{center}
\def~{\hphantom{0}}
  \begin{tabular}{cccccc}\hline\hline 
     $\phi ($\%$)$   & $Re_b$  &  $Re_p$     & $\left(h/a\right)$   & $L_x \times L_y \times L_z$       &  $\,$ $\,$ $N_x\times N_y\times N_z$\\[3pt]\hline
     $0.4$           & $144$   &  $1.7$      & $9$                  & $144a\times18a\times18a$          &  $\,$ $\,$ $1728$ $\times$ $216$ $\times$ $216$\\
     $0.4$           & $275$   &  $3.4$      & $9$                  & $72a\times18a\times18a$           &  $\,$ $\,$ $864$ $\times$ $216$ $\times$ $216$\\
     $0.4$           & $550$   &  $6.8 $     & $9$                  & $72a\times18a\times18a$           &  $\,$ $\,$ $864$ $\times$ $216$ $\times$ $216$\\
     $0.4$           & $300$   &  $1.7$      & $13$                 & $78a\times26a\times26a$           &  $\,$ $\,$ $936$ $\times$ $312$ $\times$ $312$\\
     $0.4$           & $550$   &  $3.2$      & $13$                 & $78a\times26a\times26a$           &  $\,$ $\,$ $936$ $\times$ $312$ $\times$ $312$\\
     $0.4$           & $550$   &  $1.7$      & $18$                 & $108a\times36a\times36a$          &  $\,$ $\,$ $1296$ $\times$ $432$ $\times$ $432$\\           
     $5$             & $144$   &  $1.7$      & $9$                  & $72a\times18a\times18a$           &  $\,$ $\,$ $864$ $\times$ $216$ $\times$ $216$\\
     $5$             & $550$   &  $1.7$      & $18$                 & $108a\times36a\times36a$          &  $\,$ $\,$ $1296$ $\times$ $432$ $\times$ $432$\\
     $10$            & $550$   &  $1.7$      & $18$                 & $108a\times36a\times36a$          &  $\,$ $\,$ $1296$ $\times$ $432$ $\times$ $432$\\
     $20$            & $550$   &  $1.7$      & $18$                 & $108a\times36a\times36a$          &  $\,$ $\,$ $1296$ $\times$ $432$ $\times$ $432$\\ \hline\hline 
\end{tabular}
  \end{center}
\end{table}

\section{Results}
\subsection{Validation}
The code has been validated extensively against several test cases in previous studies~\cite{picano2015,fornari2016JFM,breugem2012}.
In this study, first, we investigate how accurate it is to use the stress immersed boundary 
method to represent the duct walls. In particular we compare our results on the mean flow to the analytic result reported by Shah 
and London~\cite{shah2014}. The maximum discrepancy is found at the centerline and it is about $0.6\%$ for the resolution used in this study.\\
We then perform a validation against the experimental results reported recently by Miura et al.~\cite{miura2014} 
on the flow of dilute suspensions of neutrally buoyant spherical particles in a square duct. 
We perform a simulation to resemble the case presented in Fig.~$5(a)$ of Ref.~\cite{miura2014}. 
In particular, we consider a box of size $L_x=144a$, $L_y=L_z=18a$, a bulk Reynolds 
number $Re_b$ of $144$, duct half-width to particle radius ratio $h/a=9$  
and volume fraction $\phi$=$0.4\%$. 
\begin{figure}[t!]
   \centering
\includegraphics[width=0.492\textwidth]{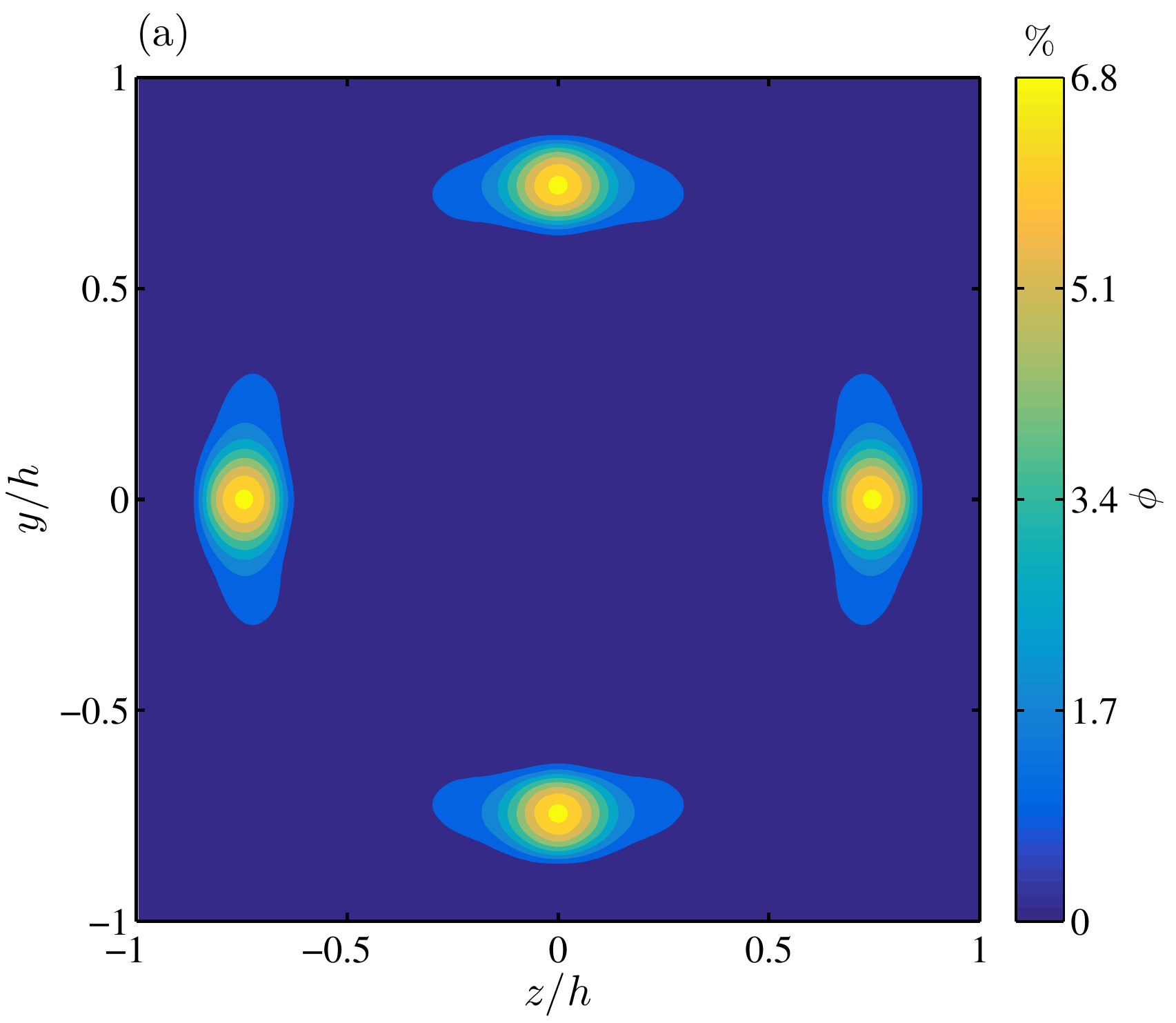}
\includegraphics[width=0.375\textwidth]{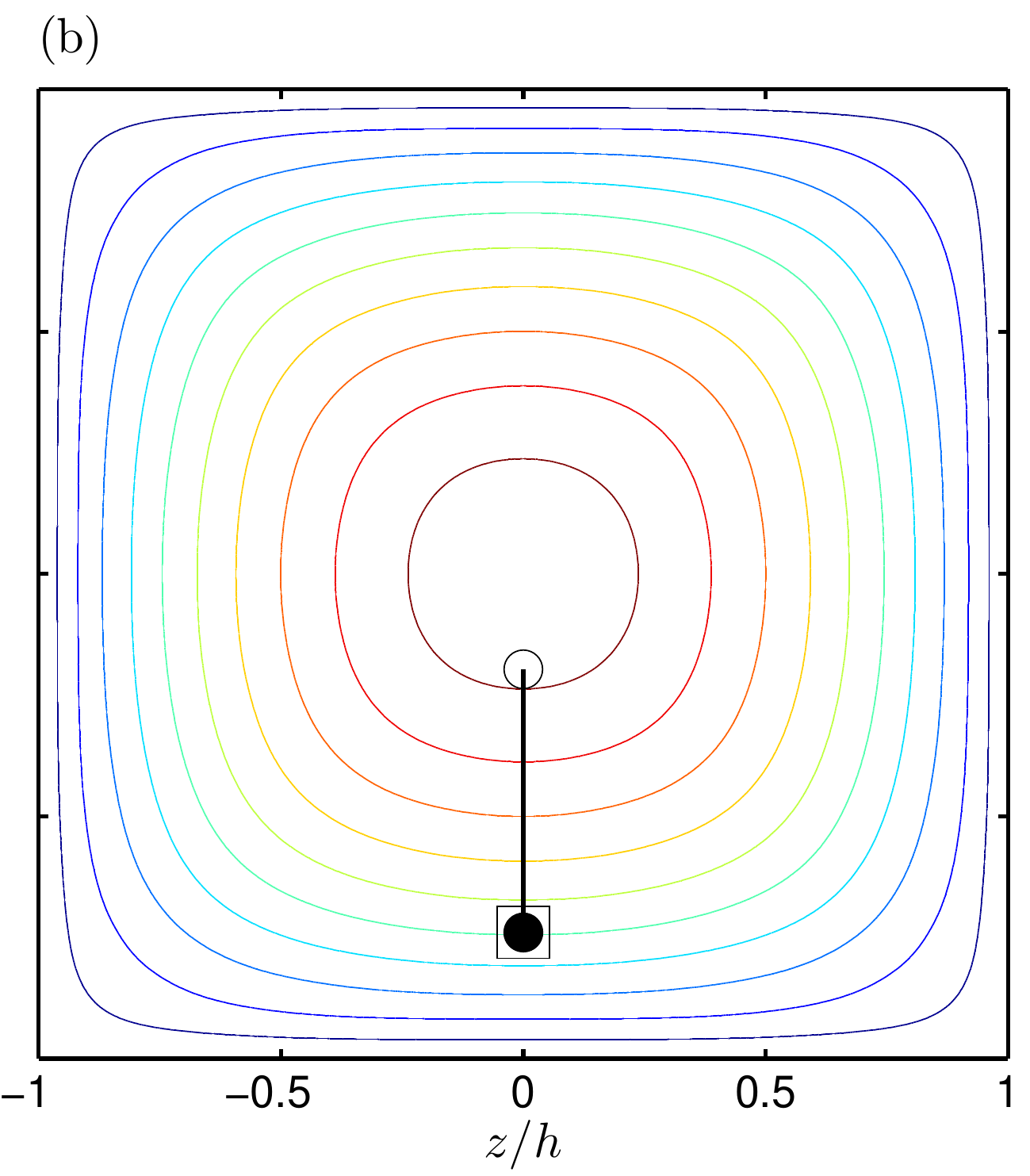}
\caption{$(a)$~Spatial distribution of particles across the duct section for $\phi=0.4\%$ and $Re_b=144~(Re_p=1.7)$. $(b)$ Equilibrium position of 
individual particle. \textcolor{black}{$\circ$}~initial position, $\bullet$~final postition, $ \square $~Chun and Ladd~\cite{chun2006} 
final equilibrium position, --- particle trajectory.}
\label{fig:Validation}
\end{figure}
After an initial transient, we compare the particle distribution across 
the duct to those found experimentally. Figure~\ref{fig:Validation}$(a)$ 
shows the particle concentration $\Phi(y,z)$ in the $(y-z)$ plane (averaged in the streamwise 
direction and over time). Excellent 
agreement can be seen between our numerical results and the experimental data 
of Miura et al.~\cite{miura2014} (see Fig.~$5(a)$ of the cited paper).\\
The dependence of the particles equilibrium position on the 
computational domain length is also checked. 
To this end, we perform a simulation in a shorter box, $L_x=72a$, and same $Re_b$ and $\phi$. The same final particle 
equilibrium position is found (not shown). 
Thus, we conclude that the results are independent of the box length for the values here considered.\\
In addition, we also examine the trajectory of an individual 
particle and compare it with the data previously reported by Chun and Ladd~\cite{chun2006}. 
We assume $Re_b=100$ and $h/a=9.1$ and compare our and their results in 
Fig.~\ref{fig:Validation}$(b)$. The particle, initially slightly below the duct centerline, slowly migrates towards the 
same focusing position found by Chun and Ladd~\cite{chun2006}. In particular, 
the focusing position is at the center of the duct wall ($z/h=0$) at a distance of $0.74h$ from the centerline.

\subsection{Semi-dilute suspensions} \label{Semi-dilute suspensions}

\begin{figure}
\centering
\includegraphics[width=0.315\textwidth]{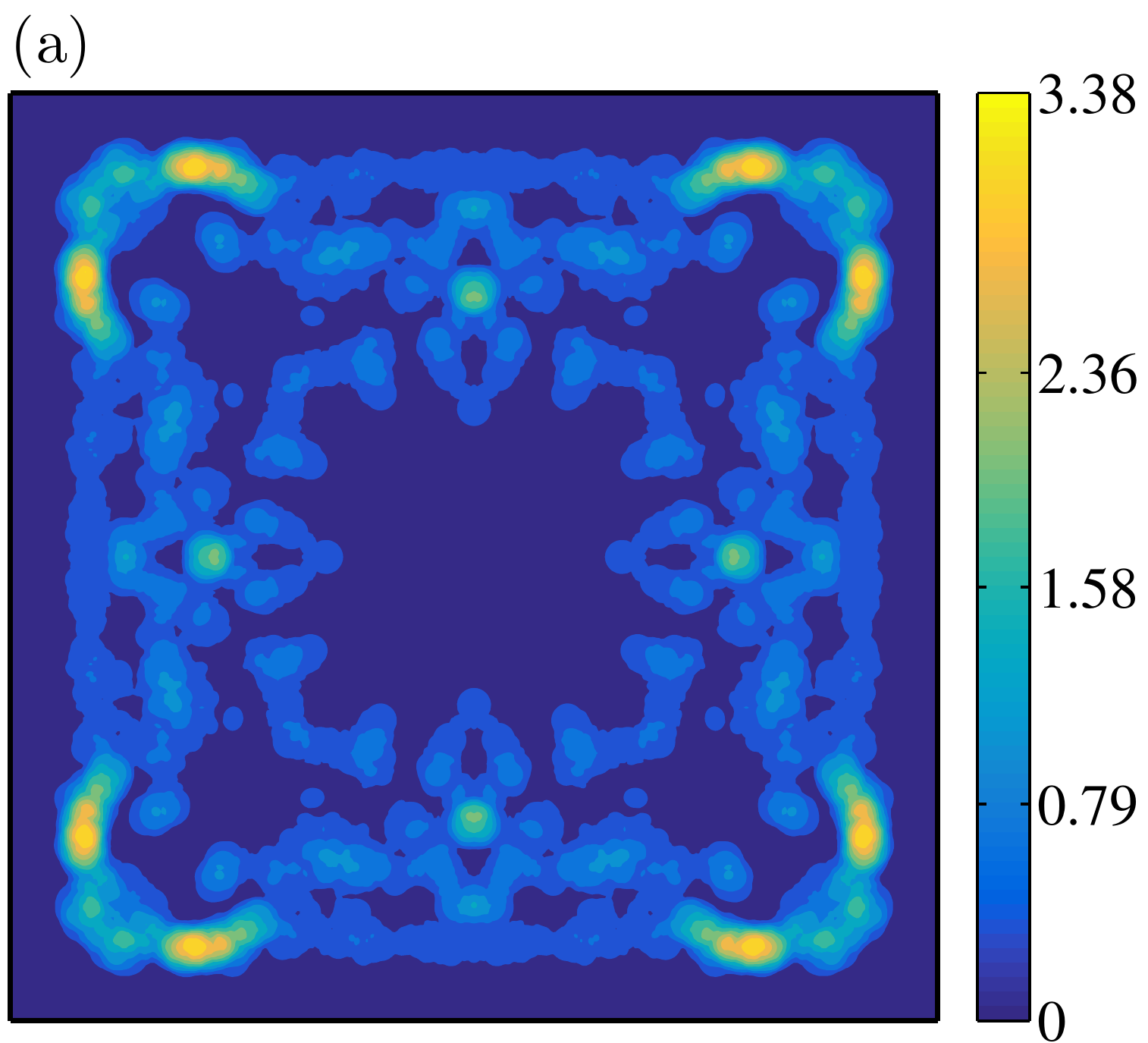}
\includegraphics[width=0.32\textwidth]{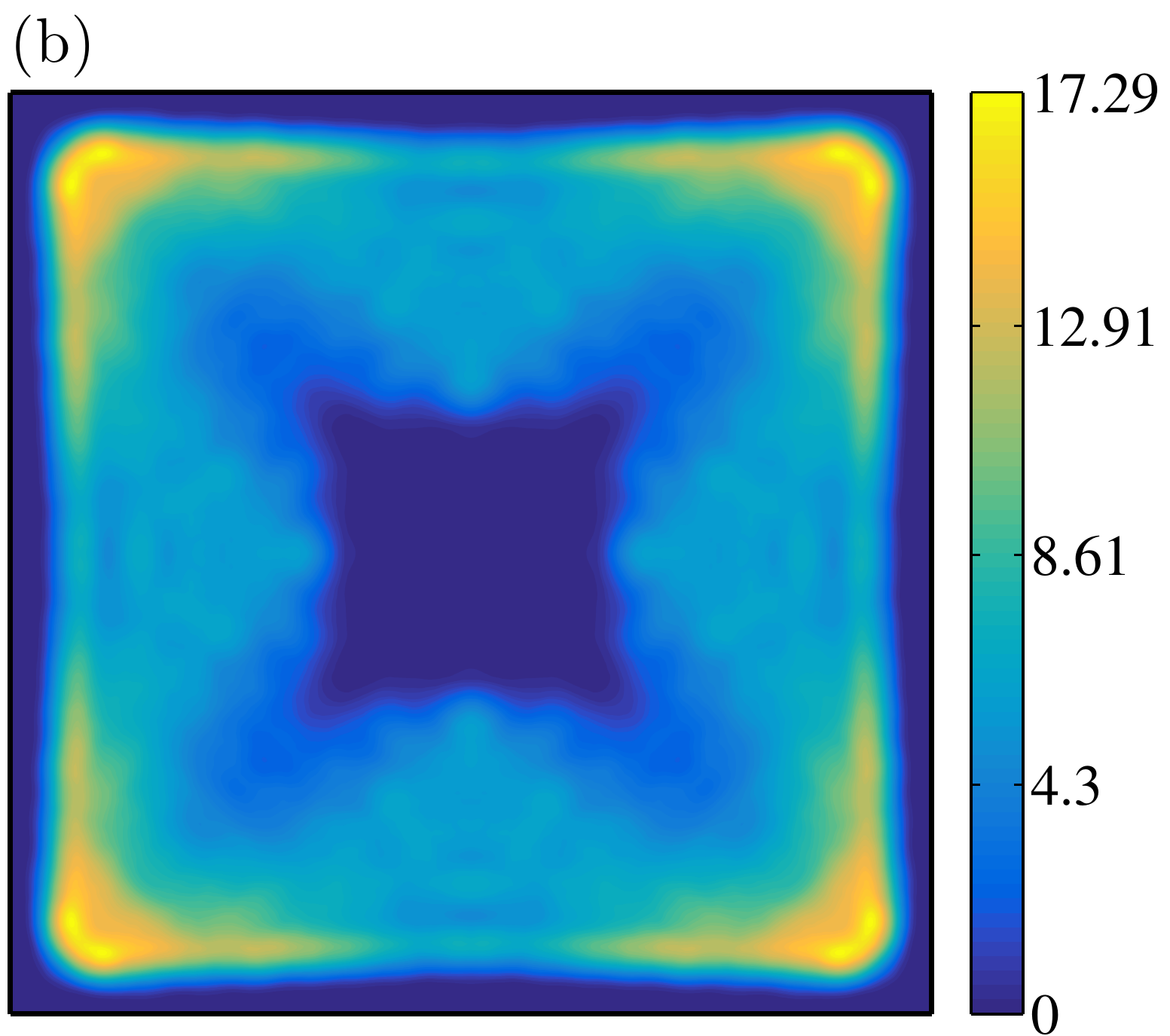}
\includegraphics[width=0.34\textwidth]{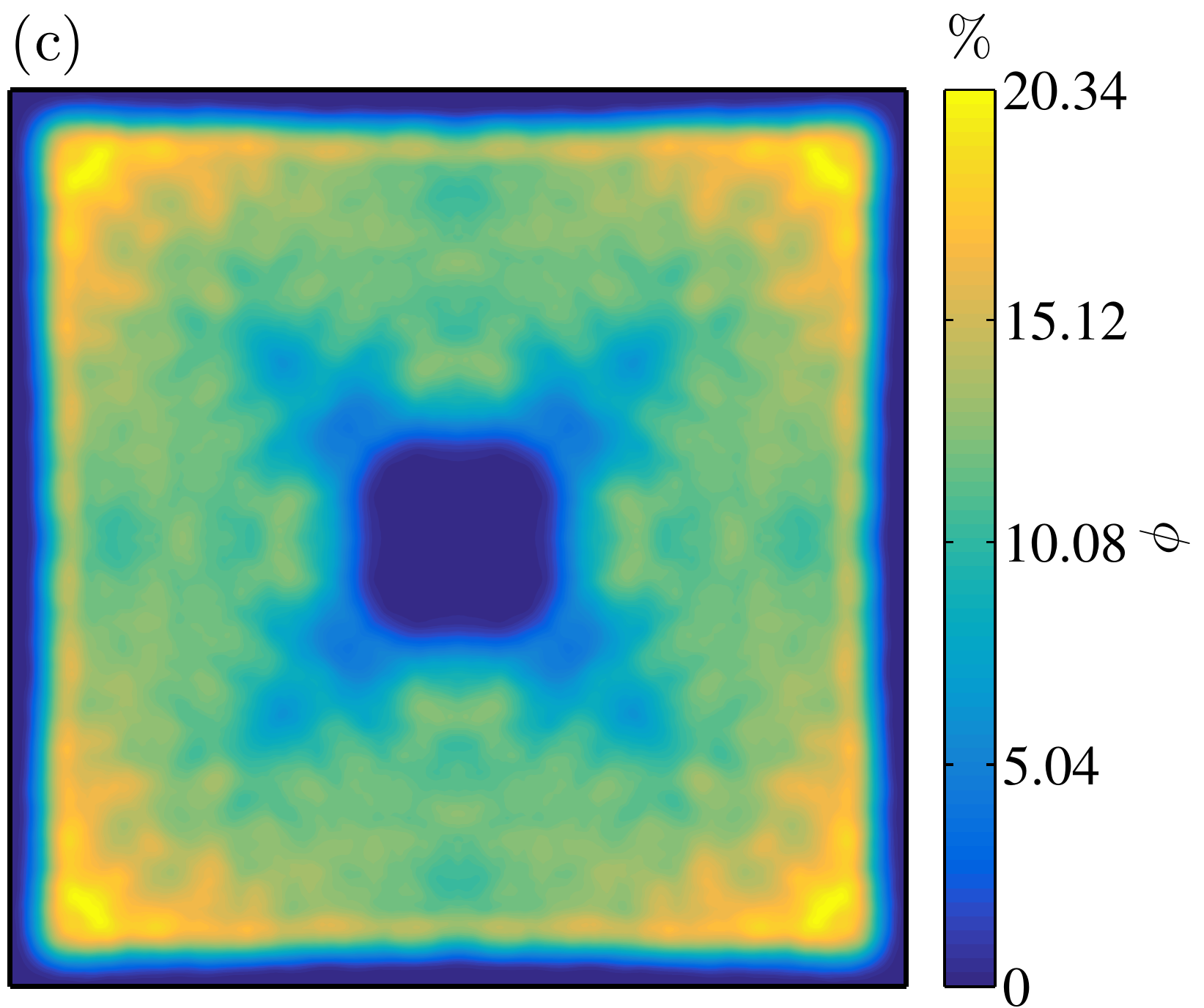}\vspace{0.5cm}

\includegraphics[width=0.32\textwidth]{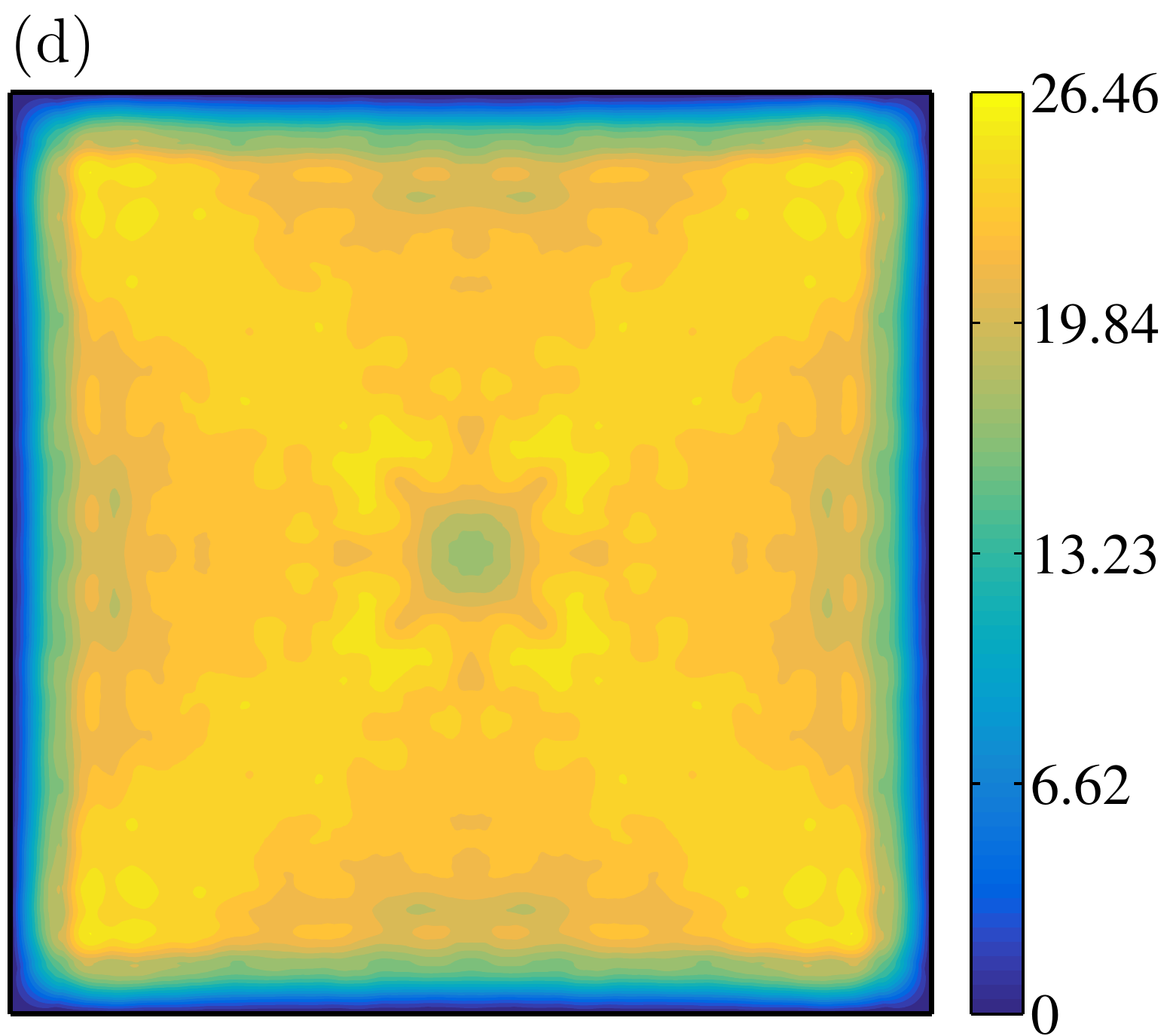}
\includegraphics[width=0.32\textwidth]{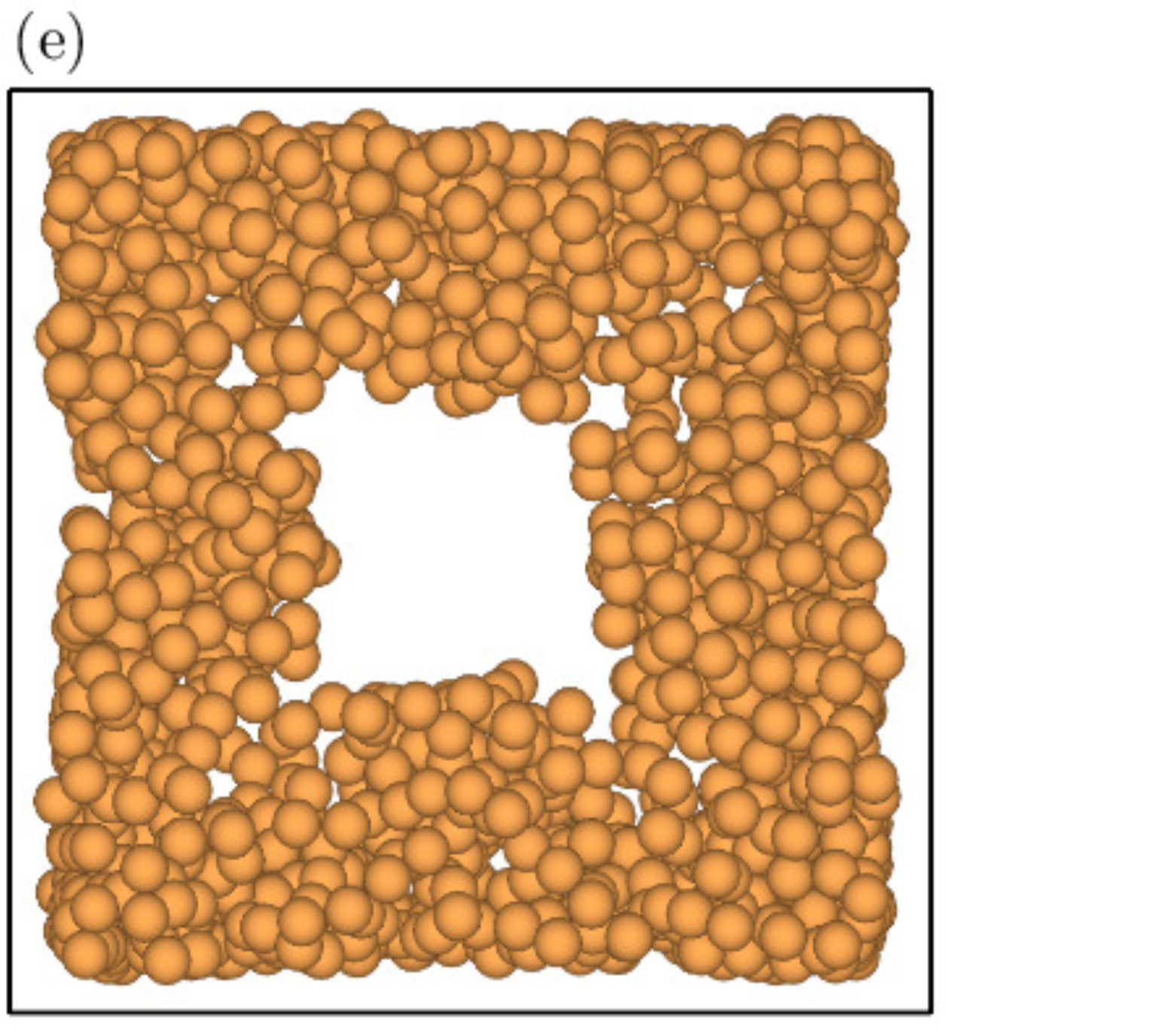}
\includegraphics[width=0.34\textwidth]{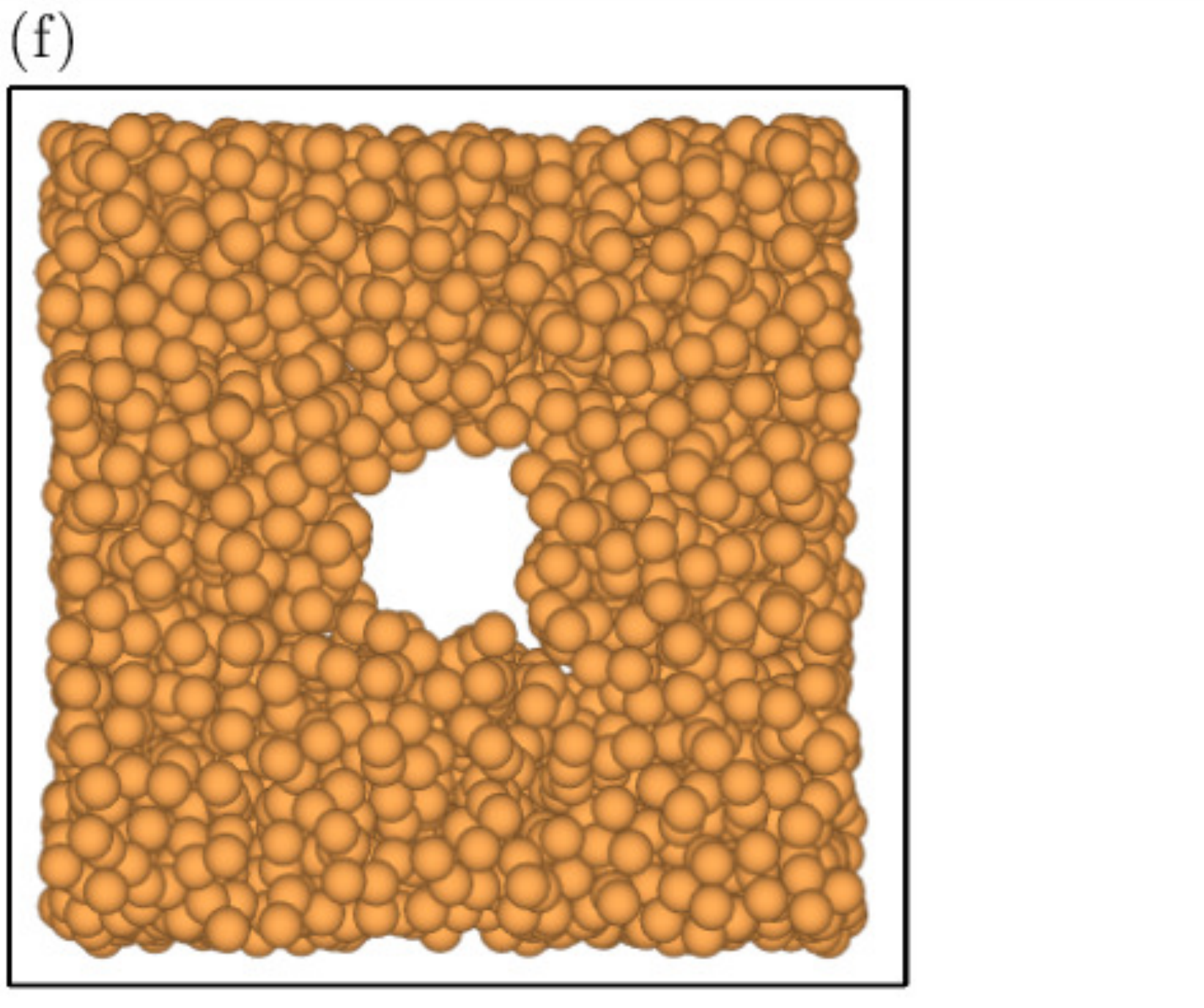}
\caption{Particle concentration distribution $\Phi(y,z)$ in the $(y-z)$ plane (duct cross-section) at $Re_b=550~(Re_p=1.7)$ and $h/a=18$: 
$(a)$ $\phi=0.4\%$, $(b)$ $\phi=5\%$, $(c)$ $\phi=10\%$, $(d)$ $\phi=20\%$, $(e)$ and $(f)$ show instantaneous snapshot of particle final positions in the $(y-z)$ 
plane for $\phi=5\%$ and $\phi=10\%$ respectively.}
\label{fig:Conc}
\end{figure}

In this section we report and discuss the results obtained 
for different solid volume fractions $\phi$ at a constant bulk Reynolds number of 
$Re_b=550~(Re_p=1.7)$ and duct to particle size ratio of $h/a=18$. Note that all 
results shown hereinafter are obtained by taking averages over the eight symmetric 
triangles that form the duct cross section.
We first show the particle concentration distribution $\Phi(y,z)$ across the duct cross section for 
$\phi= 0.4, 5, 10$ and $20\%$ in Figs.~\ref{fig:Conc}$(a)-(d)$. The particle concentration distribution is defined as
\begin{equation}
\Phi(y,z) = \frac{1}{N_t N_x} \sum_{m=1}^{N_t}\sum_{i=1}^{N_x} \xi(x_{ijk},t^m)
\end{equation}
where $N_t$ is the number of time-steps considered for the average, $t^m$ is the sampling time, and $\xi(x_{ijk},t^m)$ is 
the particle indicator function at the location $x_{ijk}$ and time $t^m$. The particle indicator function is equal to $1$ 
for points $x_{ijk}$ contained within the volume of a sphere, and $0$ in the fluid.\\
At the lowest solid volume fraction, $\phi=0.4\%$, particles migrate preferentially towards two symmetric equilibrium 
positions near the duct corners (Fig.~\ref{fig:Conc}$a$). However, some particles are also found uniformly distributed along the walls and 
in particular close to the wall centers at a distance of approximately $0.6h$ away from the duct core. 
Figures \ref{fig:Conc}$(b)$ and $(c)$ show that the particles tend to concentrate preferentially at the duct 
corners and less at the wall centers also for $\phi=5$ and $10\%$. For these three volume fractions, the core of the duct is 
completely depleted of particles. In Figs.~\ref{fig:Conc}$(e)$ and $(f)$ we 
show instantaneous snapshots of particle positions in the duct cross section for $\phi=5\%$ and $\phi=10\%$. At each instant, 
particles are always close to the duct walls. Hence, the particle concentration distribution $\Phi(y,z)$ 
at moderate $\phi$ appears to reflect the peculiar focusing positions obtained in dilute cases. However, at $\phi=20\%$, 
particles distribute almost all over the cross-section 
(see Fig.~\ref{fig:Conc}$d$). One can still notice that the particle concentration is slightly larger at the corners and that 
stable layers of particles form close to the walls (as also found for channel flows~\cite{lashg2016}). 
Qualitatively similar results have been obtained for suspensions of neutrally buoyant spheres in pipe flows at solid volume 
fractions $\phi=6, 10\%$ and $20\%$ ~\citep{han1999}. It was shown that for $\phi=6$ and $10\%$ and $Re_p\approx0.35$, particles 
migrate from the core region towards the pipe wall. At $\phi=20\%$, particles are uniformly distributed in the cross-section 
with maximum of the particle concentration at the pipe center and close to the wall. The maximum of local particle concentration at the pipe center 
for $\phi=20\%$ is not present in our results for duct flow, see figure~\ref{fig:Conc}$(d)$. The reason for this difference is 
probably the fact that the particle Reynolds number is larger ($Re_p=1.7$) 
than in the cited experiments ($Re_p=0.28$).

Despite different particle distributions across the duct cross section, residence times of particles at the duct wall center and 
duct corners are similar for $\phi=5$ and $20\%$. 
\begin{figure}[t!]
   \centering
\includegraphics[width=0.6\textwidth]{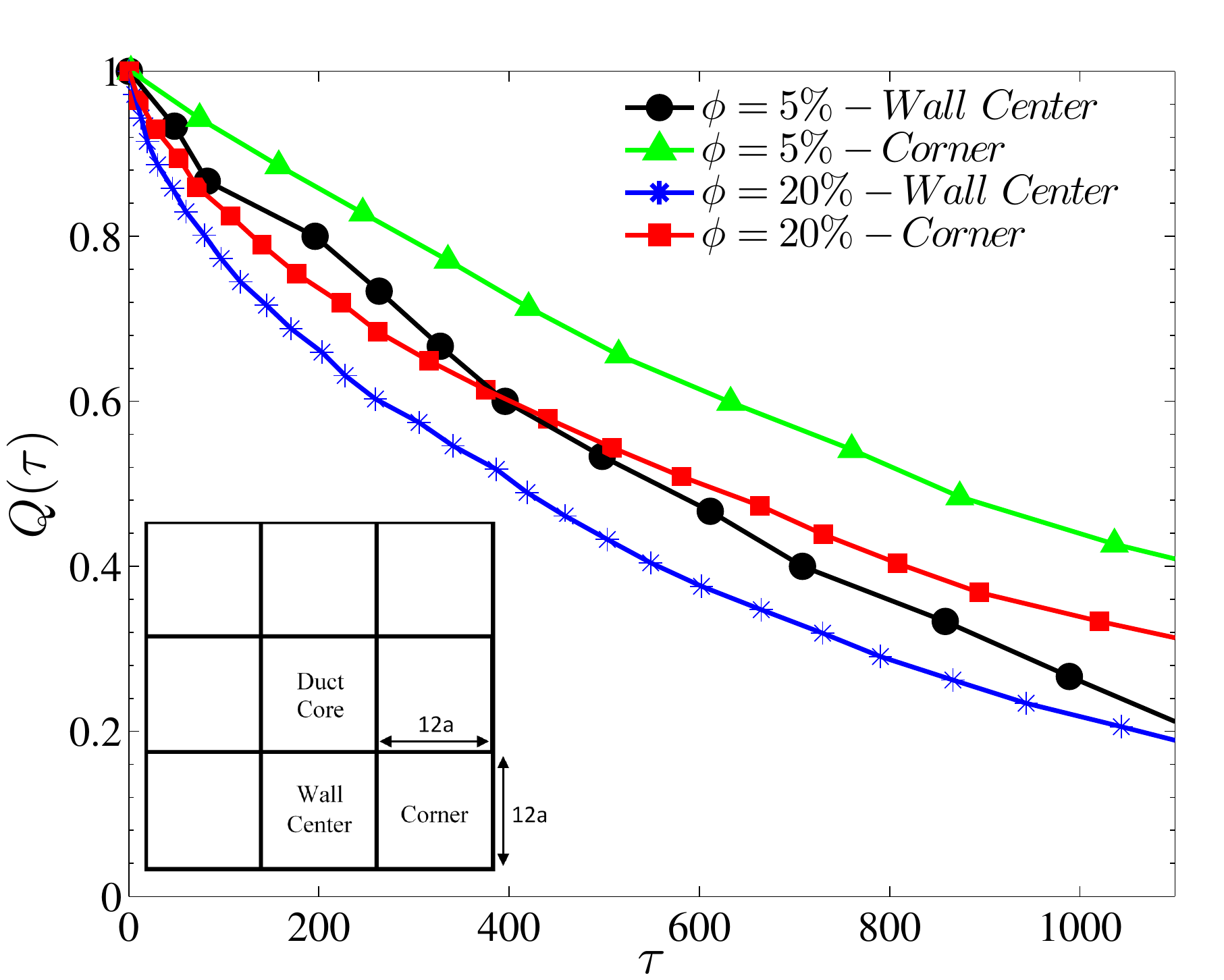}
\caption{Cumulative Density Function of residence times, Q($\tau$), in the corners or at the wall centers for $\phi=5$ and $20\%$ at $Re_b=550~(Re_p=1.7)$ and $h/a=18$. 
Inset: different regions of the duct for statistics evaluation.}
\label{fig:Cum}
\end{figure}
We demonstrate this in Fig.~\ref{fig:Cum} by calculating the cumulative probability density function $Q(\tau)$ of particles residence time $\tau$ in 
the corners or at the duct wall centers. We divide the computational domain in nine equal volumes 
of size $108a \times 12a \times 12a$: four of these volumes contain the duct corners 
while 4 contain the duct wall centers and the ninth the duct core (see inset of Fig.~\ref{fig:Cum}). The residence time $\tau$ is defined as the maximum time 
a particle stays within the boundaries of 1 specific volume. The cumulative probability density function $Q(\tau)$ is 
calculated via the rank-order method~\cite{mitra} which is free from binning errors. The statistics are collected using the 
last $1200$ nondimensional times. The results for $\phi=5$ and $20\%$ are shown in Fig.~\ref{fig:Cum} 
where we see that $Q(\tau)$ is larger at the corners than at the wall 
centers (for sake of clarity, 
the curves for $\phi=10\%$ are not reported). 

\begin{figure}[t!]
  \centering
\includegraphics[width=0.33\textwidth]{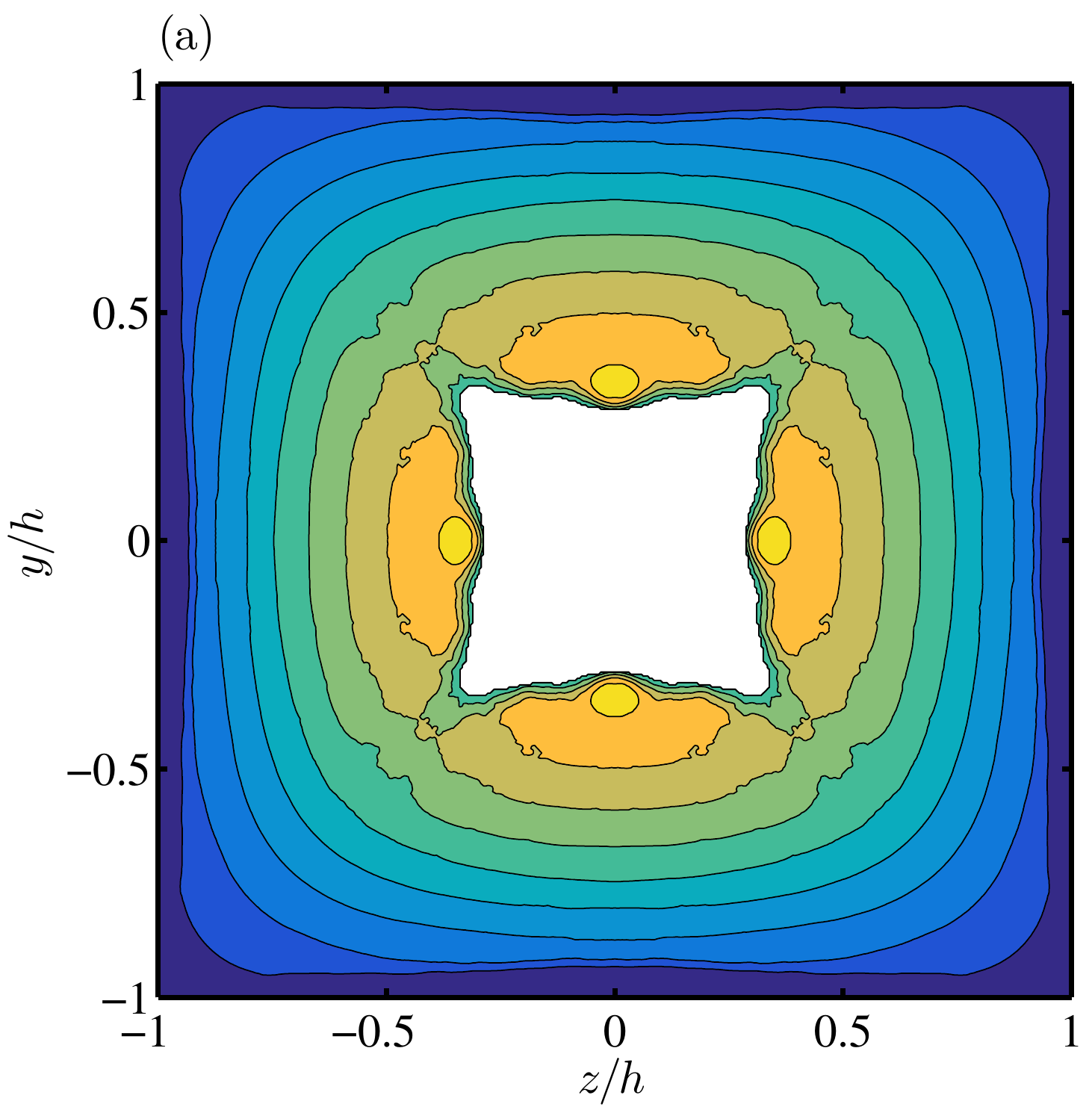}
\includegraphics[width=0.295\textwidth]{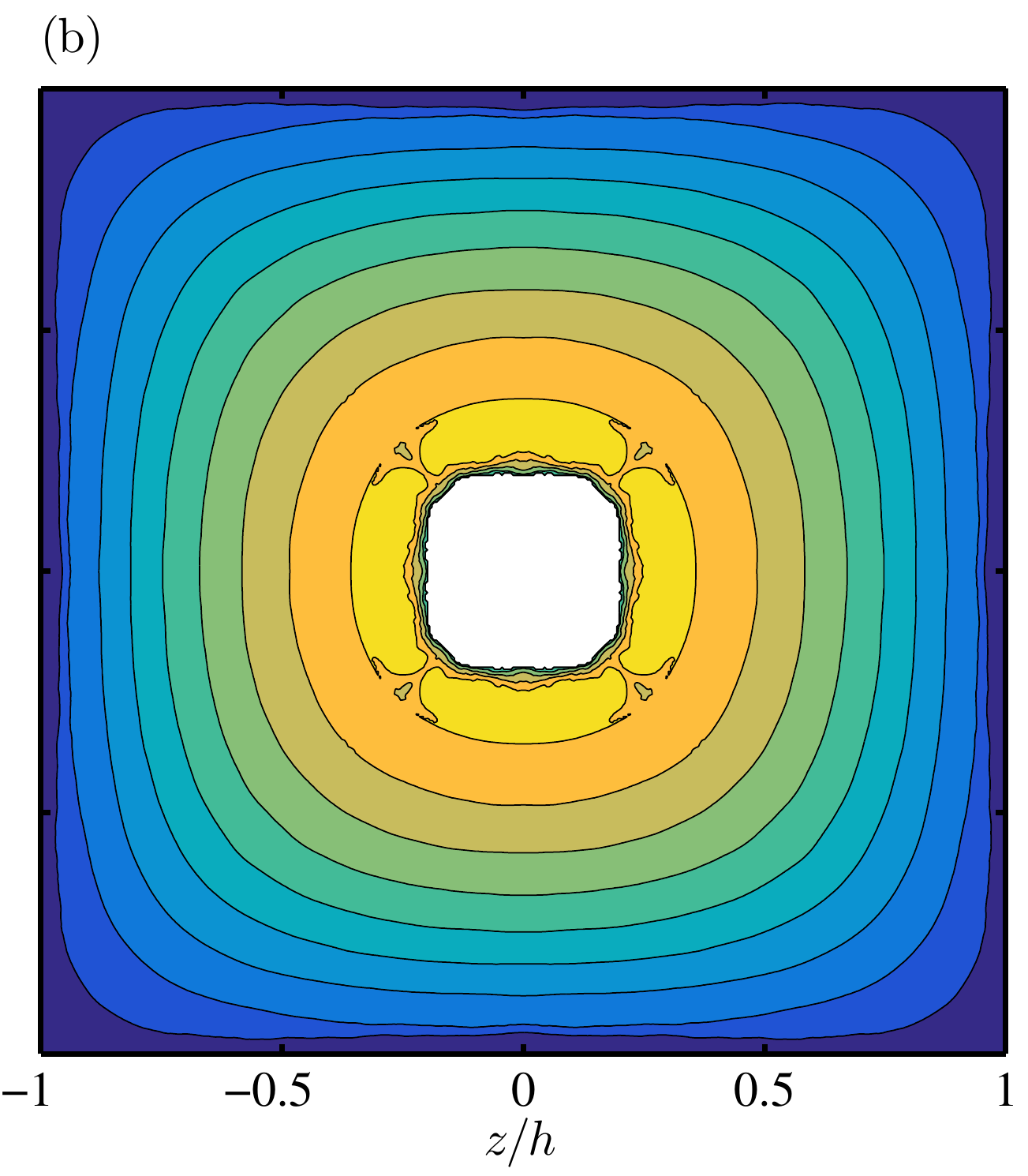}
\includegraphics[width=0.345\textwidth]{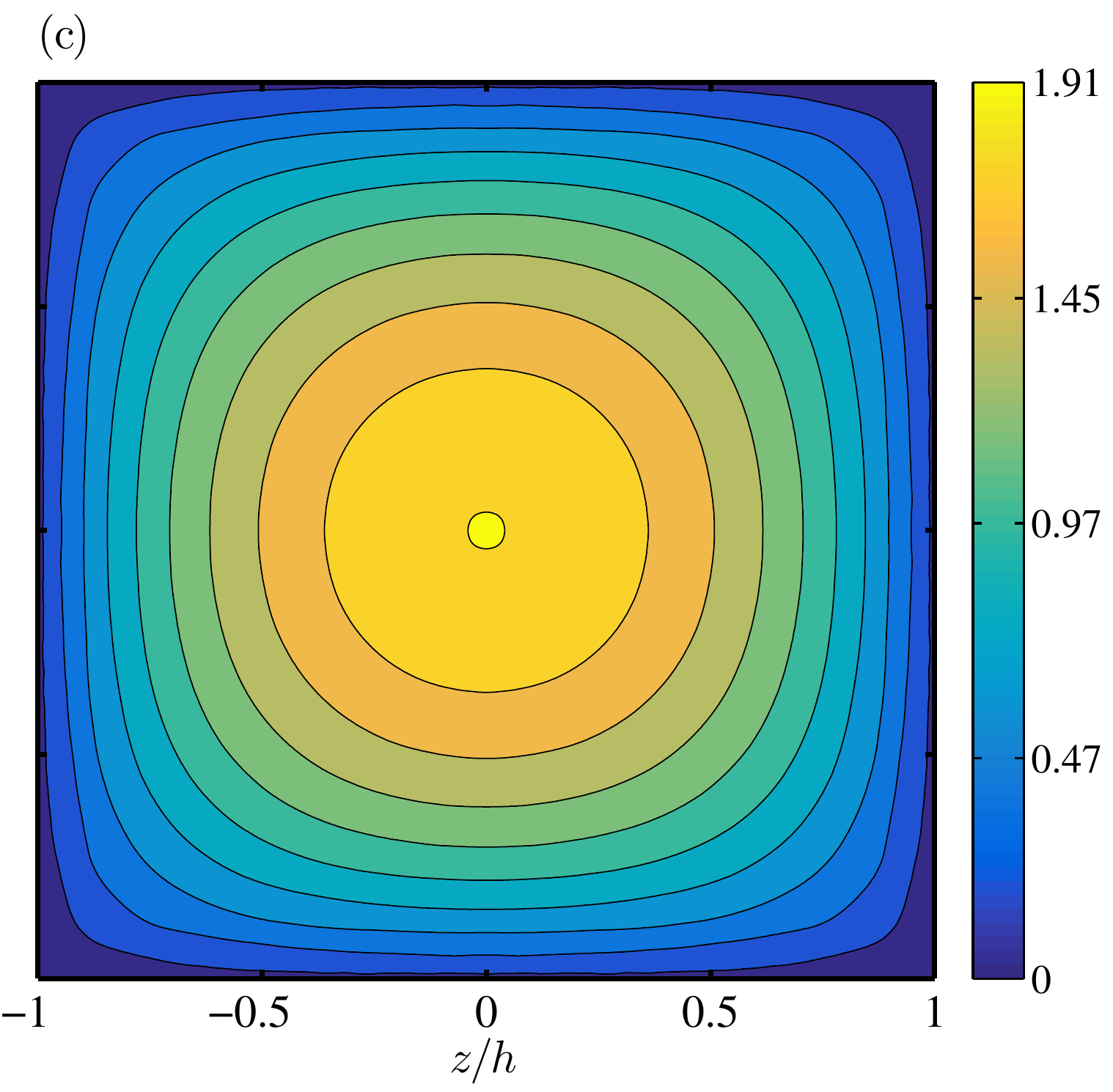}
\caption{ Streamwise mean particle velocity contours, $U_p/U_b$, in the cross-stream $(y-z)$ plane at $Re_b=550~(Re_p=1.7)$ and $h/a=18$ for: $(a)$ $\phi=5\%$, $(b)$ $\phi=10\%$, $(c)$ $\phi=20\%$.}
\label{fig:U_P}
\end{figure}

The streamwise mean particle 
velocity, $U_p(y,z)$, normalized by the bulk velocity $U_b$, is illustrated in Figs.~\ref{fig:U_P}$(a)-(c)$ over the duct cross section for the different volume fractions $\phi$. 
The contours resemble closely those of the streamwise fluid velocity except at the duct core where it is fully depleted of 
particles for $\phi=5$ and $10\%$. The uniform distribution of particles across the duct cross section for $\phi=20\%$ results instead 
in an uniform streamwise mean velocity contour in the duct core (Fig.~\ref{fig:U_P}$c$).\\
\begin{figure}[h!]
   \centering
\includegraphics[width=0.4\textwidth]{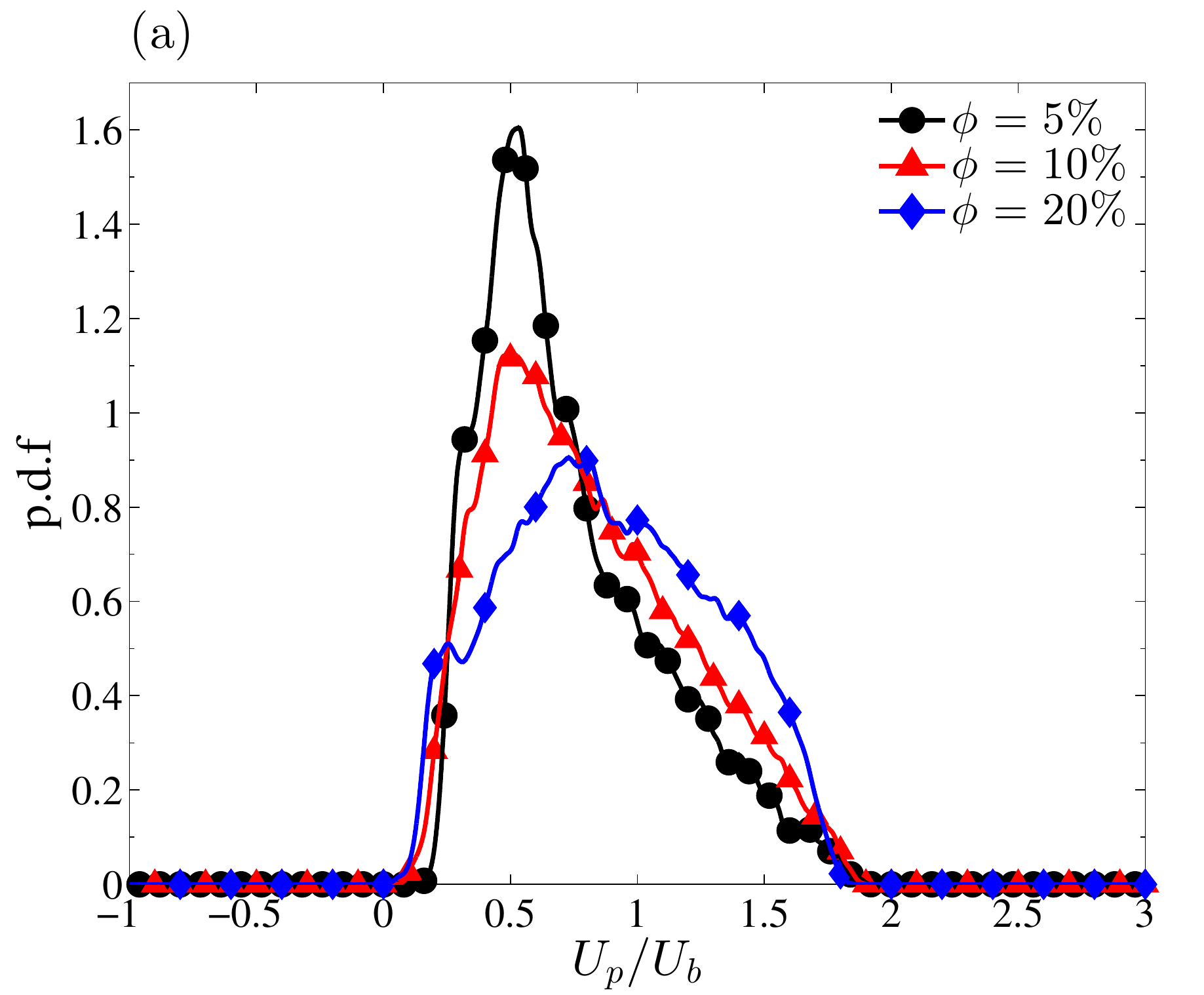}
\includegraphics[width=0.4\textwidth]{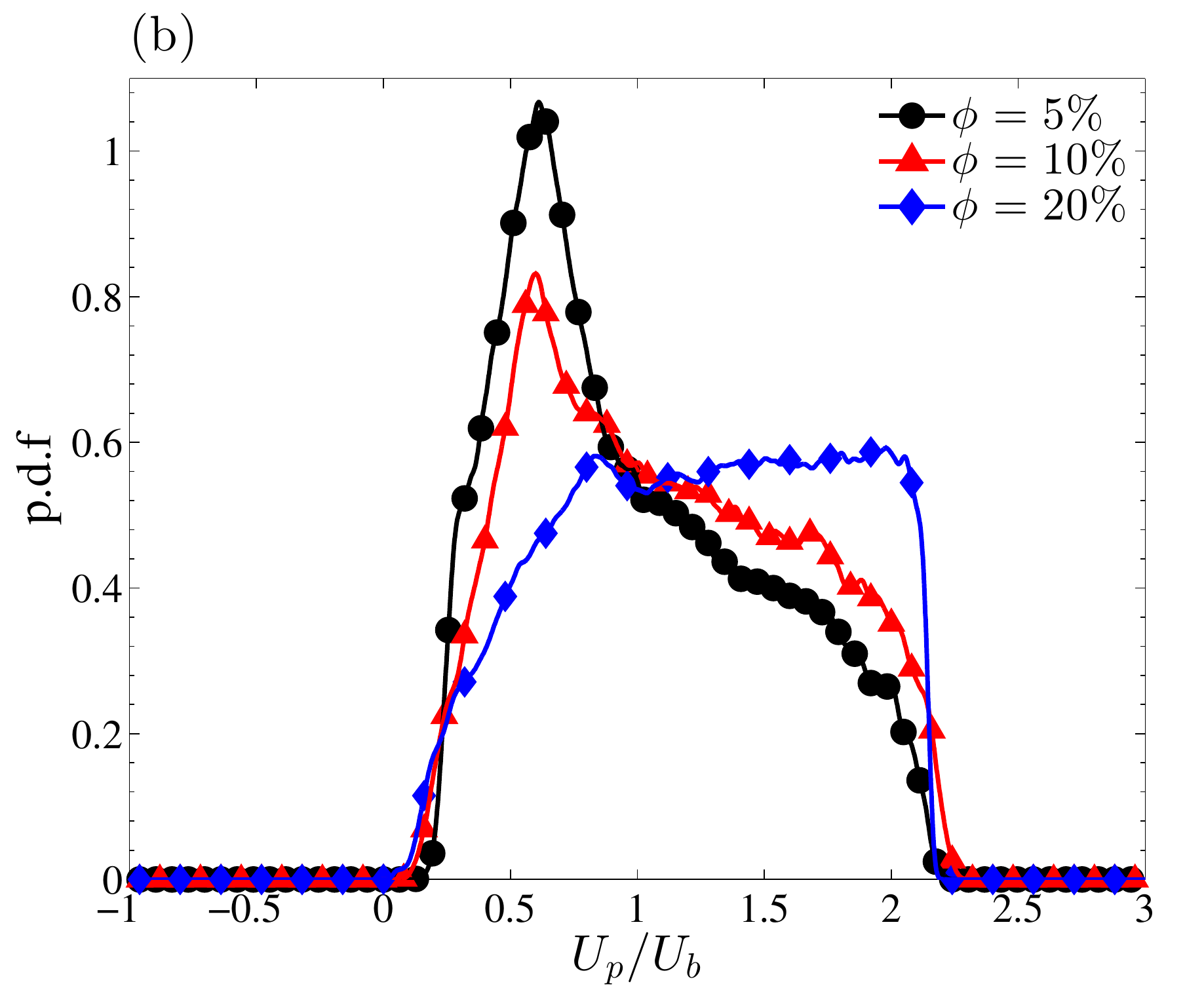}

\caption{ Probability density function p.d.f. of particle velocities in streamwise direction for different solid volume fractions at $Re_b=550~(Re_p=1.7)$ and $h/a=18$: $(a)$ at the duct corner, $(b)$ over the whole duct cross section.}
\label{fig:pdf_U_P}
\end{figure}
\begin{figure}[h!]

\centering
\includegraphics[width=0.33\textwidth]{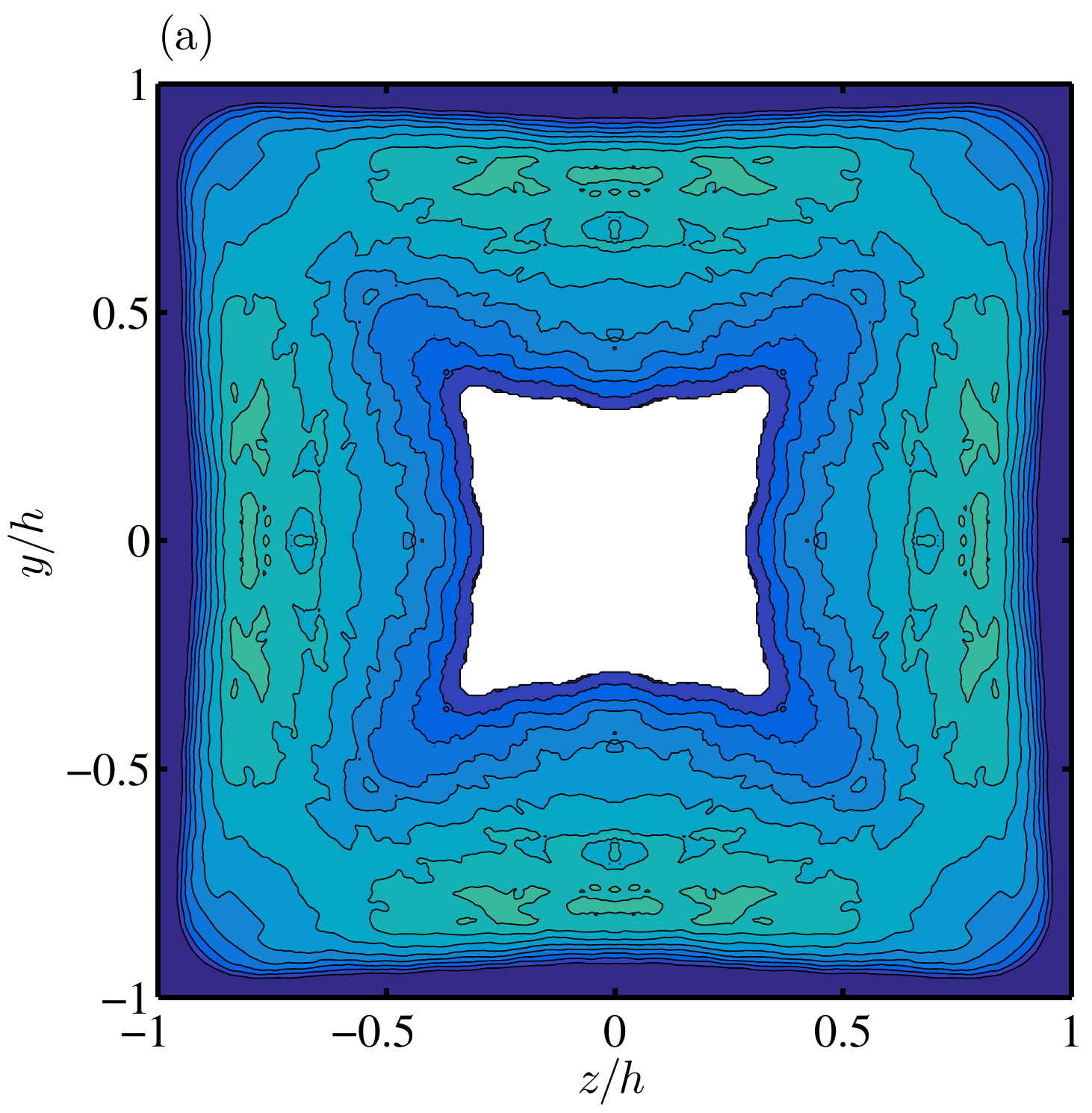}
\includegraphics[width=0.295\textwidth]{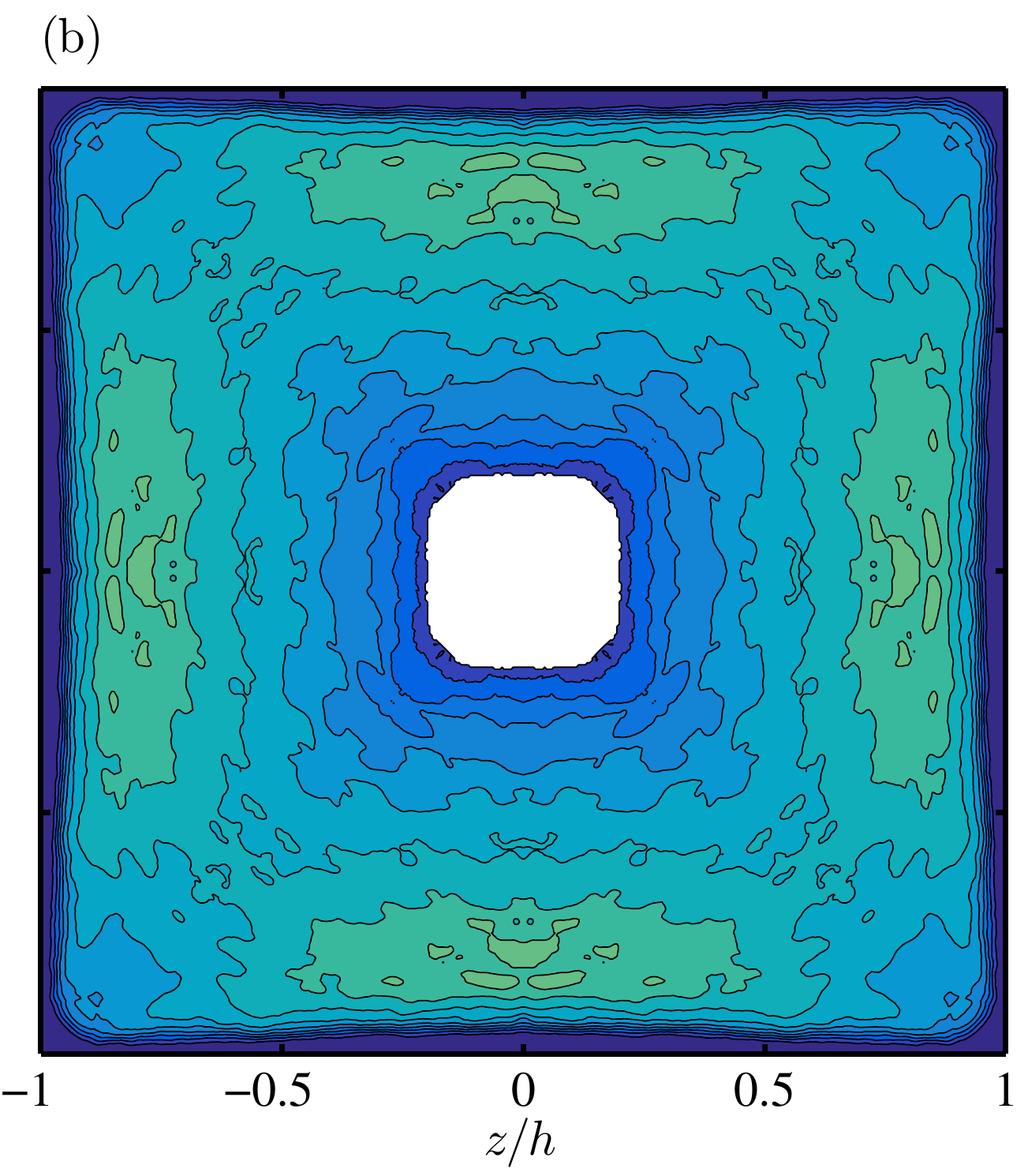}
\includegraphics[width=0.345\textwidth]{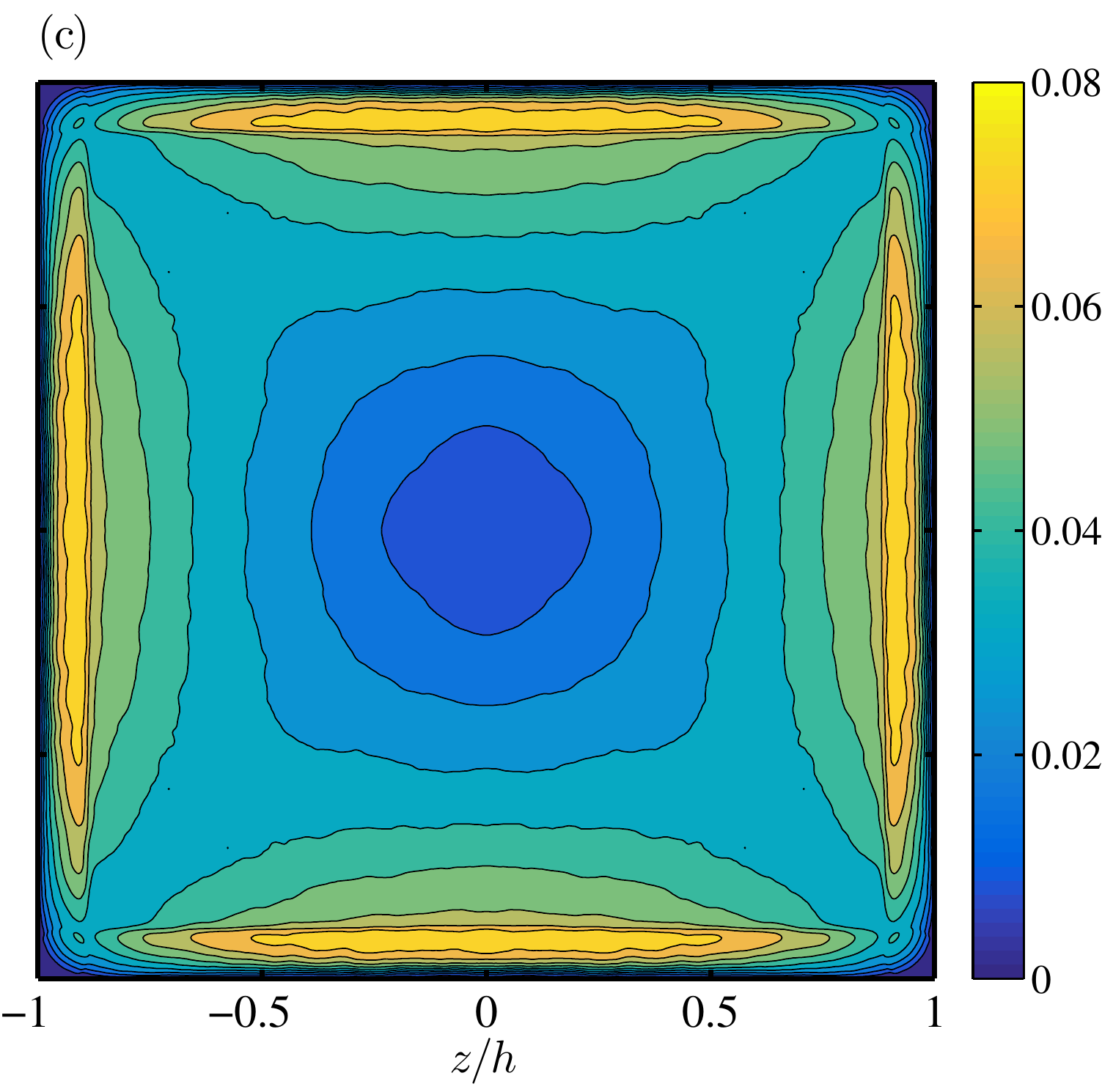}

\caption{ Streamwise particle velocity fluctuation contours, $u_{p,rms}'/U_b$, in the cross-stream $(y-z)$ plane at $Re_b=550~(Re_p=1.7)$ and $h/a=18$ for: $(a)$ $\phi=5\%$, $(b)$ $\phi=10\%$, $(c)$ $\phi=20\%$.}
\label{fig:Urms_P}
\end{figure}

\begin{figure}[h!]
  \centering
\includegraphics[width=0.33\textwidth]{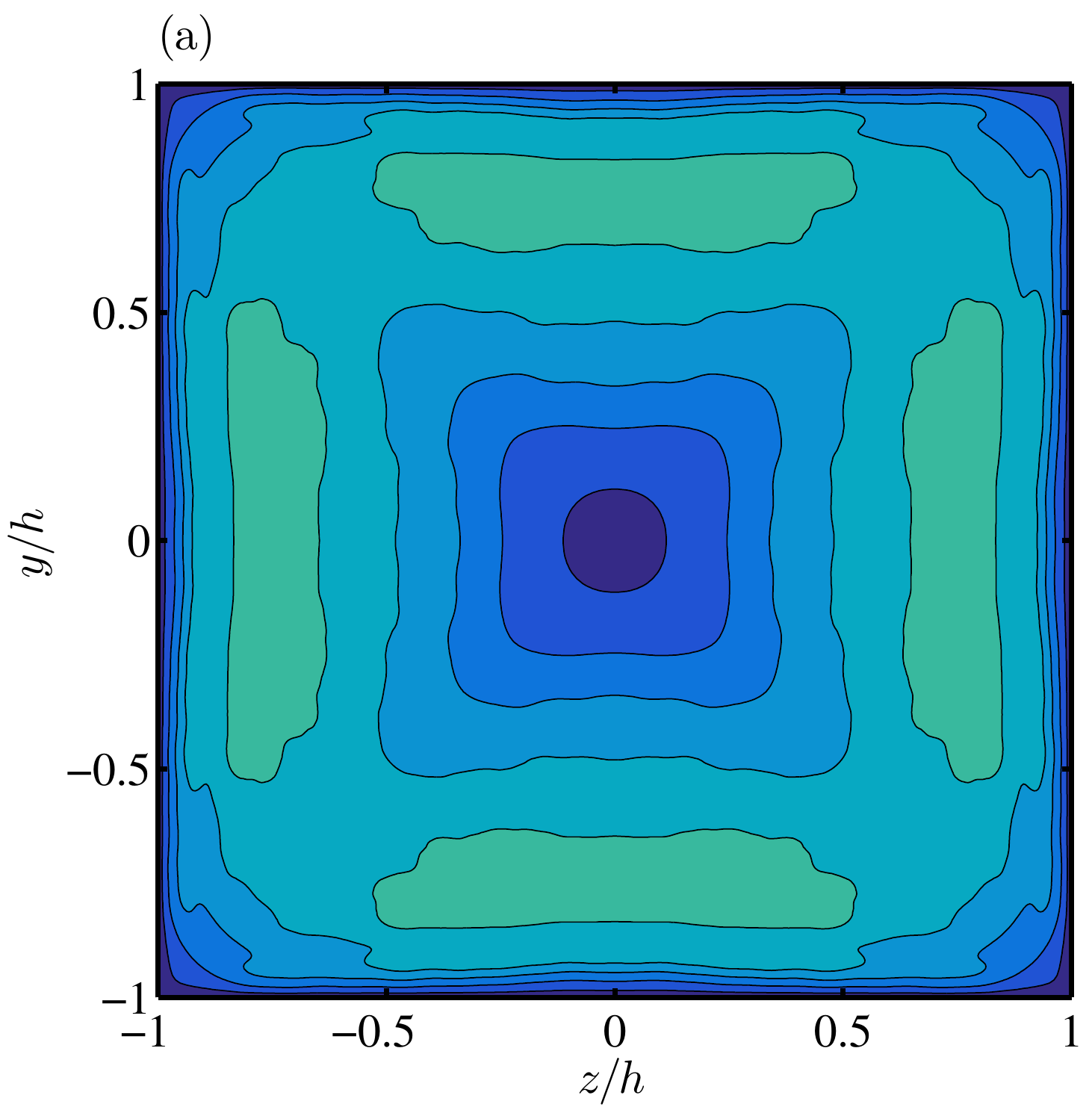}
\includegraphics[width=0.295\textwidth]{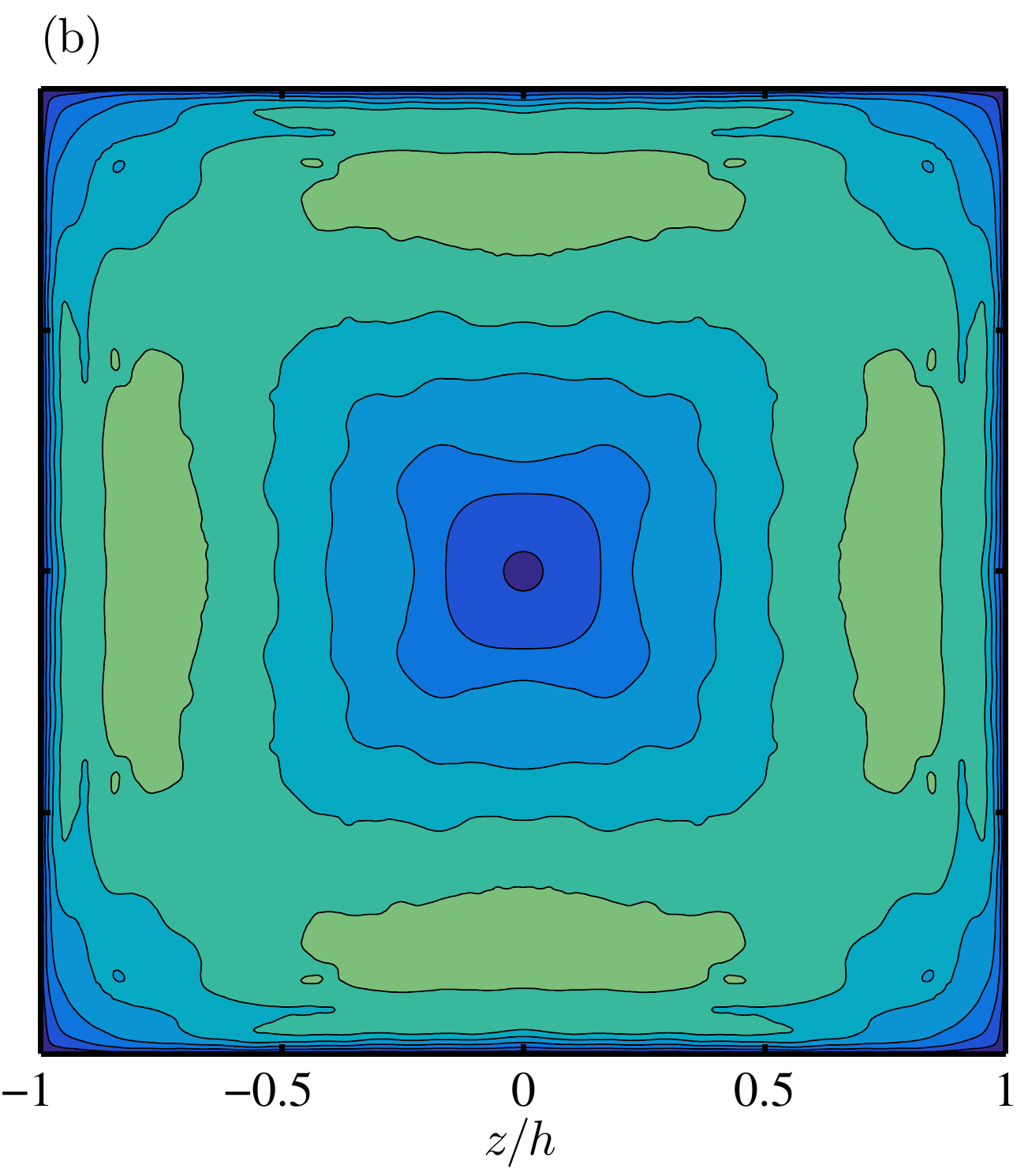}
\includegraphics[width=0.345\textwidth]{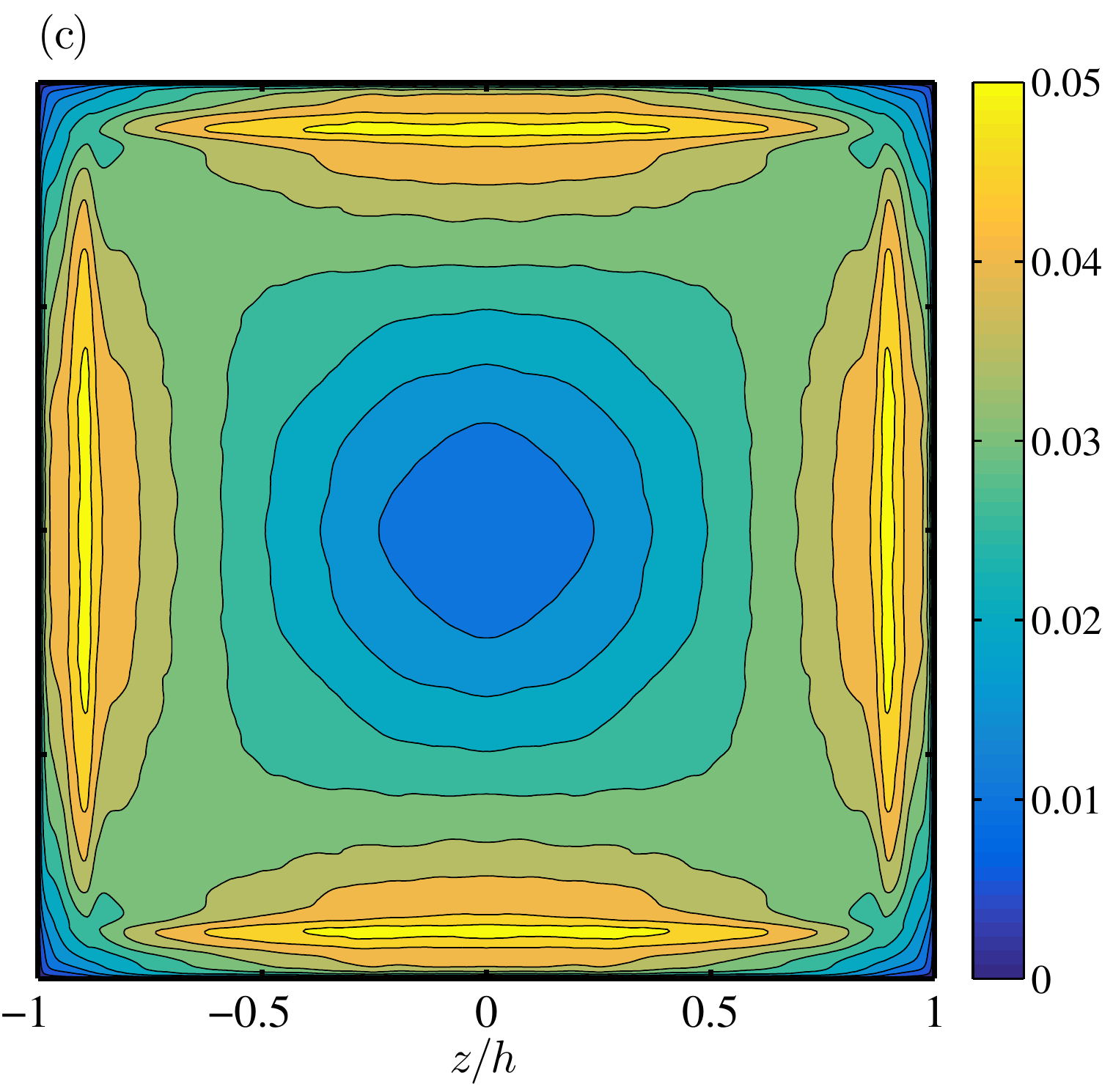}

\caption{ Streamwise fluid velocity fluctuation contours, $u_{f,rms}'/U_b$, in the cross-stream $(y-z)$ plane at $Re_b=550~(Re_p=1.7)$ and $h/a=18$ for: $(a)$ $\phi=5\%$, $(b)$ $\phi=10\%$, $(c)$ $\phi=20\%$.}
\label{fig:Urms_F}
\end{figure}

The probability density functions (p.d.f.) of particle streamwise velocities $U_p$ calculated at the duct corners and over the whole 
duct cross section are reported in Fig.~\ref{fig:pdf_U_P}. 
At the corners, we observe that the variance of the p.d.f.s increases as the volume fraction increases (see Fig.~\ref{fig:pdf_U_P}$a$).
The mean value of the streamwise particle velocity is also found to increase with the volume fraction. As will be shown 
later, this behavior may be due to the fact that the streamwise mean fluid velocity at the duct corner increases 
with the solid volume fraction.\\ 
The variance of the p.d.f. of the particle streamwise velocity $U_p$ in the whole duct is instead similar for all $\phi$ 
(see Fig.~\ref{fig:pdf_U_P}$b$). However, as $\phi$ increases, particles are forced to distribute more uniformly across the duct 
and hence exhibit higher velocities towards the centerline. This leads to a progressive enhancement of the mean particle streamwise 
velocity and to a substantial change of the shape of the p.d.f.s. Indeed, for $\phi=20\%$ the mean streamwise particle 
velocity is $23\%$ larger than that for $\phi=5\%$; the different, more flattened, shape of the p.d.f. is due to the uniform distribution 
of particles across the duct.

Figures~\ref{fig:Urms_P}$(a)-(c)$ show the streamwise velocity fluctuation contours $u_{p,rms}'(y,z)$ of the solid phase. 
We observe that the maxima of $u_{p,rms}'(y,z)$ are located close to 
the wall centers for all $\phi$. At the highest $\phi$, the maxima of $u_{p,rms}'(y,z)$ is almost twice that for $\phi=5\%$. 
As expected, the fluctuations are lower at the duct corners where particles reside longer. 

Fluid velocity fluctuations, absent in laminar regimes, are induced by the solid phase. Contours of the streamwise mean 
velocity fluctuations of the fluid phase, $u_{f,rms}'$, are shown in Figs.~\ref{fig:Urms_F}$(a),(b),(c)$ for $\phi=5, 10$ and $20\%$. 
As for the particle velocity fluctuations, the maxima of $u_{f,rms}'$ are located in the proximity of the duct walls and increase by increasing the solid volume fraction. Spatially averaged streamwise fluid velocity fluctuations 
increase by $25$ and $64\%$ for $\phi=10\%$ and $\phi=20\%$ compared to the case $\phi=5\%$. Comparing the data with the particle fluctuation velocities it 
can be seen that $u_{p,rms}'/u_{f,rms}'$ is larger than 1 for all volume fractions under investigations.

\begin{figure}[t!]
   \centering
\includegraphics[width=0.355\textwidth]{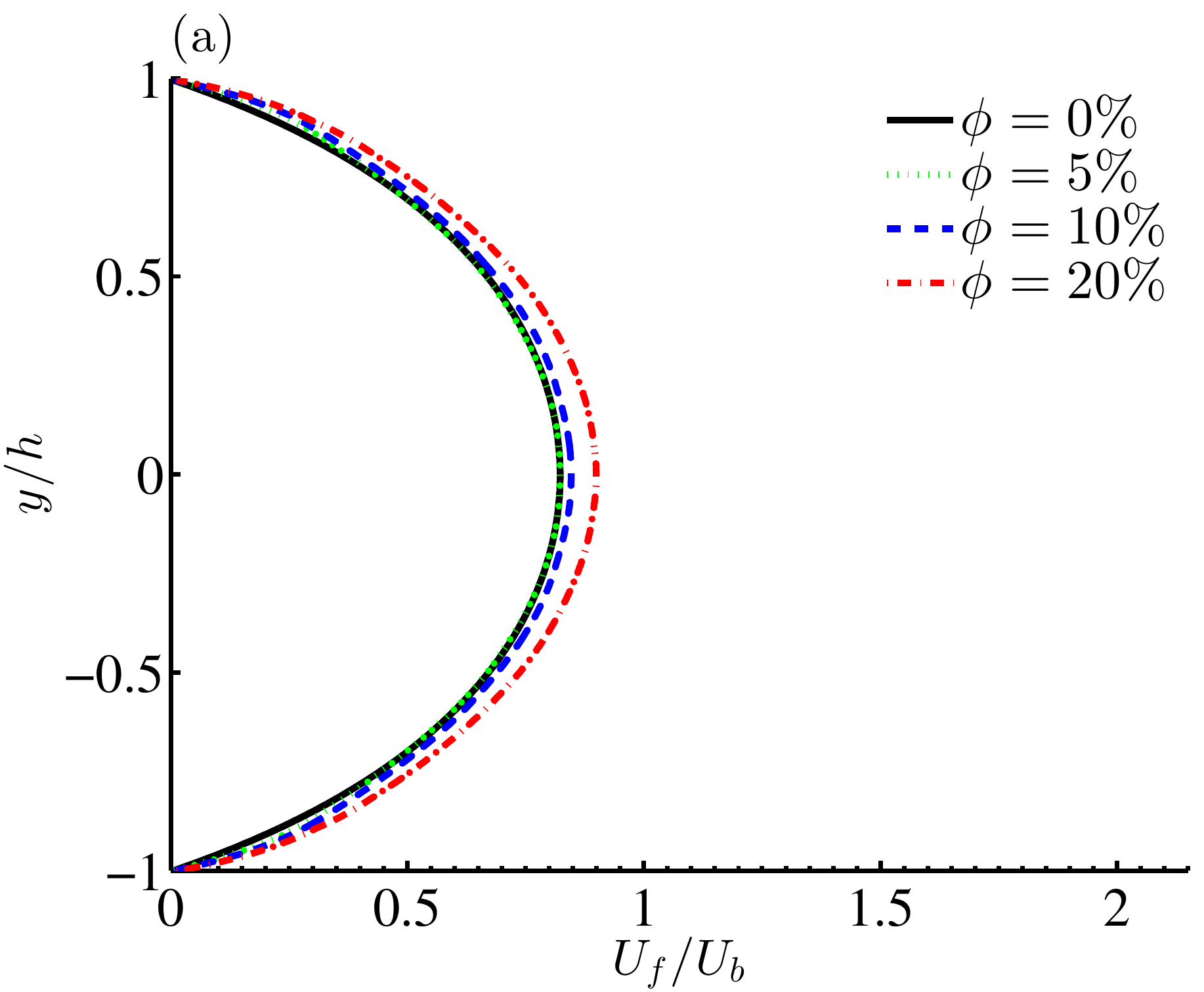}
\includegraphics[width=0.31\textwidth]{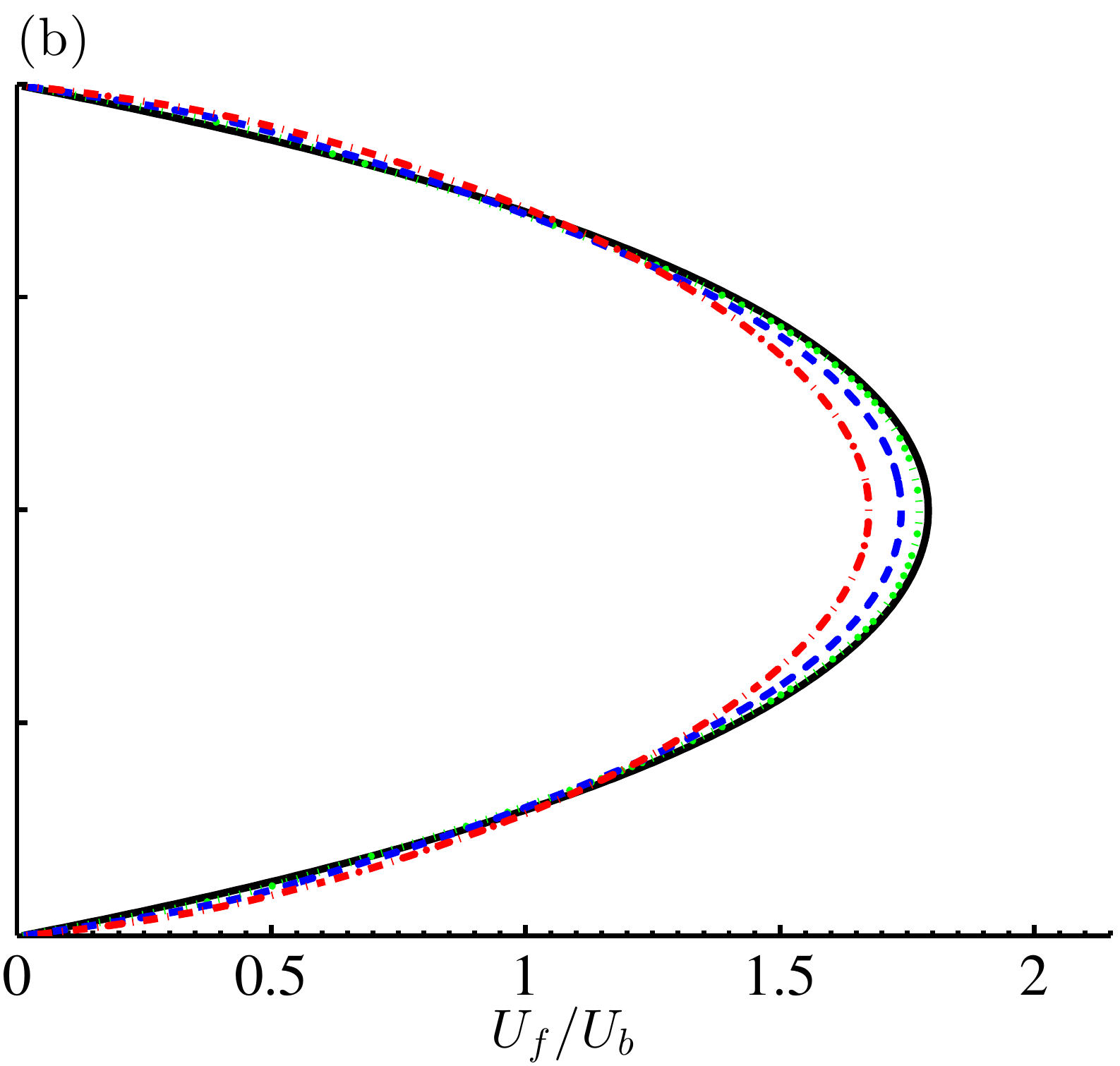}
\includegraphics[width=0.31\textwidth]{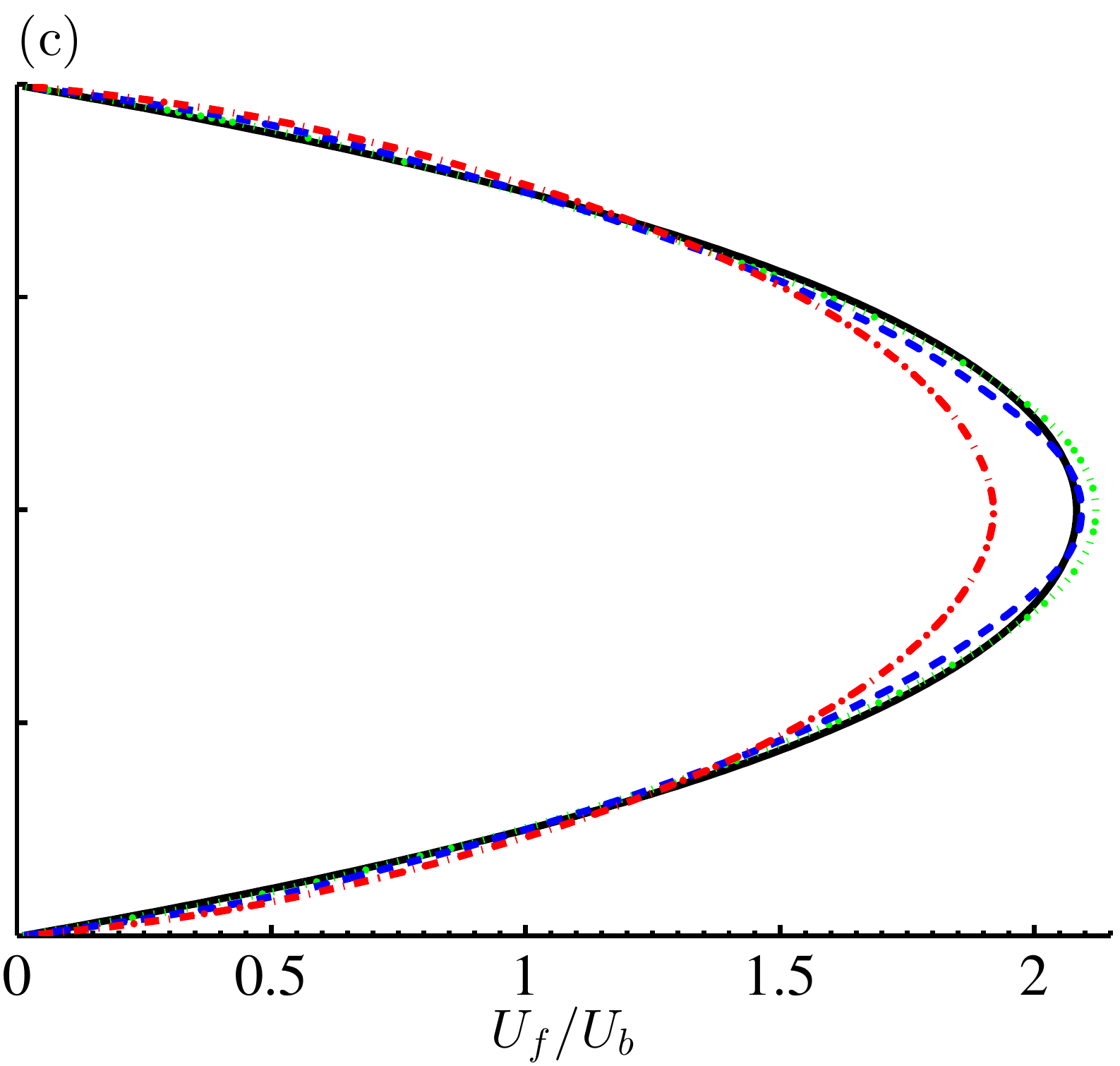}

\caption{ Streamwise mean fluid velocity profiles, $U_f/U_b$, for different solid volume fractions at $Re_b=550~(Re_p=1.7)$ and $h/a=18$ at different spanwise locations along the $y$ axis: 
$(a)$ $z/h=-0.8$, $(b)$ $z/h=-0.4.$, $(c)$ $z/h=0$.}
\label{fig:Veloc}
\end{figure}

Due to the solid phase, the mean fluid velocity profiles are altered with respect to the unladen case. In 
Fig.~\ref{fig:Veloc} we compare the streamwise mean fluid velocity profiles $U_f(y)$, normalized
by the bulk velocity $U_b$, for each solid volume fraction $\phi$ at different spanwise locations, $z/h$. 
One can see in Fig.~\ref{fig:Veloc}$(c)$ that at the wall bisector, $z/h=0$, the maximum velocity first increases for $\phi=5\%$ and $\phi=10\%$ and then decreases for $\phi=20\%$. 
Blunted velocity profiles for pipe and channel flows of dense suspensions 
at low bulk Reynolds numbers have also been reported by other authors~\cite{karnis1966,hampton1997,koh1994,lyon1998}.
At $\phi=5\%$ and $10\%$, particles migrate to the duct corners and  depletion 
is seen at the duct center (Figs.~\ref{fig:Conc}$e$ and $f$). As the bulk velocity $U_b$ is kept constant in our simulations, 
this results in a slight increase of the streamwise fluid velocity $U_f$ around the centerline. 
On the other hand, at the highest volume fraction considered ($\phi=20\%$) particles 
are homogeneously distributed across the duct cross section. 
Hence, the local viscosity of the suspension increases everywhere and this leads to the blunted velocity profile. 
Close to the duct corners, $z/h=-0.8$, the maximum streamwise velocity  is largest for $\phi=20\%$ 
(see Fig.~\ref{fig:Veloc}$a$). Indeed, since $U_b$ is constant, the reduction 
of the mean fluid velocity at the centerline for $\phi=20\%$ leads to an expansion of the three-dimensional 
paraboloid describing the fluid velocity and hence to an increase of the fluid velocity at the corners.

\begin{figure}[t!]
\centering
\includegraphics[width=0.495\textwidth]{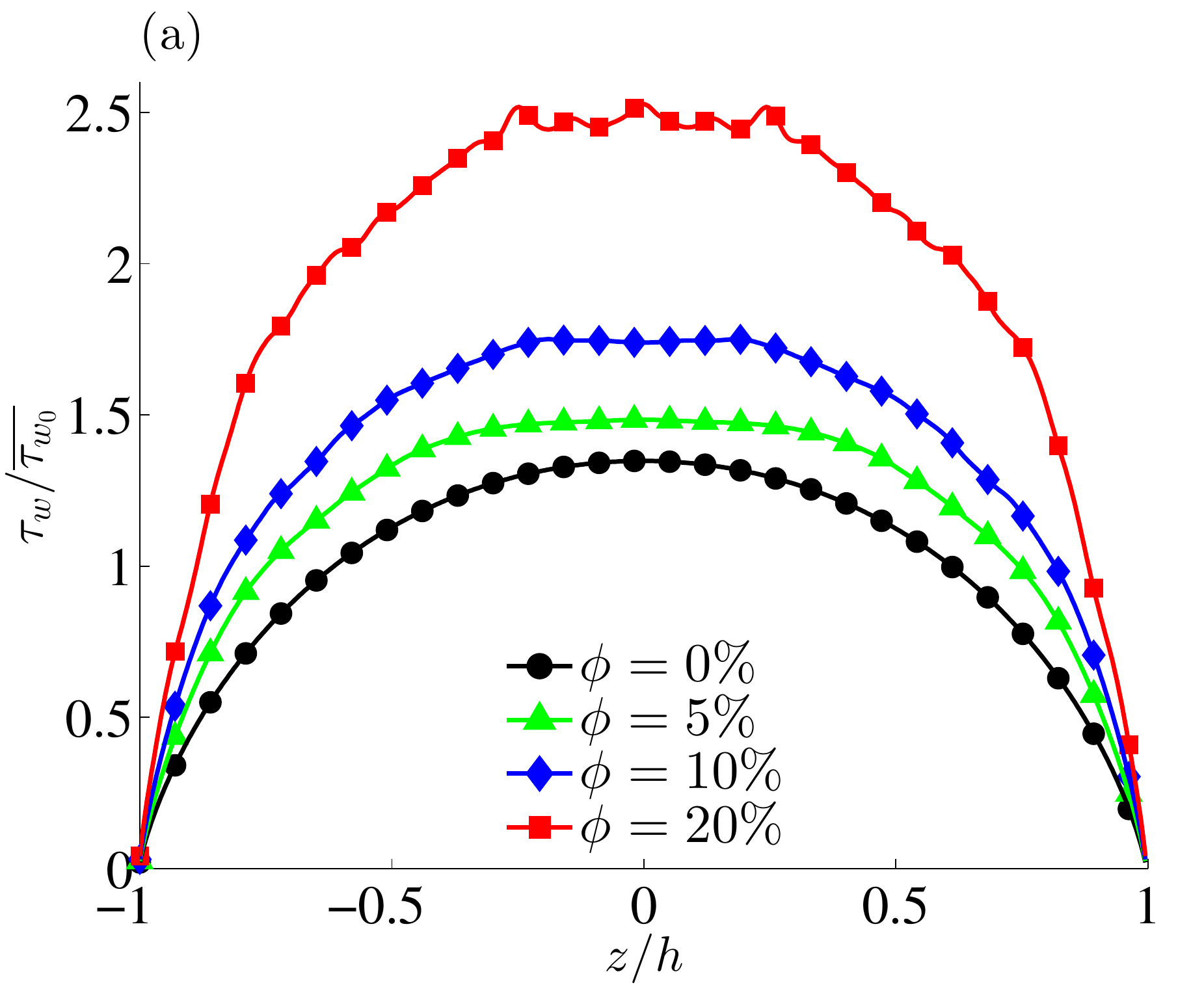}
\includegraphics[width=0.495\textwidth]{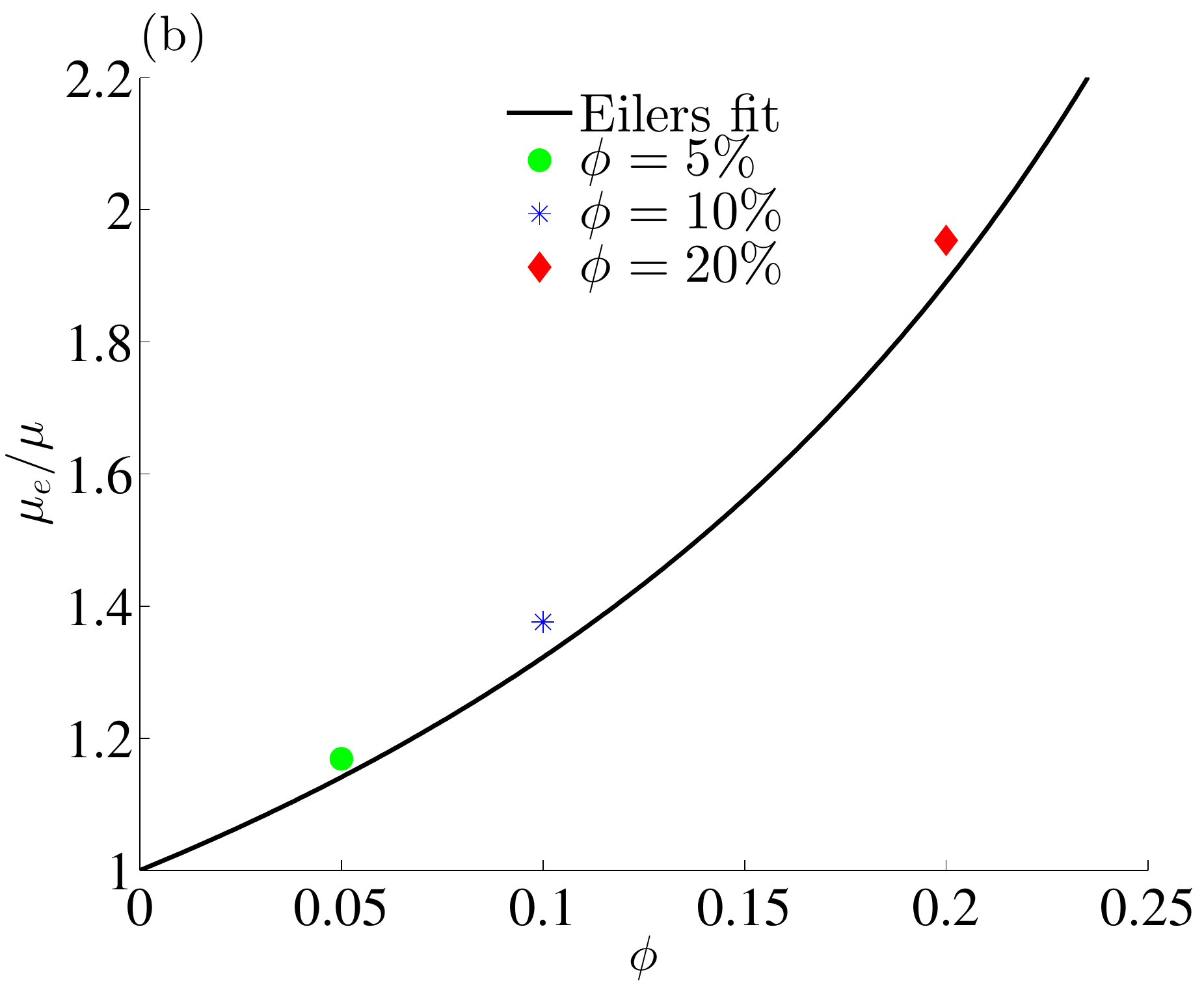}

\caption{$(a)$ Distribution of shear stress along the duct wall, $(b)$ relative viscosity of the suspensions under investigation
in comparison with Eilers fit: $\mu_r=\mu_e / \mu = {[1 + 1.25 \phi / (1-\phi/0.6)]}^2$. All cases presented here have the same bulk Reynolds number $Re_b=550~(Re_p=1.7)$ and $h/a=18$.}
\label{fig:shear}
\end{figure}

The presence of particles in the flow increases the rate of energy dissipation and consequently the suspension viscosity 
and the wall shear stress. Figure~\ref{fig:shear}$(a)$ shows the distribution of the normalized shear stress $\tau_{w}/\overline{\tau_{w0}}$ along the 
duct wall where $\overline{\tau_{w0}}$ denotes the mean value of wall shear stress pertaining the unladen flow. 
The results show a $16\%$, $36\%$ and $93\%$ 
increase in the mean value of wall shear stress $\overline{\tau_{w}}$ for $\phi=5\%$, $10\%$ and $20\%$ with respect to the unladen case. Taking the ratio 
between $\overline{\tau_w}$ and $\overline{\tau_{w0}}$, the relative viscosity, $\mu_r=\mu_e / \mu$, 
(i.e. the ratio between the effective viscosity of the suspension and the viscosity of the fluid phase) can be determined.
When plotted as function of the solid volume fraction $\phi$, we see that our results match empirical predictions 
given by the Eilers fit~\cite{stickel2005}. It should be noted 
that $Re_p=Re_b (a/h)^2 = 1.7$. 
Therefore, inertial effects are expected to be weak and this may explain the accuracy of the fit valid for vanishing inertia. Indeed, 
inertial shear thickening~\cite{picano2013} is still weak in this case and we report just a weak 
underprediction of the effective viscosity.

\subsection{Effect of geometry, bulk and particle Reynolds numbers on the migration in dilute suspensions}
\begin{figure}[h!]
  \centering
\includegraphics[width=0.31\textwidth]{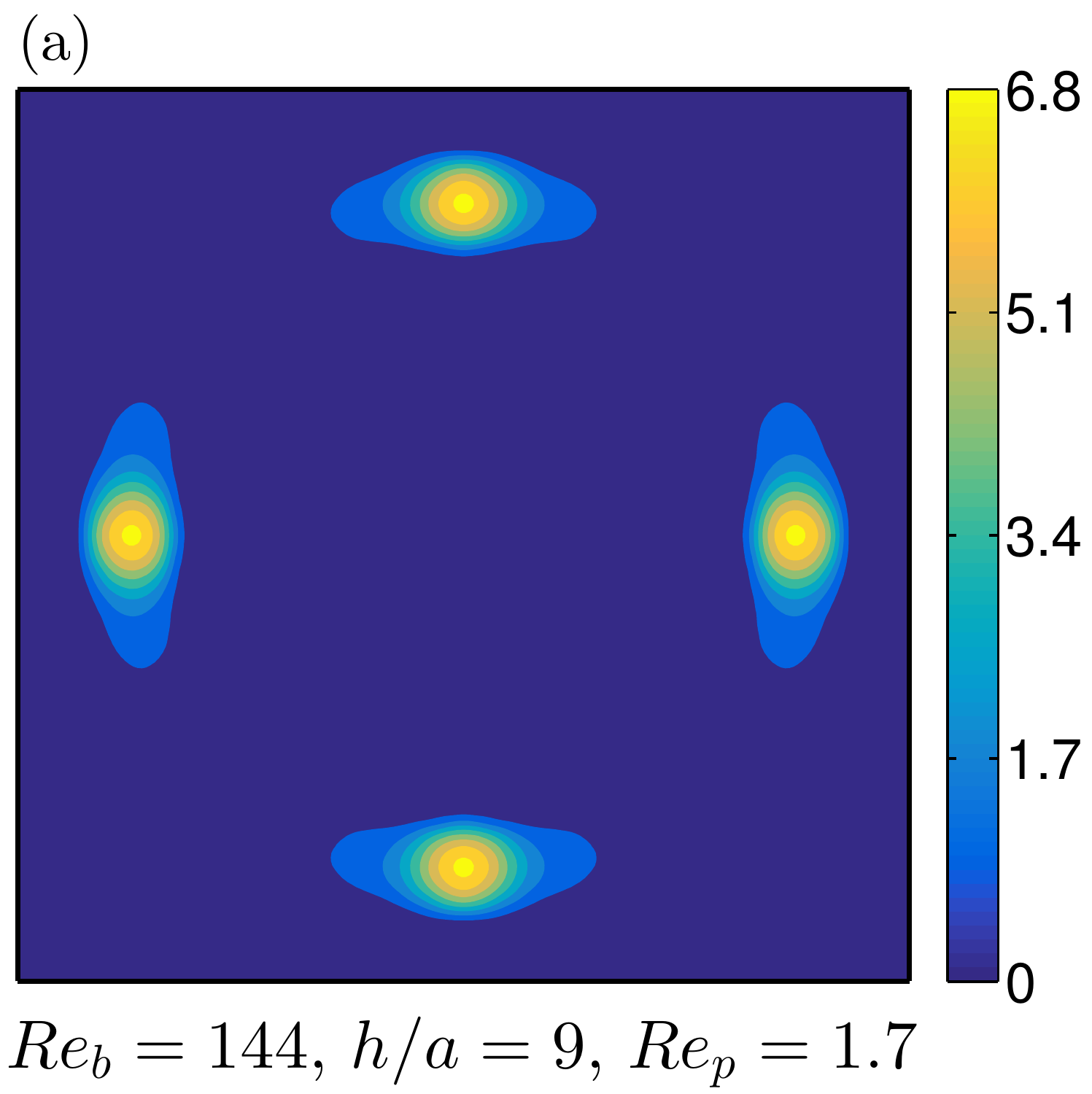}
\includegraphics[width=0.32\textwidth]{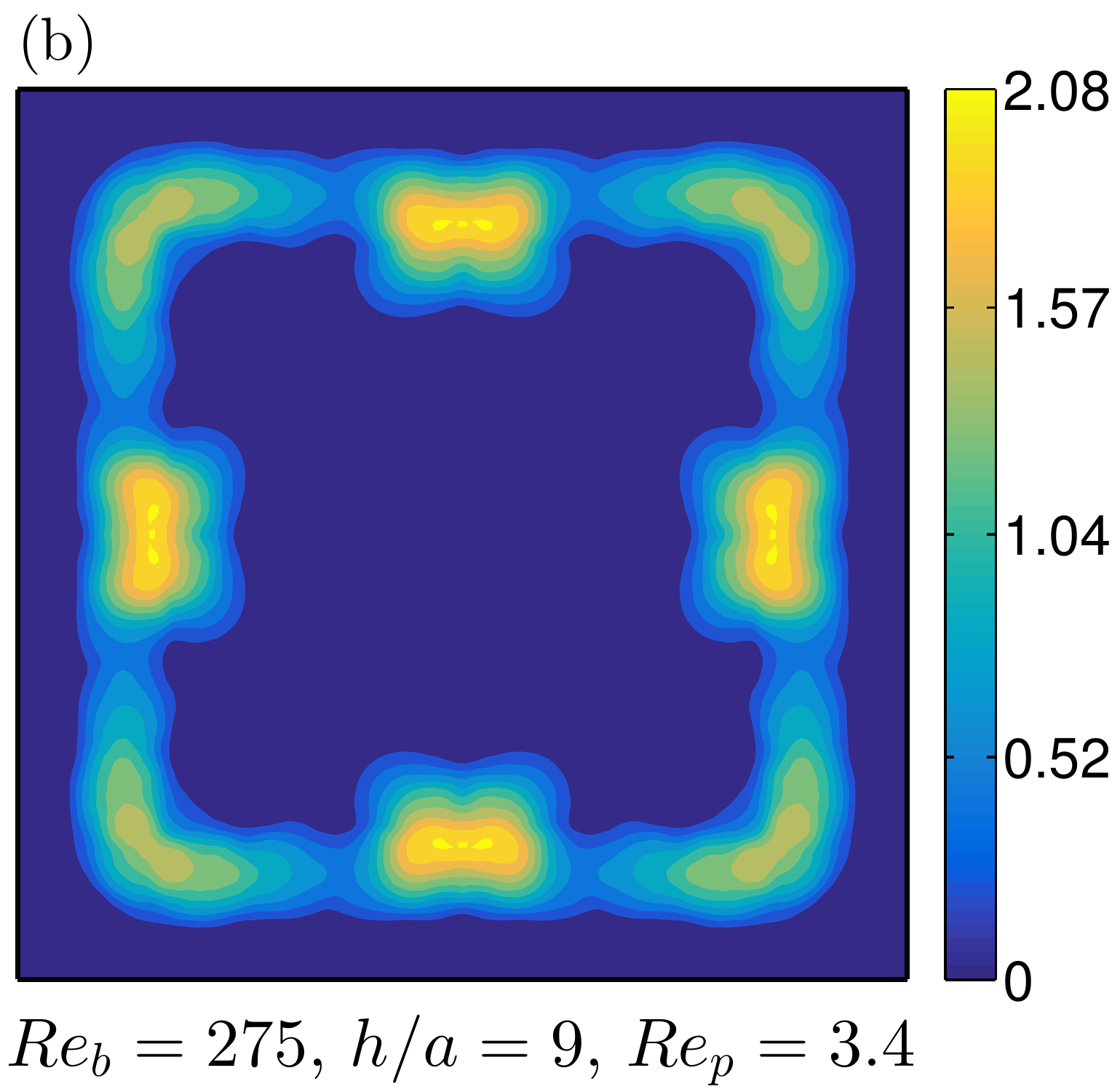}
\includegraphics[width=0.34\textwidth]{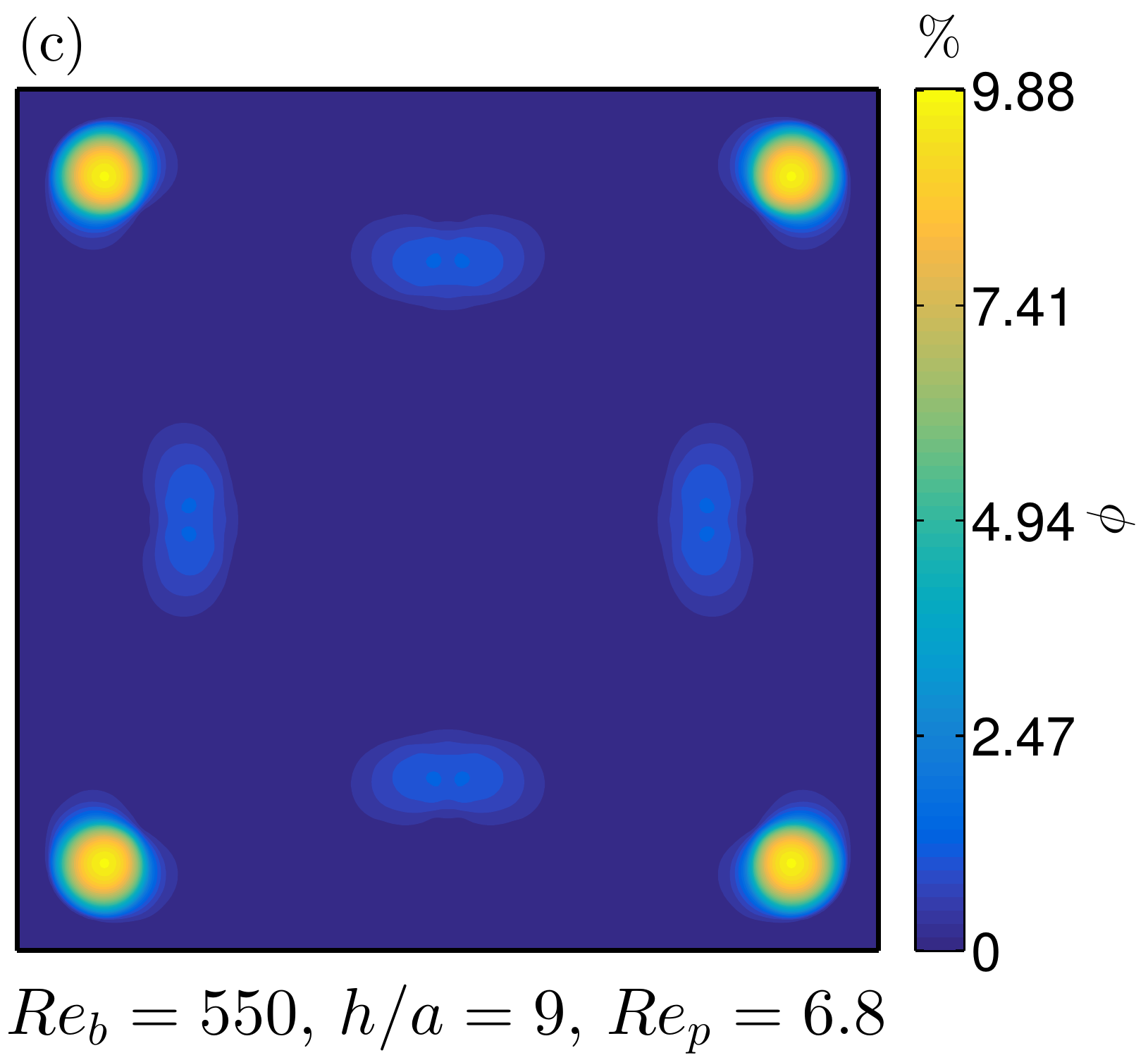}\vspace{1cm}

\includegraphics[width=0.31\textwidth]{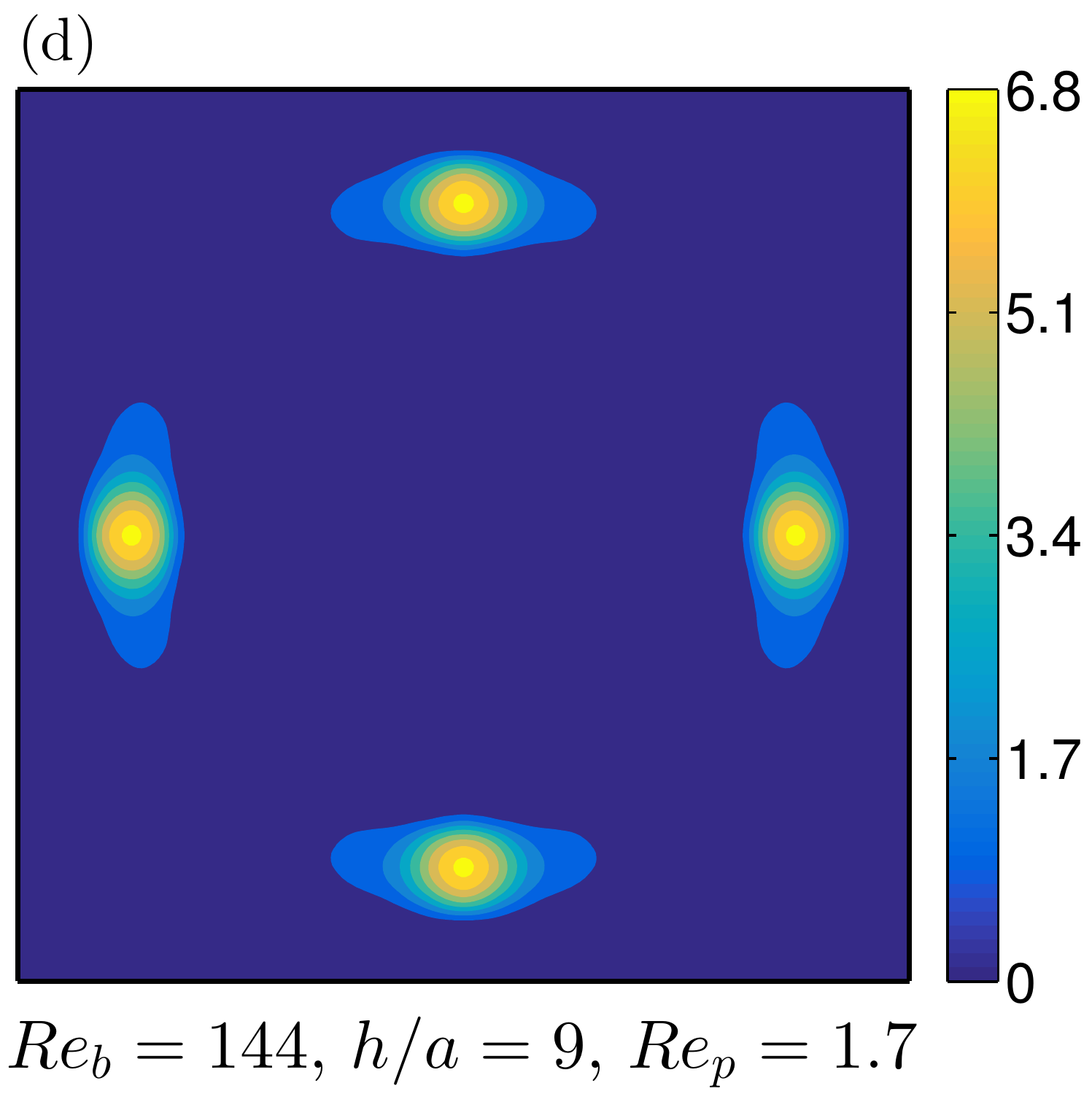}
\includegraphics[width=0.32\textwidth]{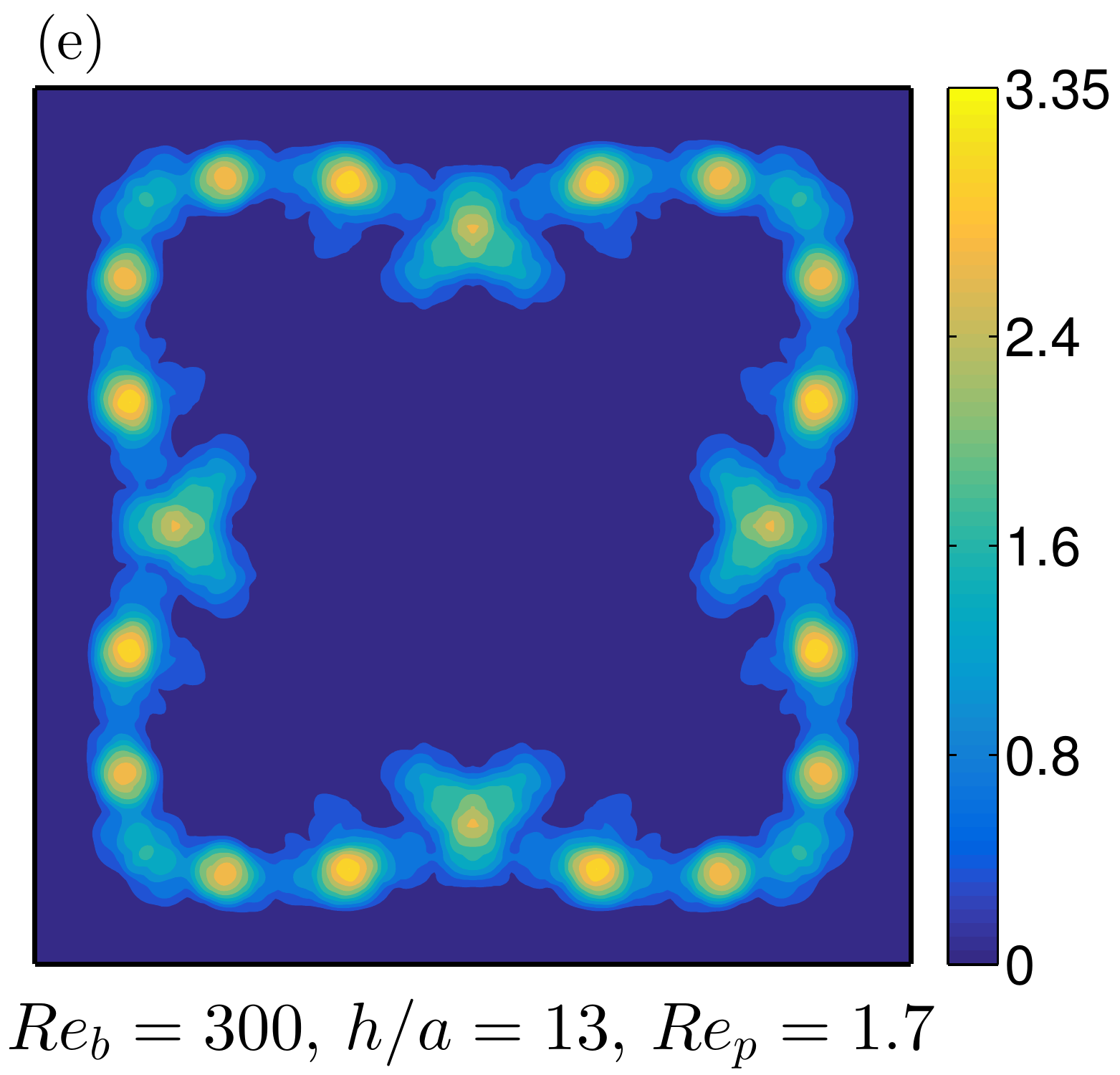}
\includegraphics[width=0.34\textwidth]{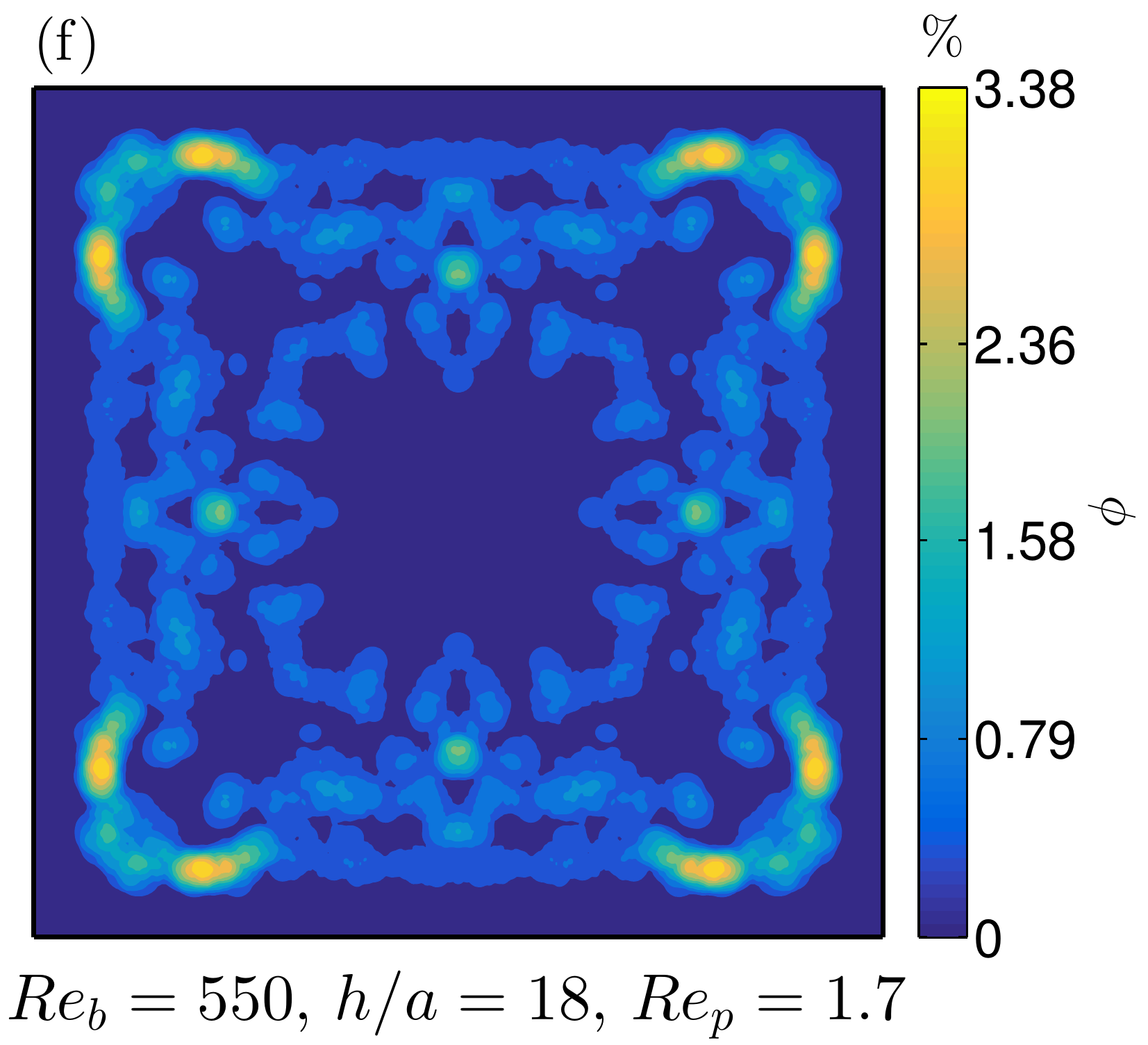}\vspace{1cm}

\includegraphics[width=0.319\textwidth]{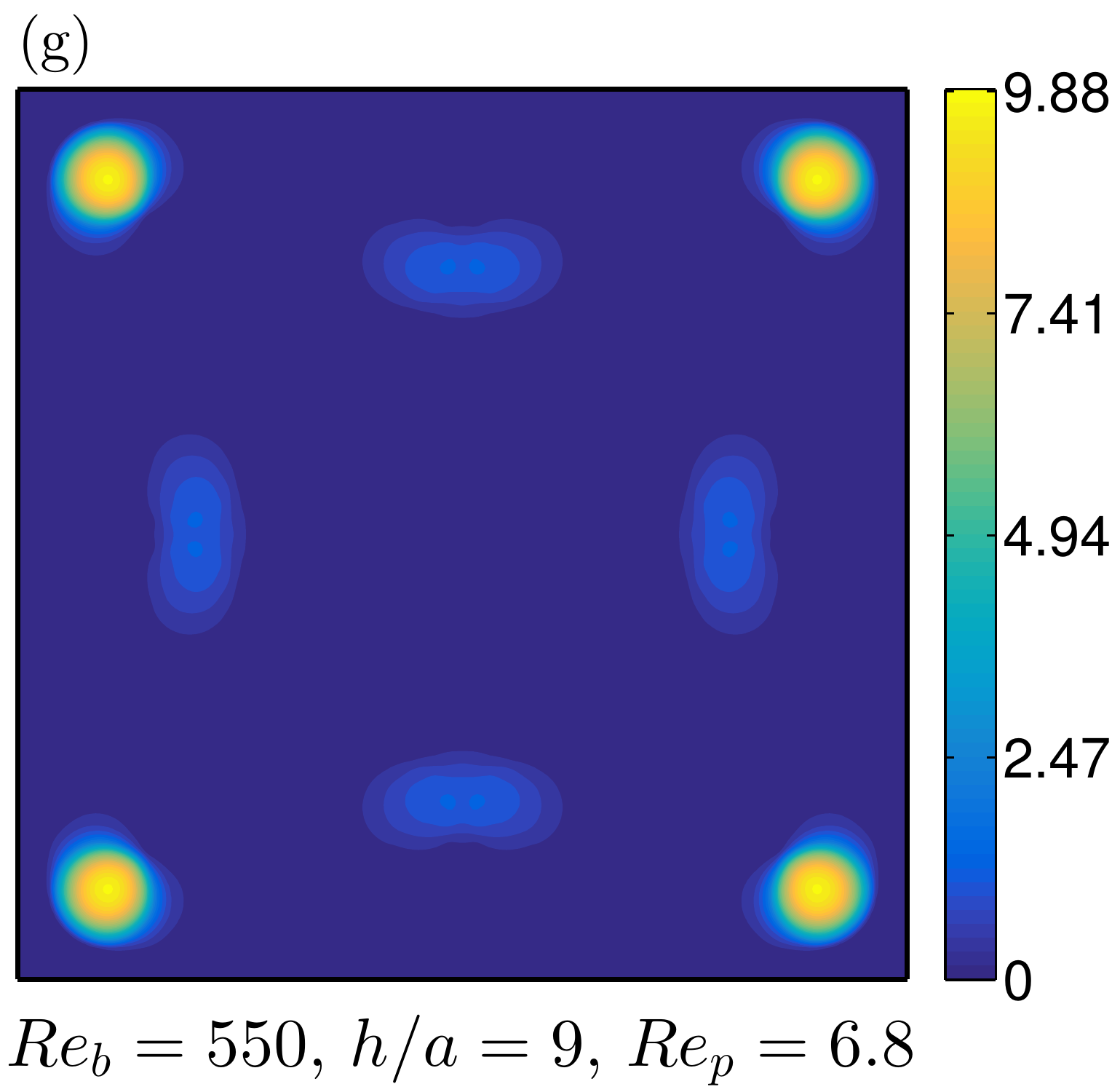}
\includegraphics[width=0.324\textwidth]{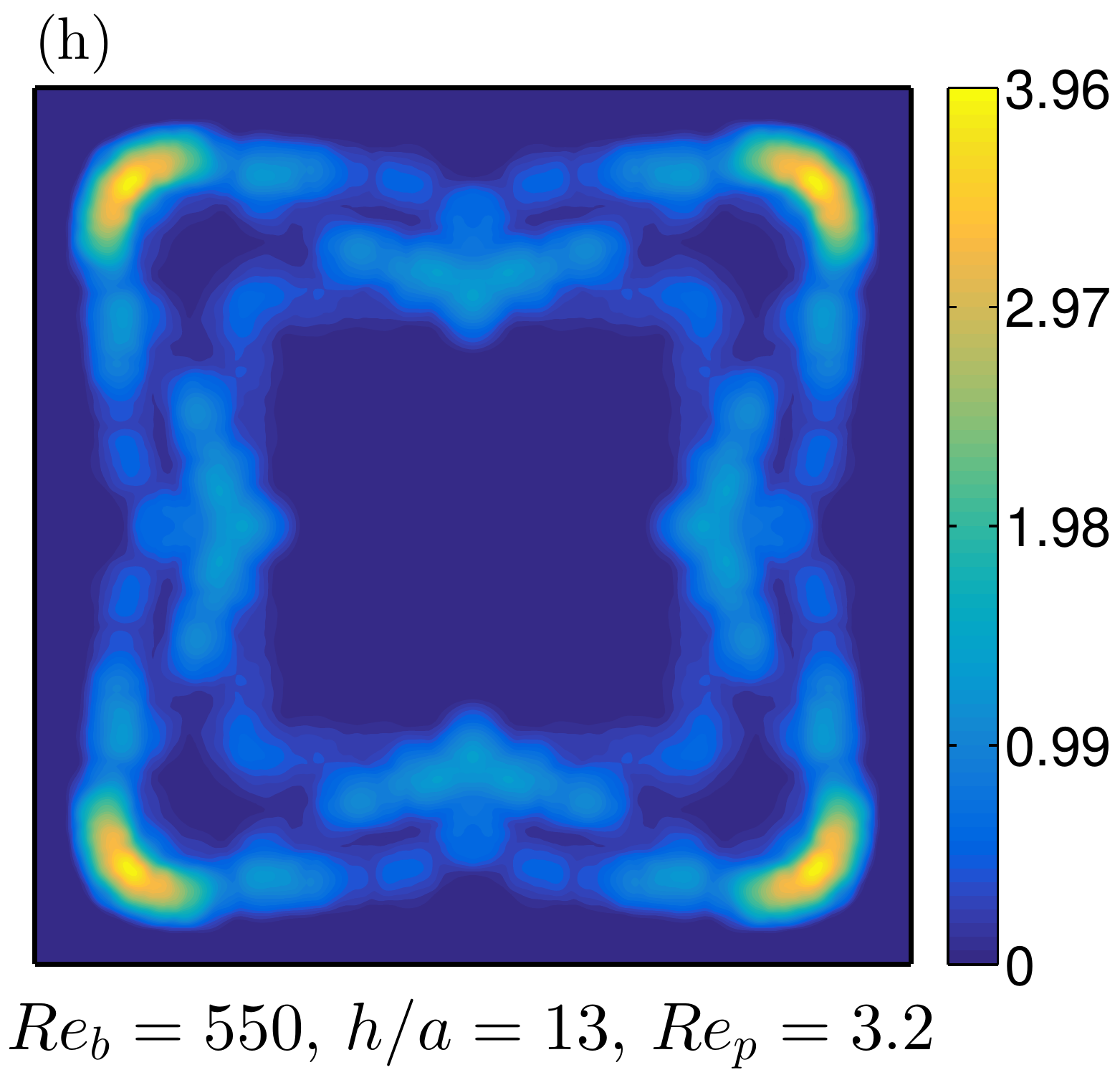}
\includegraphics[width=0.342\textwidth]{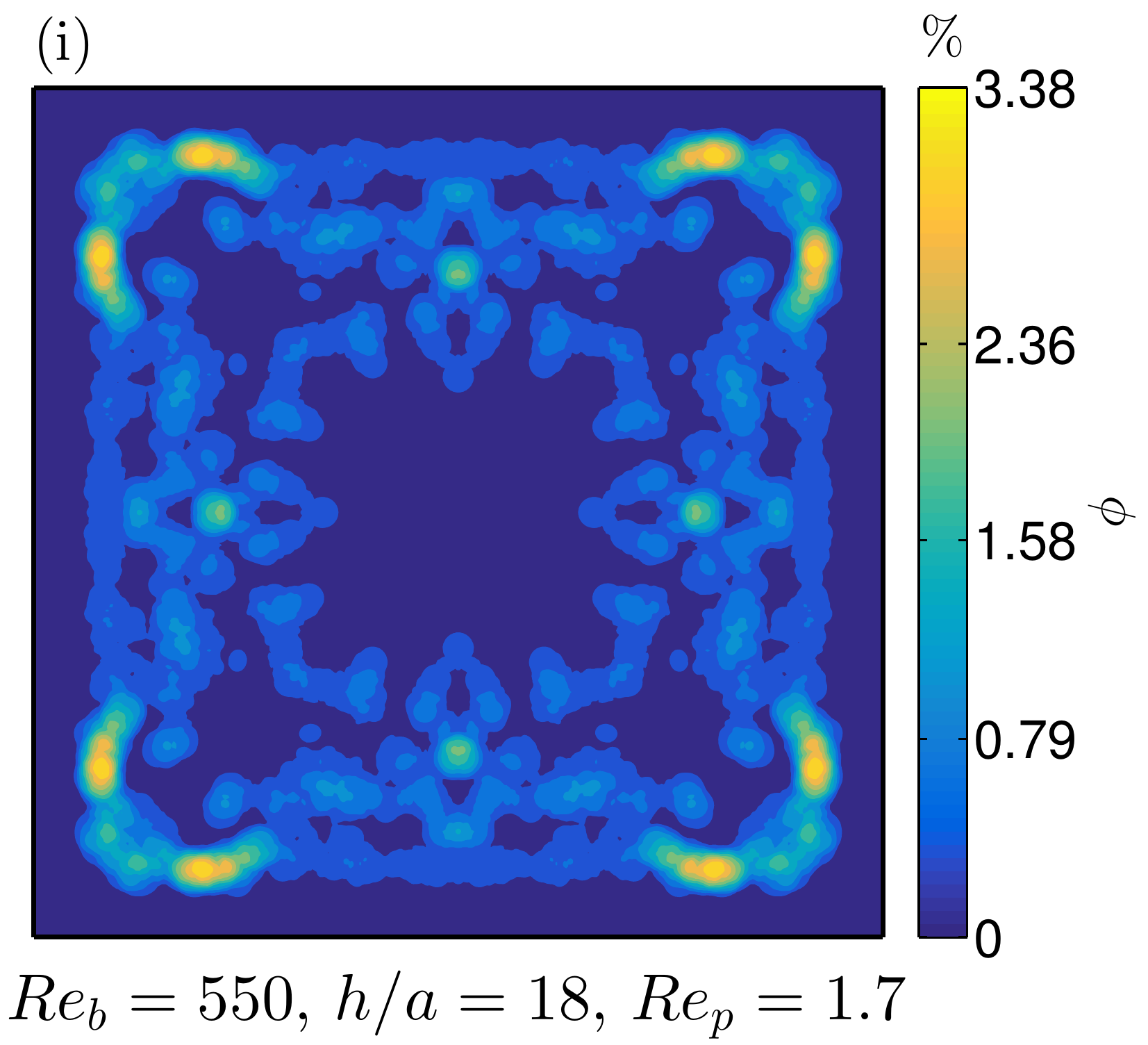}
\caption{ Particle concentration distribution $\Phi(y,z)$ in the duct cross section for $\phi=0.4\%$:
$(a)-(c)$: constant particle relative size, $h/a=9$, and increasing bulk and particle Reynolds numbers, $Re_b$ and $Re_p$.
$(d)-(f)$: constant $Re_p=1.7$ for increasing  $Re_b$ and $h/a$.
$(g)-(i)$: constant $Re_b=550$ for $Re_p$ and $h/a$.}
\label{fig:L_Conc}
\end{figure}

In this section we discuss the influence of the bulk Reynolds number $Re_b$, duct to particle size 
ratio $h/a$ and particle Reynolds number $Re_p$ on the spatial distribution of particles across the duct. 
To this aim, we focus on dilute suspensions of spheres with solid volume fraction $\phi=0.4\%$. 
We consider three different duct to particle size ratios $9$, $13$ and $18$. As the particle Reynolds number is function of the bulk Reynolds number
and the duct to particle size ratio, $Re_p=Re_b\left(a/h\right)^2$, two parameters are steadily changed in 
three steps while one parameter is fixed. The results are shown in Figs.~\ref{fig:L_Conc}$(a)-(i)$, 
where we only report Reynolds numbers for laminar duct flow as turbulent diffusion would alter the particle distribution across the duct and no direct comparison could be made. 

Figures~\ref{fig:L_Conc}$(a)-(c)$ show the particle concentration distribution $\Phi(y,z)$ 
at constant duct to particle size ratio $h/a=9$ while increasing the bulk and consequently particle Reynolds numbers.
As the bulk Reynolds number $Re_b$ is increased from $144$ to $275$, particles that were initially focused at the wall centers (Fig.~\ref{fig:L_Conc}$a$) 
start to spread on a ring parallel to the duct walls with slightly larger concentration at the duct wall centers (see 
Fig.~\ref{fig:L_Conc}$b$). A further increase in the bulk Reynolds number leads to the rupture of the ring. At $Re_b=550$, 
we observe the appearance of equilibrium positions at the duct corners and a low concentration focusing point close to the 
duct wall centers (see Fig.~\ref{fig:L_Conc}$c$). 
These results are consistent with those by Nakagawa et al.~\cite{nakagawa2015} who studied the migration of single particles in 
square duct flows. For $Re_b=260$ and $h/a=9$ these authors reported the existence of two equilibrium positions close to the duct 
corners in addition to stable equilibrium position at the duct wall centers. One of the two equilibrium positions close to the duct 
corner is located on the diagonal, whereas the second appears between the diagonal and the wall center equilibrium position 
(see Fig 6 of Ref.~\cite{nakagawa2015}). A similar pattern can be seen in Fig.~\ref{fig:L_Conc}$(b)$ of the present study for 
$\phi=0.4\%$, $h/a=9$ and $Re_b=275$, where two symmetric equilibrium positions emerge close to the duct corners. In addition, 
Nakagawa et al.~\cite{nakagawa2015} showed that by increasing the bulk Reynolds number from $260$ to $514$, the equilibrium position 
at the wall center moves towards the duct core. The same behavior is observed in our simulations for $Re_b=550$ where the 
equilibrium position at the duct wall center is closer to the duct center in comparison to the case with $Re_b=275$. 
In fact, the location of maximum local particle concentration changes from about $0.7h$ for $Re_b=275$ to 
approximately $0.6h$ for $Re_b=550$. Similar results were found experimentally by Miura et al.~\cite{miura2014}. 
Interestingly, these results are in contrast to what has been observed in pipe flow where the particle equilibrium position 
has been shown to approach the wall as the bulk Reynolds number $Re_b$ is increased (see Segre and 
Silberberg~\cite{segre1962} and Matas et al.~\cite{matas2004}).   

Figures~\ref{fig:L_Conc}$(d)-(f)$ illustrate the particle spatial distribution across the duct cross section 
when the particle Reynolds number is kept constant at $Re_p=1.7$ while adjusting the bulk Reynolds number $Re_b$ and duct to particle 
size ratio $h/a$. Qualitatively, we observe a particle distribution similar to that shown in Figs.~\ref{fig:L_Conc}$(a)-(c)$ for 
constant $h/a$. 
For $Re_b=300$ and $h/a=13$ as shown in Fig.~\ref{fig:L_Conc}$(e)$, we observe multiple equilibrium positions on the 
heteroclinic orbits that connect the wall center and corner equilibrium positions. The existence of additional equilibrium positions 
on the heteroclinic orbits for $Re_b\ge260$ had also been hypothesized by Nakagawa et al.~\cite{nakagawa2015}.
Comparing Figs.~\ref{fig:L_Conc}$(c)$ and $(f)$, we note that particles distribute more uniformly across the cross 
section for larger $h/a$ and same bulk Reynolds number of $Re_b=550$. Moreover, the regions of higher concentration are located 
at two symmetric points around each corner. 
Therefore, for larger $h/a$ these symmetric equilibrium positions appear at higher bulk Reynolds numbers as particles experience 
less inertia (i.e. smaller $Re_p$) in comparison to the case with $h/a=9$.

Finally, to  further explore the role of the bulk Reynolds number $Re_b$, 
we show in Figs.~\ref{fig:L_Conc}$(g)-(i)$ the particle concentration $\Phi(y,z)$ at constant 
bulk Reynolds number $Re_b=550$ for different particle Reynolds numbers $Re_p$ and duct to particle size ratios $h/a$. 
The
results show similar patterns  at the same bulk Reynolds number $Re_b$: all cases in Figs.~\ref{fig:L_Conc}$(g)-(i)$ display
clear equilibrium positions at the corners and weaker focusing at the wall center. This shows that changes in the bulk Reynolds numbers lead to most significant variations of the particle distribution. 
Therefore, the bulk Reynolds number appears to be 
the dominant parameter in the system. However, increasing the duct to particle size ratio, the particle concentration at the duct corner 
broadens until two separate equilibrium positions appear for $h/a=18$ (Fig.~\ref{fig:L_Conc}$i$). 
By increasing the duct to particle size ratio $h/a$ at the same bulk Reynolds number $Re_b$, 
particles feel weaker velocity gradients (i.e.\ inertial effects) resulting in a more uniform distribution across the duct cross section.
\clearpage

\begin{figure}[h!]
   \centering
\includegraphics[width=0.55\textwidth]{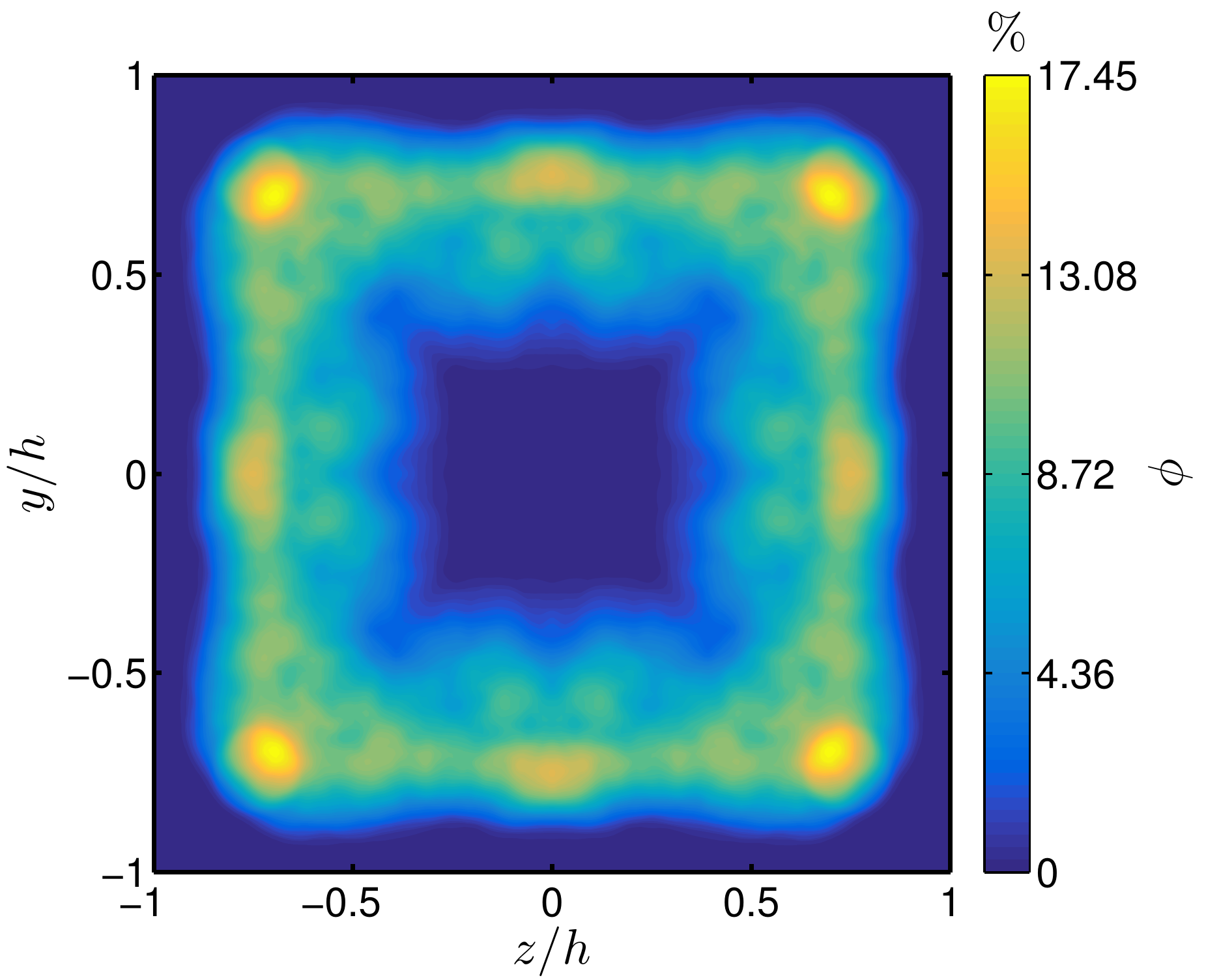}
\caption{Particle concentration distribution, $\Phi(y,z)$, in the duct cross-section for $\phi=5\%$, $h/a=9$ and $Re_b=144$.} 
\label{fig:L_Concb}
\end{figure}

We have shown in Fig.~\ref{fig:L_Conc}$(a)$ that particles 
accumulate at the wall centers at low bulk Reynolds ($Re_b=144$), $\phi=0.4\%$ and $h/a=9$. 
Moreover, in Sec.~\ref{Semi-dilute suspensions}, we observed similar particle distribution pattern 
for dilute ($\phi=0.4\%$) and semi-dilute ($\phi=5\%$) suspensions at $Re_b=550$ and $h/a=18$. Indeed, in both cases, particles are preferentially distributed at the duct corners rather than at the wall centers (Figs.~\ref{fig:Conc}$a$ and $b$). Hence, we would expect a similar behavior for $\phi=0.4\%$ and $\phi=5\%$
also at $Re_b=144$ and $h/a=9$. To test this conjecture, we have performed an additional simulation with volume fraction $\phi=5\%$, $h/a=9$ and $Re_b=144$. The resulting particle distribution at steady state is shown in Fig.~\ref{fig:L_Concb}. 
Surprisingly, however, particles accumulate preferentially closer to the corners, and less at the wall centers. Therefore, we conclude that in semi-dilute suspensions ($\sim 5\%$), the exact particle concentration distribution depends mainly on the nominal volume fraction $\phi$ and only
partially on $Re_b$ (particles still undergo inertial migration away from the core). This observation can have implications for inertial microfluidics 
at high throughput.

To summarise, the results in this section indicate that the key parameter in defining particle migration and focusing positions at low $\phi$ is the bulk Reynolds number $Re_b$. Similar  particle distributions across the duct are obtained for equal and finite bulk Reynolds numbers ($> 100$) and different $h/a$. The small 
differences in the distributions are due to the duct to particle size ratio $h/a$ (and hence to particle inertia). In particular, our systematic 
study shows that at lower $Re_b$, particles focus at the wall centers. Increasing the bulk Reynolds number, particles first form a ring-like structure 
close to the four walls, and finally accumulate mostly at the duct corners (higher $Re_b$).

At a constant $Re_b$, however, larger particles feel stronger velocity gradients than smaller particles. The particle Reynolds number is hence 
different for larger and smaller particles and this results in a slight modification of the exact particle distribution across the duct. Indeed, 
for $Re_b=550$ and $h/a=9$ we see that almost all particles are precisely at the corners while fewer focus closer to the wall centers. For $h/a=18$, 
the results are similar. However, particles do not accumulate precisely at the corners (i.e.\ on the diagonal), and the distribution is slightly more 
uniform close to the walls than for the previous case.
For semi-dilute suspensions ($\sim 5\%$), on the contrary, the final particle concentration distribution depends mostly 
on the excluded volume. The effect of $Re_b$ is to induce the inertial migration away from the duct core.

\subsection{Secondary flows}

No secondary motions are present in a laminar duct flow. Typically secondary flows
appear at high bulk Reynolds numbers once the flow becomes turbulent. According
to Prandtl~\cite{prandtl1926} there are two kinds of secondary flows: skew-induced and Reynolds stress
induced. The former are absent in fully developed turbulent duct flows while the latter are
produced by the deviatoric Reynolds shear stress $\langle v_f'w_f' \rangle$ and the cross-stream Reynolds
stress difference $\langle v_f'^2 \rangle - \langle w_f'^2 
\rangle$ (where $\langle \cdot \rangle$ denotes averaged quantities). When a solid phase is
dispersed in the liquid, particle-induced stresses generate cross-stream secondary motions also in originally laminar flows.

\begin{figure}[t]
  \centering
\includegraphics[width=0.328\textwidth]{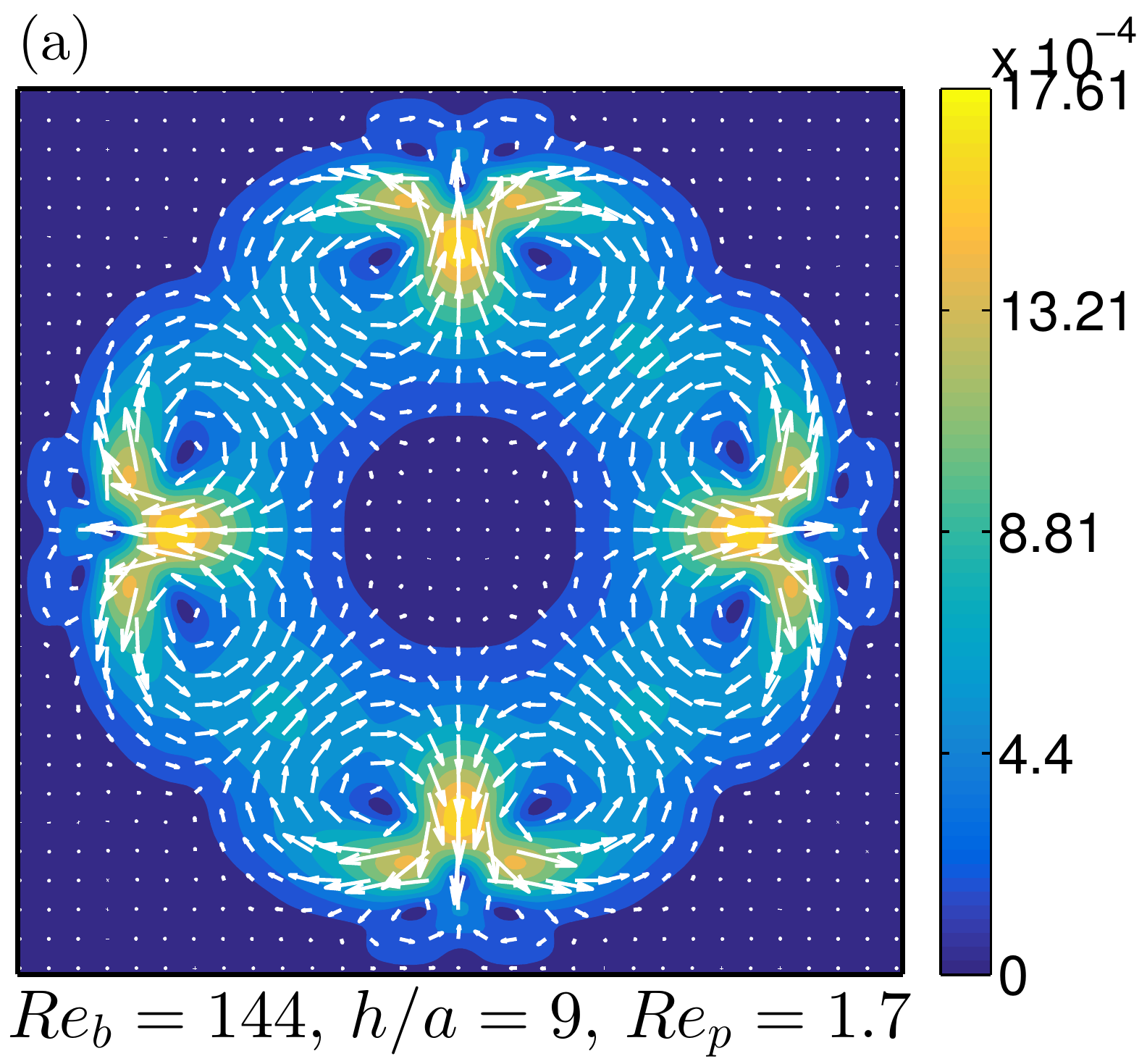}
\includegraphics[width=0.328\textwidth]{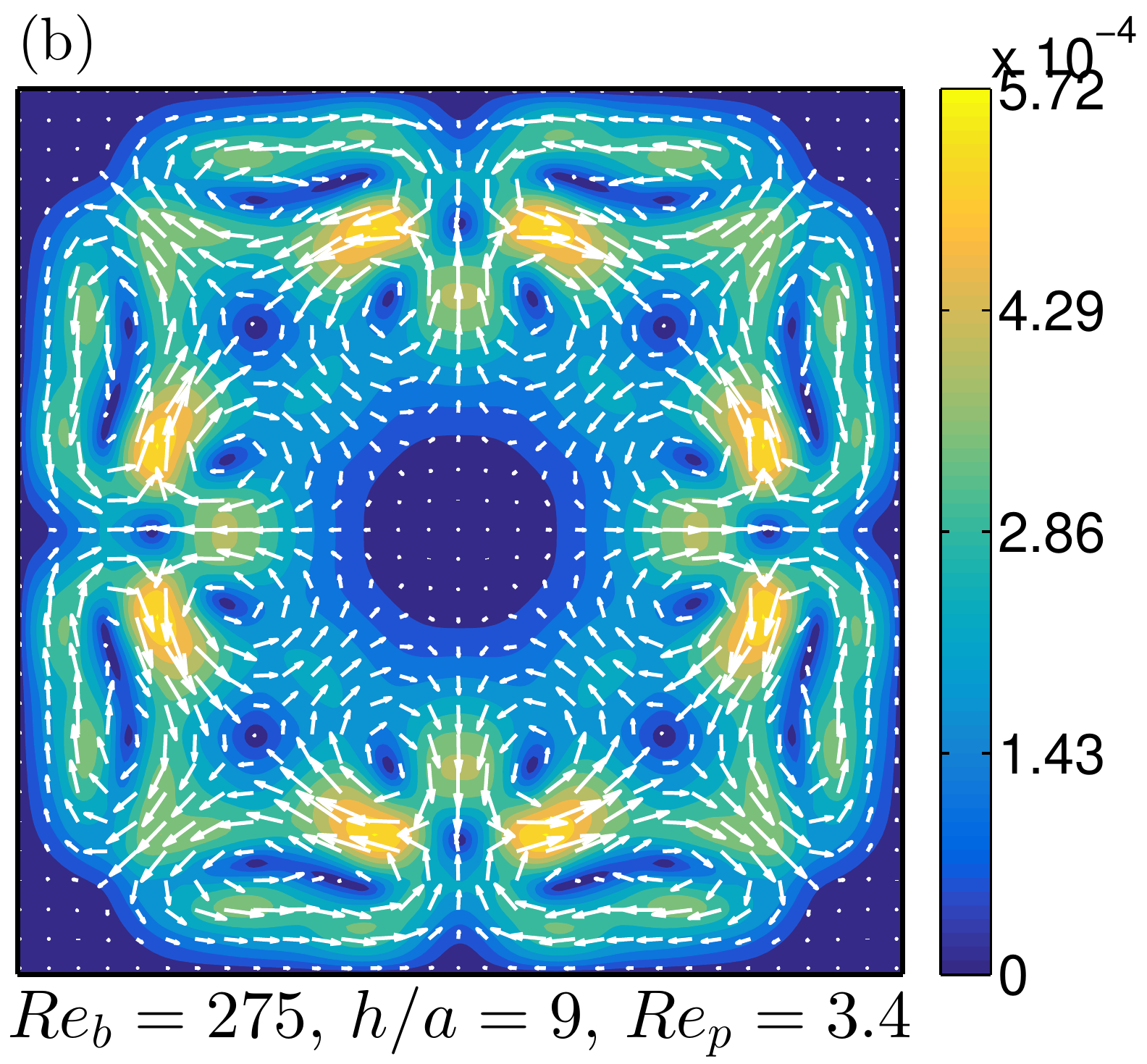}
\includegraphics[width=0.328\textwidth]{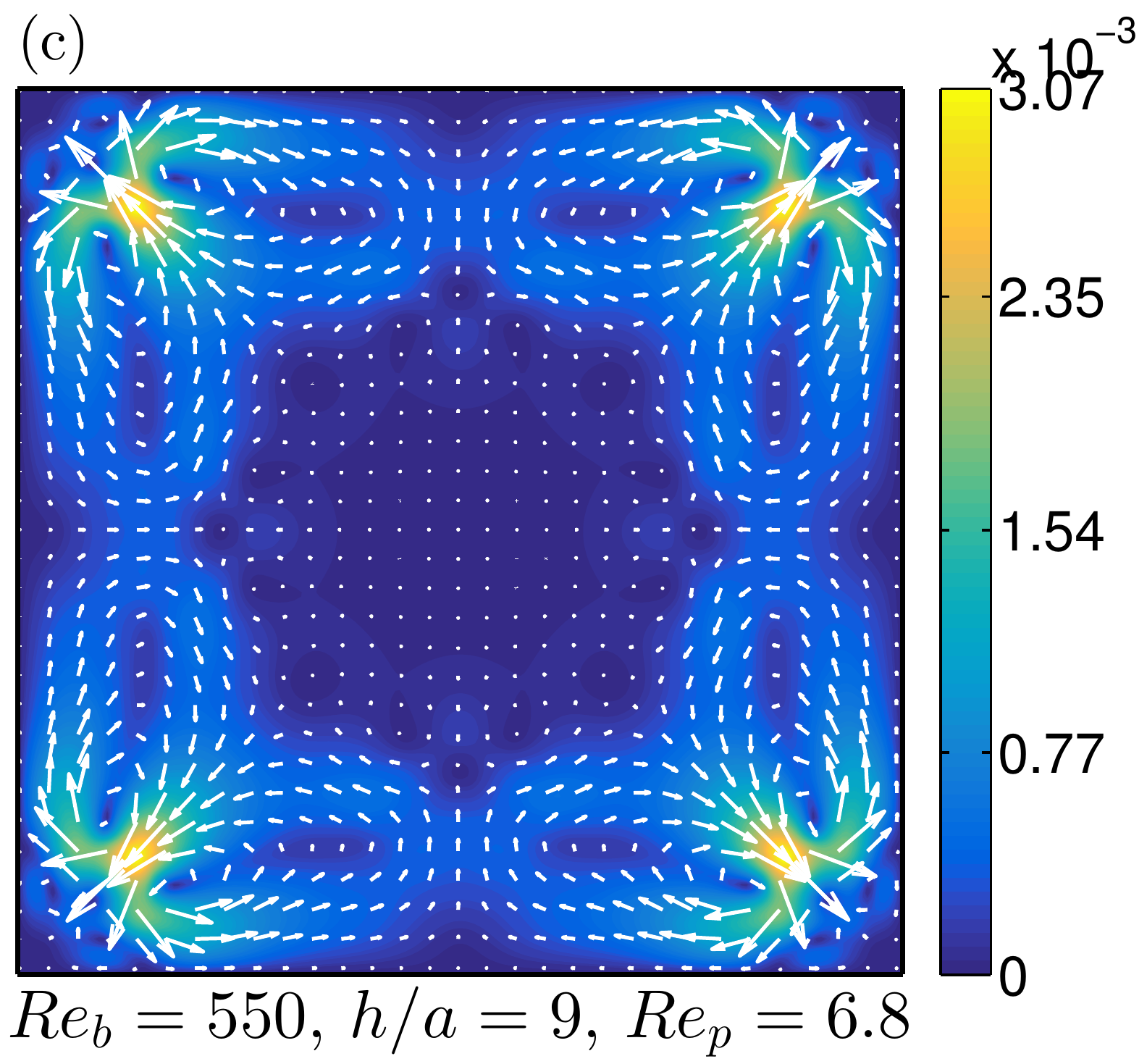}\vspace{1cm}

\includegraphics[width=0.328\textwidth]{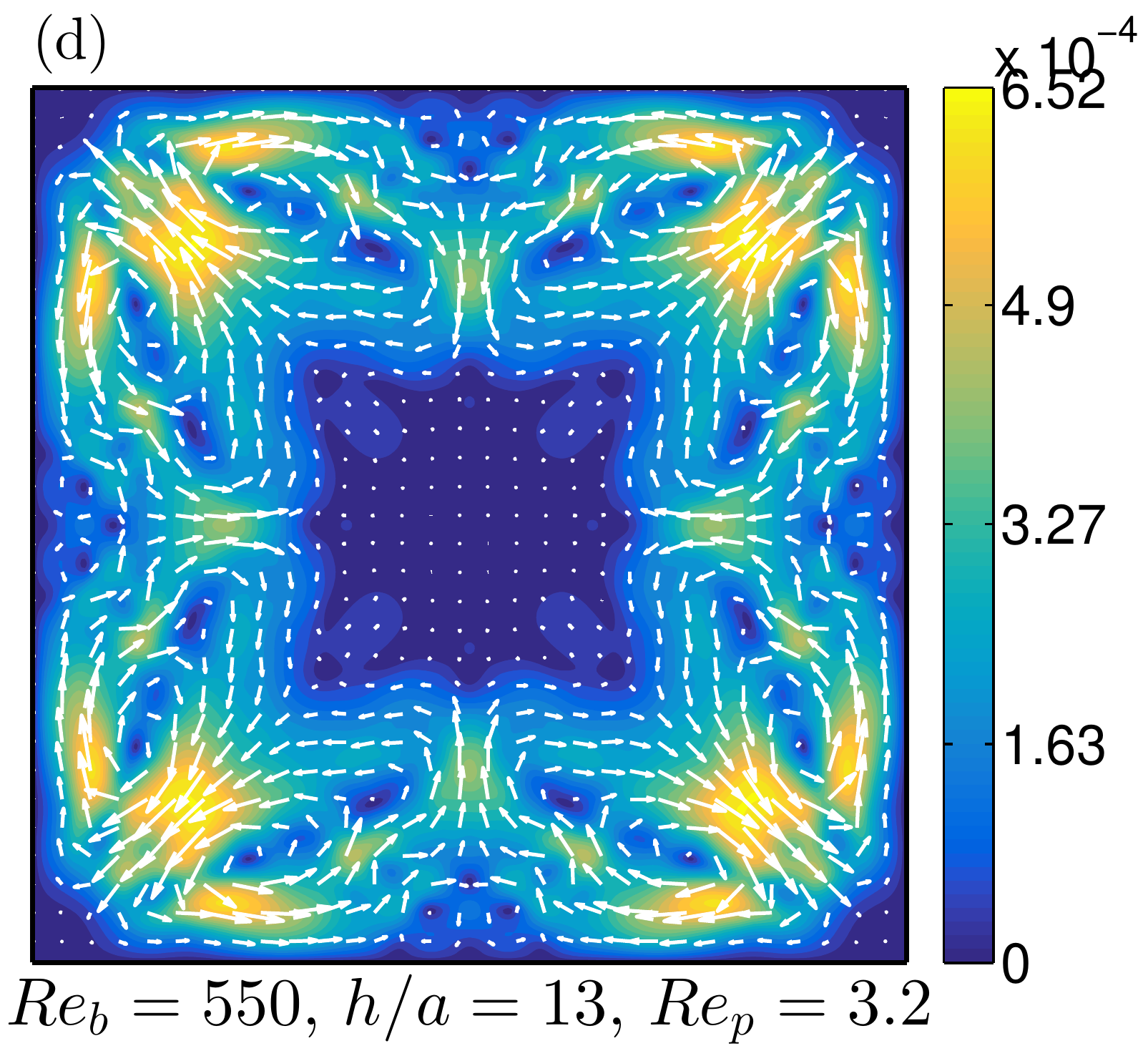}
\includegraphics[width=0.328\textwidth]{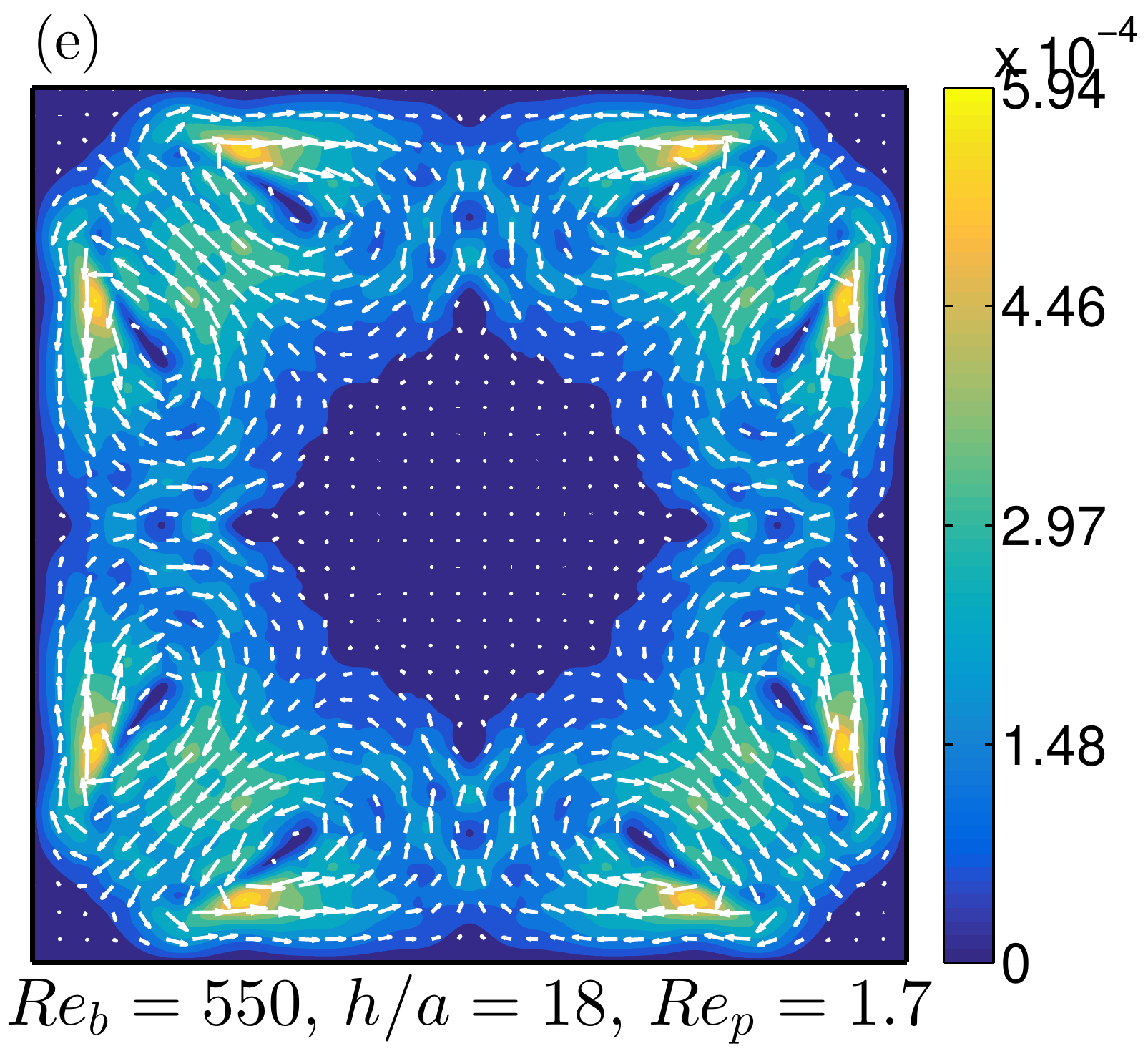}
\includegraphics[width=0.328\textwidth]{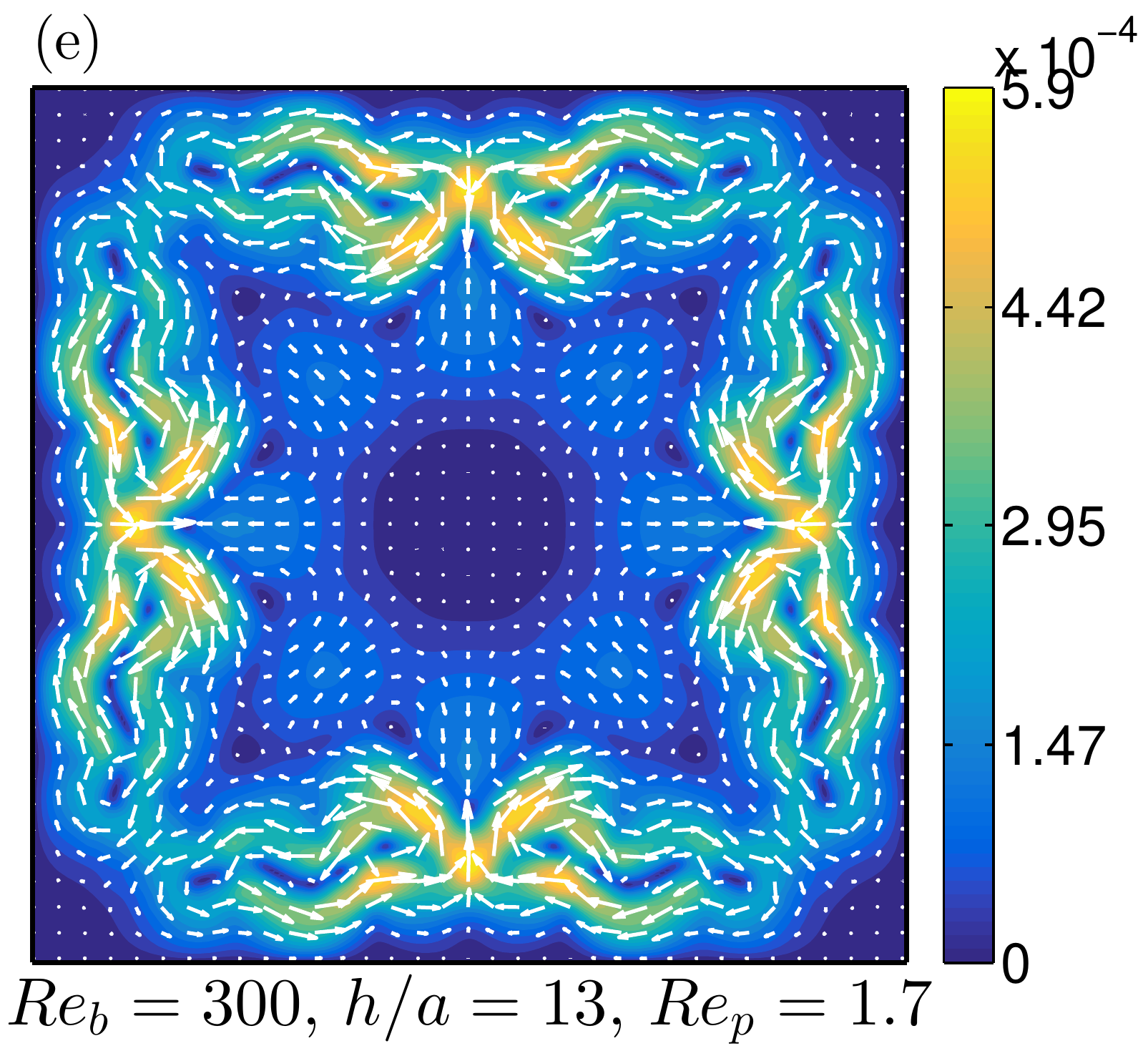}

\caption{Contour plot of cross flow velocity magnitude $\sqrt{V_f^2+W_f^2}$ and vector field for $\phi=0.4\%$ and Reynolds numbers and particle size reported in the legend. For comparison, $(a)-(c)$ correspond 
to the cases presented in Figs.~\ref{fig:L_Conc}$(a)-(c)$, $(d)$ and $(e)$ to the cases in Figs.~\ref{fig:L_Conc}$(h)$ and $(i)$,
$(f)$ to the case in Fig.~\ref{fig:L_Conc}$(e)$.}
\label{fig:Second_L}
\end{figure}

The results of the present study shows the existence of secondary flows induced by particles in dilute suspensions. 
In Figs.~\ref{fig:Second_L} $(a)$-$(f)$ we report the cross-flow 
velocity magnitude $\sqrt{V_f^2+W_f^2}$ and vector fields for different cases with $\phi=0.4\%$. 
We clearly see that the intensity of these secondary flows is stronger close to the particle focusing positions. 
For $h/a=9$ and $Re_b=144$ particles focus at the wall centers (cf.\ Fig.~\ref{fig:L_Conc}$a$) and, 
accordingly,
secondary motions are stronger around the focusing positions and point from the core towards the wall centers (Fig.~\ref{fig:Second_L}$a$). 
As documented in Sec.~\ref{Semi-dilute suspensions}, when the bulk Reynolds number $Re_b$ is increased, 
these four focusing positions are lost and particles form a ring-like 
structure close to the walls (Fig.~\ref{fig:L_Conc}$b$). 
The corresponding secondary motions are displayed in Figure~\ref{fig:Second_L}$(b)$: 
their intensity reduces significantly due to the more uniform particle distributions across the 
duct cross section.
Further increasing the bulk Reynolds number, $Re_b=550$, particles focus at the duct corners (Fig.~\ref{fig:L_Conc}$c$). 
Consequently, the secondary motion is more evident at these locations, now directed towards the corners along the bisectors (Fig.~\ref{fig:Second_L}$c$). 
Comparing Fig.~\ref{fig:Second_L}$a$ and $c$, we note that the cross-stream motions are directed from the duct core to the locations of particle focusing.

Further, we observe in Fig.~\ref{fig:L_Conc}$(c)$ that for $h/a=9$ and $Re_b=550$, the local particle concentration at the duct 
corners is higher than that found at the wall centers for $Re_b=144$ (Fig.~\ref{fig:L_Conc}$a$). Moreover, 
Fig.~\ref{fig:Second_L}$(c)$ shows that the secondary flows intensity increases  for $Re_b=550$ with respect to the cases at 
$Re_b=144$ and $275$ (Figs.~\ref{fig:Second_L}$a$ and $b$). These observations suggest
that the intensity of the secondary flows is determined by the local particle concentration. 
However, the intensity of these secondary flows is small and less than $0.4\%$ of $U_b$.

Figures.~\ref{fig:Second_L}$(d)$ and $(e)$ show the cross-flow velocity magnitude $\sqrt{V_f^2+W_f^2}$ and vector fields for two 
cases with the same bulk Reynolds number $Re_b=550$ and different duct to particle size ratios $h/a$ of 13 and 18. 
We see that both the maximum value of the local particle concentration (Fig.~\ref{fig:L_Conc}$i$) and the secondary flow intensity 
are higher for the duct with $h/a=13$. At smaller $h/a$ and constant $Re_b$, when the particle inertia (i.e. the particle 
Reynolds number $Re_p$) and particle-induced stresses are higher, stronger secondary flows are generated. Finally, 
we report the secondary flow pattern for bulk Reynolds number $Re_b=300$ and $h/a=13$ in Fig.~\ref{fig:L_Conc}$(f)$. 
This configuration has same particle Reynolds number ($Re_p=1.7$) and similar maximum value of local particle concentration as 
the case with $Re_b=550$ and $h/a=18$ (see Figs.~\ref{fig:L_Conc}$e$ and $i$). In agreement with the previous results, we observe similar 
secondary flow intensity in Figs.~\ref{fig:Second_L}$(e)$ and $(f)$. 

\begin{figure}[t!]
  \centering
\includegraphics[width=0.328\textwidth]{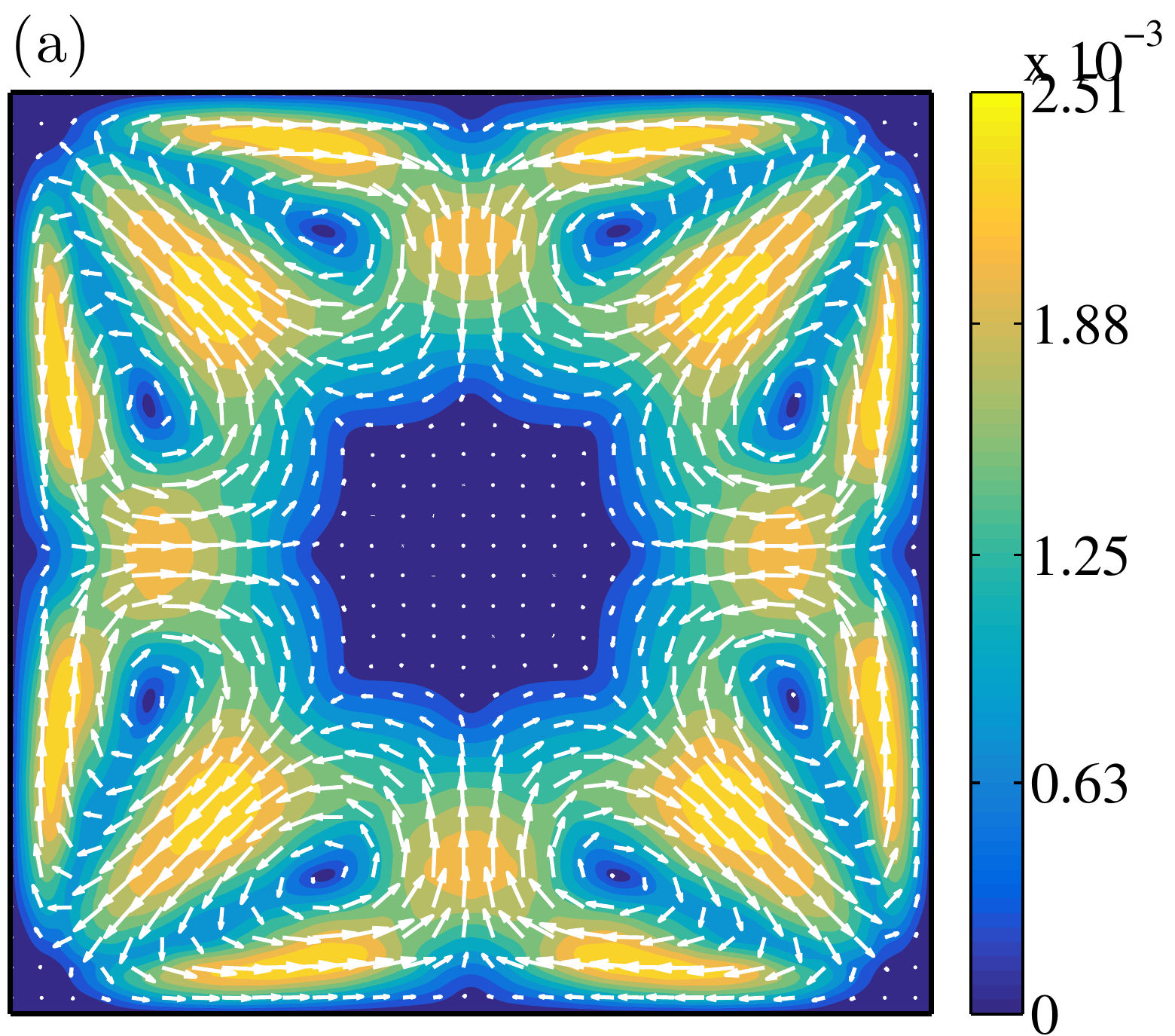}
\includegraphics[width=0.328\textwidth]{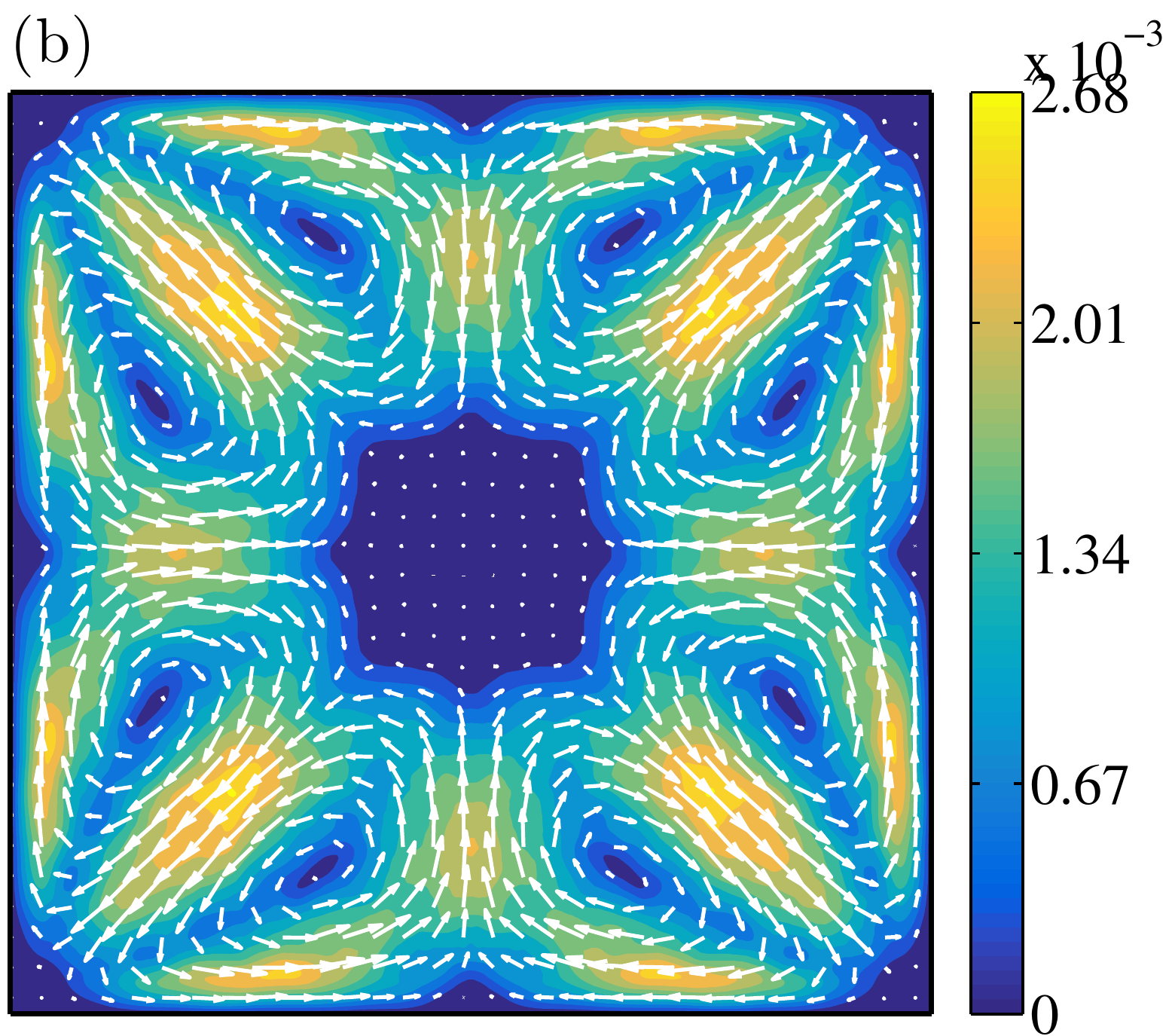}
\includegraphics[width=0.328\textwidth]{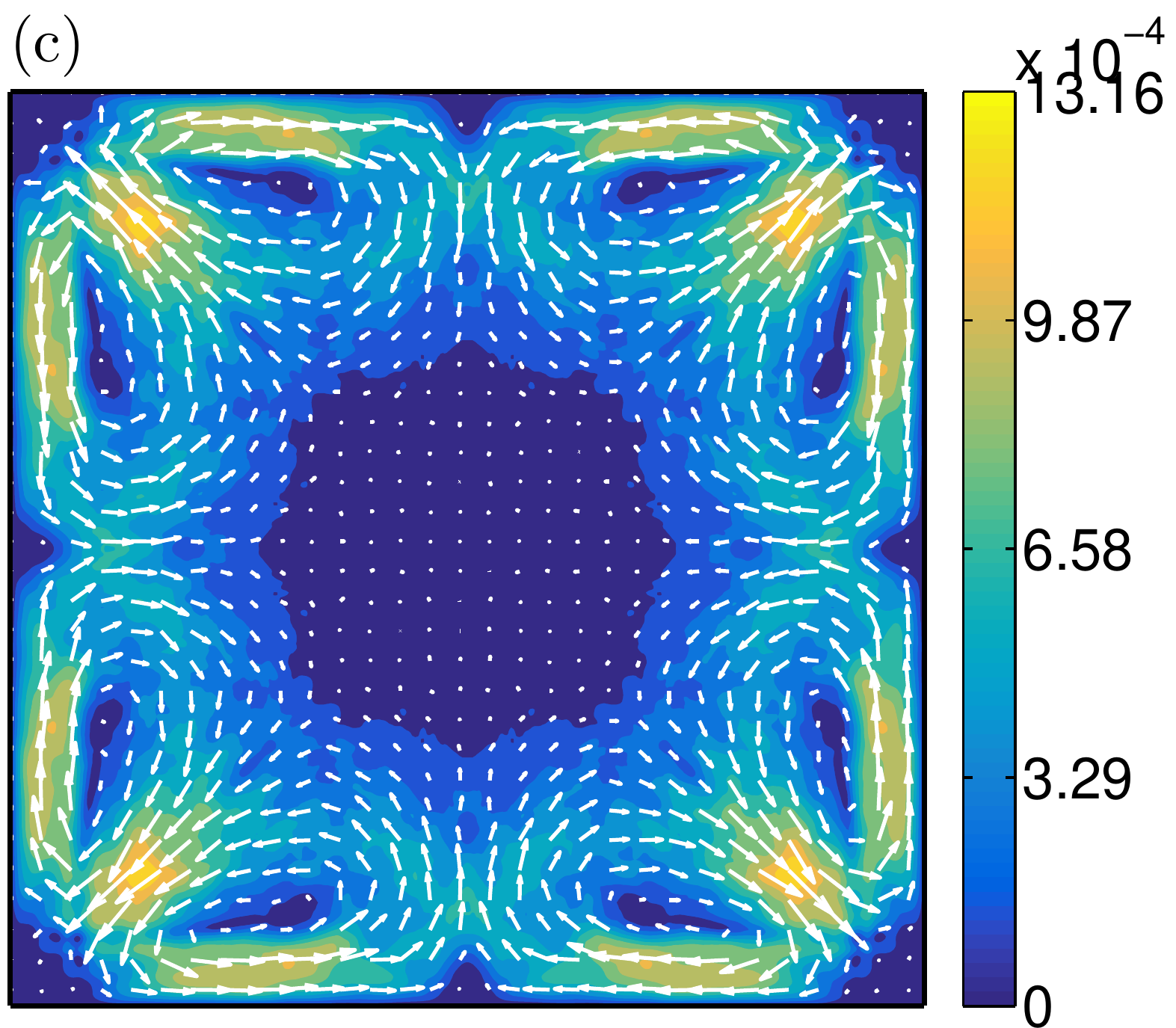}

\caption{ Contour plots of the cross-flow velocity magnitude $\sqrt{V_f^2+W_f^2}$ and velocity vectors for $Re_b=550$, $Re_p=1.7$ and $h/a=18$: $(a)$ $\phi=5\%$, 
$(b)$ $\phi=10\%$, $(c)$ $\phi=20\%$.}
\label{fig:Sec_H}
\end{figure}

\begin{figure}[t!]
   \centering
\includegraphics[width=0.8\textwidth]{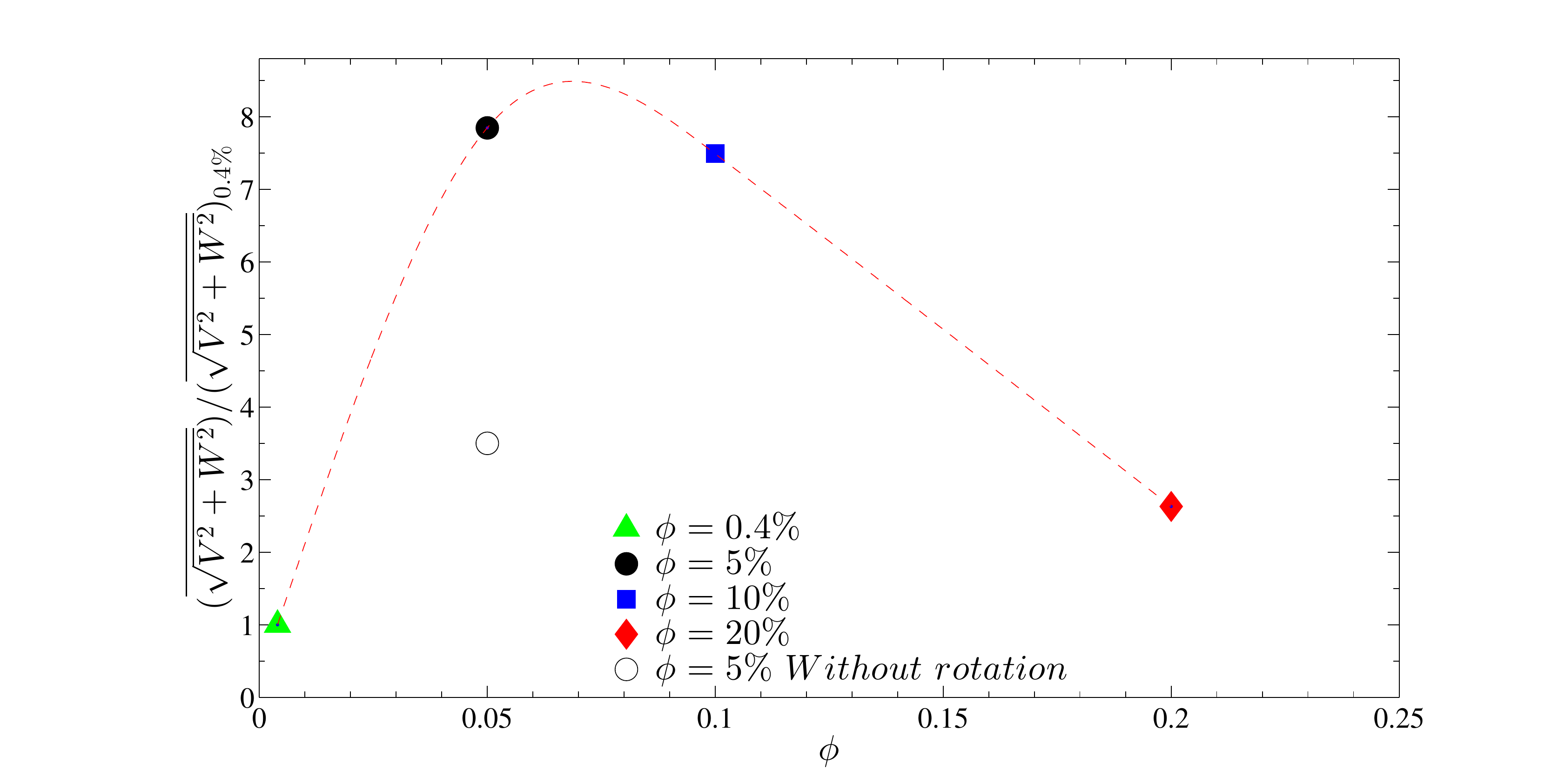}

\caption{Mean value of $\sqrt{V_f^2+W_f^2}$ normalized by the mean value of $\Big(\sqrt{V_f^2+W_f^2}\Big)_{0.4\%}$ 
for semi-dilute suspensions with $\phi=5, 10$ and $20\%$. The result from 
a simulation with $\phi=5\%$, in which particles are constrained not to rotate is also shown. Dashed line is for visualization.}
\label{fig:Kinetic_Energy}
\end{figure}

Next, we explore the dependence of the fluid secondary motions on the solid volume fraction $\phi$. To this aim, contours of the cross-flow velocity magnitude $\sqrt{V_f^2+W_f^2}$ 
and velocity vectors are reported in Figs.~\ref{fig:Sec_H}$(a)$-$(c)$ for the semi-dilute cases under investigation, $\phi=5, 10$ and 
$20\%$ ($h/a=18$). The maximum intensity of these secondary flows is still low, about $0.2\%$ of the bulk velocity 
$U_b$ (circa $1/10$ of that found in turbulent duct flows). The maximum of 
the secondary cross-stream velocity is similar for $\phi=5$ and $10\%$ while it substantially decreases for $\phi=20\%$. 
The mean of cross-flow velocity magnitude, $\sqrt{V_f^2+W_f^2}$, initially increases and then 
decreases as the volume fraction $\phi$ increases as shown in Fig.~\ref{fig:Kinetic_Energy} where we report the mean values of 
$\sqrt{V_f^2+W_f^2}$ for each $\phi$. All results are normalized by the mean value obtained for 
$\phi=0.4\%$.

It is also interesting to note that when particles accumulate at the corners, the secondary flow patterns are similar to those found in turbulent flows. 
Particle-induced stresses act in a similar fashion to Reynolds stresses 
and consequently lead to similar secondary flows. These secondary flows, although weak, convect mean velocity from regions of 
large shear along the walls towards regions of low shear. This convection occurs along the corner bisectors resulting in a 
lower mean streamwise velocity at the walls (and particularly at the wall centers)~\cite{prandtl1952}. This effect can also be seen in the contours of 
mean particle streamwise velocity $U_p$ (which closely resemble the contours of the mean fluid streamwise velocity), see Fig.~\ref{fig:U_P}. 
This behavior is attenuated as the solid volume fraction increases and the secondary flows are progressively damped.

To gain further insight into the role of particle angular velocities on the intensity of secondary flows, we 
performed an additional simulation at $Re_b=550$, $h/a=18$ and $\phi=5\%$ in which we artificially impose zero 
particle rotation (i.e.\ constant null angular velocities) while allowing translations. The result of this simulation is reported in 
Fig.~\ref{fig:Kinetic_Energy}: 
the mean value of the cross-flow velocity magnitude reduces significantly ($\sim 55\%$)
with respect to the reference case at $\phi=5\%$. This confirms that the intensity of secondary motions strongly depends on the 
particle angular velocities. 

\begin{figure}[t!]
   \centering
\includegraphics[width=0.415\textwidth]{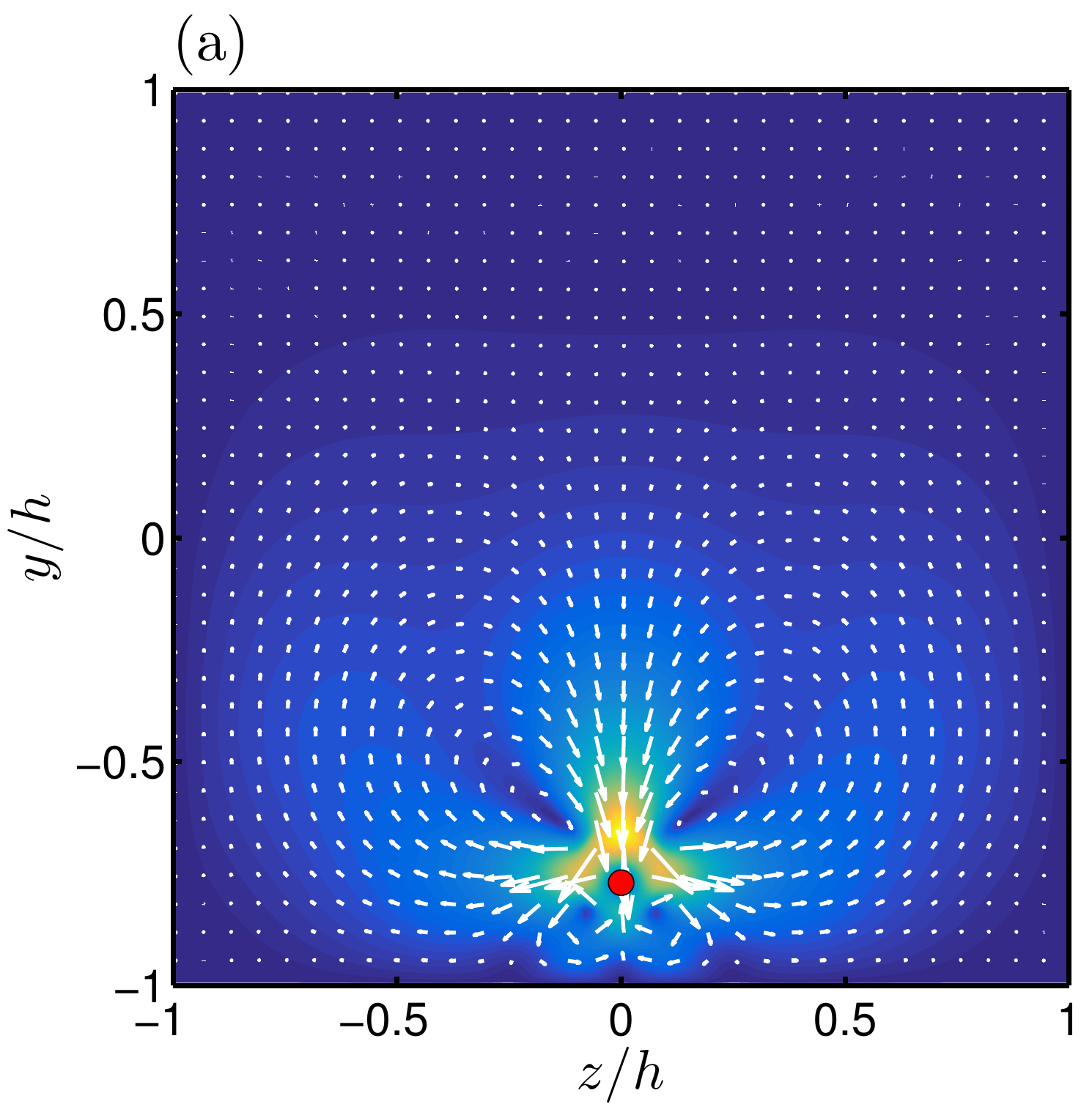}
\includegraphics[width=0.45\textwidth]{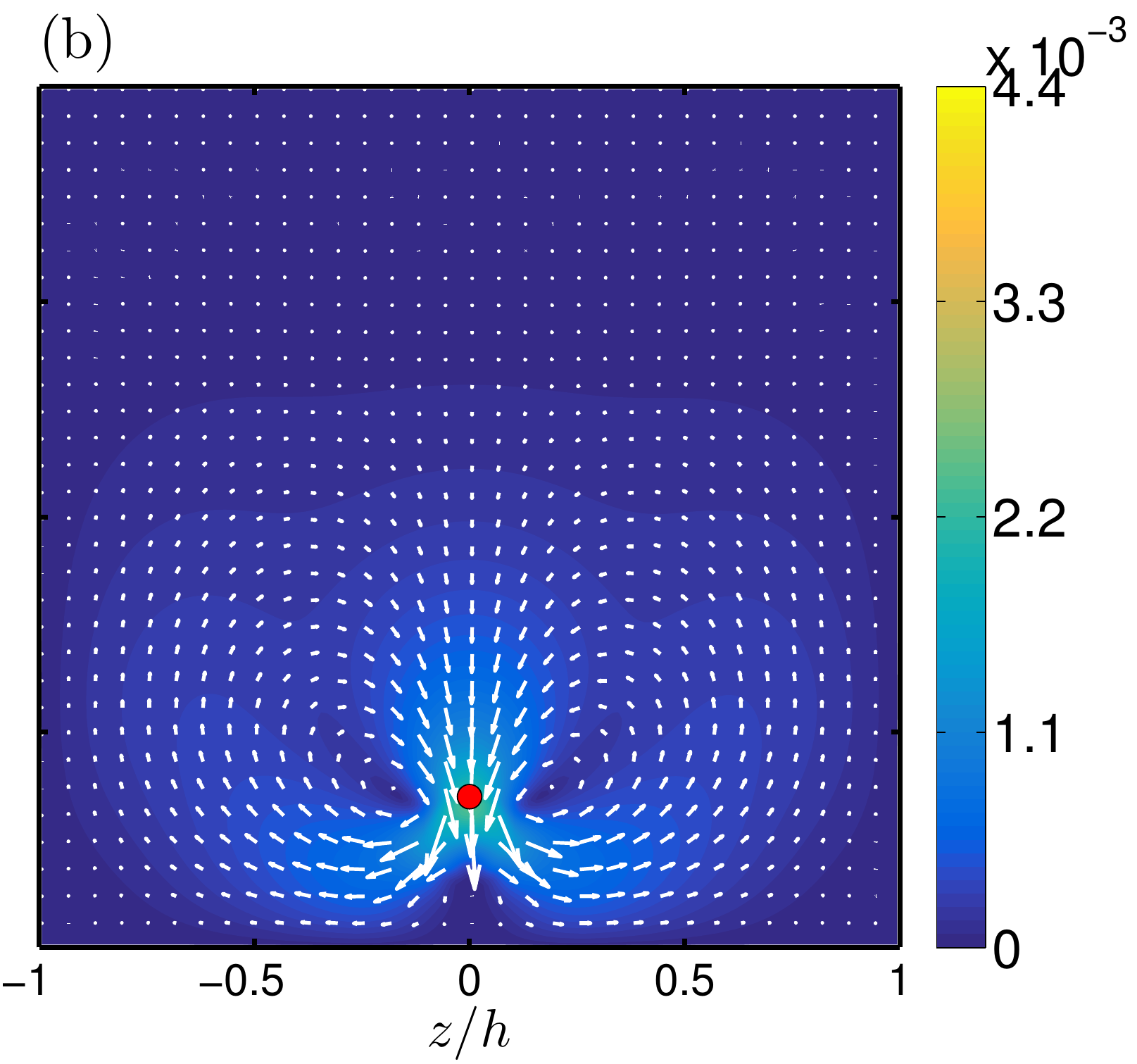}
\caption{Contour plot of the cross-flow velocity magnitude $\sqrt{V_f^2+W_f^2}$ and velocity vectors around a particle: $(a)$ particle moves and rotates freely through the duct,
$(b)$ particle moves downstream with spanwise angular velocity set to zero. The red dot shows the position of the particle center.}
\label{fig:vortex}
\end{figure}

At finite particle Reynolds number, $Re_p$, when inertia plays an important role, the flow field around a particle is 
altered and the fore-aft symmetry of the streamlines is lost~\cite{subramanian2006}. Amini et al.~\cite{amini2012} 
investigated the flow field around a translating and rotating spherical particle in Poiseuille flow at finite particle 
Reynolds number. These authors showed the existence of a pair of recirculating zones perpendicular to the primary flow 
in the vicinity of the particle. Here, we investigate the flow field around an individual particle moving through a duct at its 
equilibrium position at $Re_b=100$ and $h/a=10$. 
We first consider a particle free to move and rotate and then artificially set the spanwise particle angular velocity to zero, $\omega_z=0$, to quantify the 
effect of particle rotation on the intensity of the recirculating flows (calculated as in Ref.~\onlinecite{amini2012}). As shown in Fig.~\ref{fig:vortex}, 
the intensity of the flow around the particle is directly related to the particle 
angular velocity and drastically decreases by setting $\omega_z=0$. Moreover, the particle focusing position 
changes in the absence of rotation and moves slightly towards the duct core. The 
presence of this local secondary flow near the particle is also reported by Shao et al.~\cite{shao2008} at bulk Reynolds number 
$Re_b=1000$ in pipe flow. 

\section{Conclusions}

We present results from direct numerical simulations of laminar duct flow of suspensions of finite-size neutrally buoyant 
spherical particles at different solid volume fractions. We use an immersed boundary method for the fluid-solid interactions 
with lubrication and collision models for the short-range particle-particle (particle-wall) interactions. The stress immersed 
boundary method is applied to generate the duct walls.
Initially we investigate excluded volume effects in dilute and semi-dilute suspensions with $\phi=0.4, 5, 10$ and $20\%$, for 
duct to particle size ratio $h/a=18$ and bulk Reynolds number $Re_b=550$. 
We show that for solid volume fractions $\phi=5$ and $10\%$, particles mostly accumulate at the duct corners and particle 
depletion can be seen at the core of the duct. For $\phi=20\%$, particles distribute uniformly over the whole domain with slightly 
higher concentration at the diagonal of the the duct. For all $\phi$, particles reside longer at the corners than at the wall 
centers. An effective viscosity increase leads to a blunted streamwise fluid velocity profile at the duct center 
at a solid volume fraction $\phi=20\%$. Nonetheless, Eiler's fit is able to predict the increase of dissipation 
in the duct as inertial effects (at the particle scale) are small, $Re_p=1.7$.

We then investigate the interactions and role of $Re_b$, $Re_p$ and $h/a$ on the behavior of dilute suspensions with $\phi=0.4\%$. 
Initially, we keep the duct to particle size ratio constant at $h/a=9$ and increase the bulk and particle Reynolds number. For $Re_b=144$ particles focus 
at the walls centers. Increasing $Re_b$, particles initially form a ring close to the walls and finally, for $Re_b=550$, 
 accumulate preferentially at the duct corners and partially closer to the wall centers, at the distance of $0.6h$ away from 
the core. The particle equilibrium position at the wall center moves towards the duct core when $Re_b$ is increased from $144$ to $550$.
The same behavior of the evolution of the particle 
local volume fraction is observed at constant particle Reynolds number $Re_p=1.7$ when increasing bulk Reynolds number $Re_b$ and duct to particle size ratio $h/a$.
Finally, for constant bulk Reynolds number $Re_b=550$ and different particle Reynolds numbers $Re_p$ and duct to particle size ratios $h/a$, 
we find high concentration around the duct corners and less at the wall centers.
Therefore, for the range of $h/a$ and $Re_p$ investigated here, we conclude that in dilute suspensions the particle focusing position is mainly governed by the bulk Reynolds number.

For $Re_b=144$, $h/a=9$ and $\phi=5\%$ we would have expected particles to accumulate at the walls and particularly at the wall centers (since for 
$\phi=0.4\%$ the focusing positions are located there). Instead, we find that particles accumulate mostly around the corners, with a lower 
concentration at the wall centers. Excluded volume therefore seems to be the the key parameter in determining the particle concentration distribution 
in semi-dilute suspensions. Thus, trivially extending results for single particles to semi-dilute suspensions may lead to wrong predictions.

Secondary flows are generated in the duct due to the presence of particles. At low volume fractions, $\phi=0.4\%$,  secondary flows 
appear around the particle focusing positions and the corresponding vorticity strength is dominated by the local particle concentration. 
We show that the secondary flow intensity initially increases with the solid volume 
fraction from $\phi=0.4\%$ to $\phi=5\%$ while it decreases for $\phi>5\%$. In the semi-dilute regime ($\phi \ge 5\%$), the secondary flows appear as a 
pair of counter-rotating vortices directed towards the corners, along the bisectors. 
Since they resemble closely the secondary flows found in turbulent duct flows, 
particle-induced stresses generate secondary flows in a similar fashion to Reynolds stresses. Their intensity is however 
$1/10$ of that found in turbulent duct flows. We see that the mean intensity of these secondary flows decreases especially 
above $\phi=10\%$. When many particles are injected in the duct, the particle concentration distribution becomes more uniform across the cross-section, and 
cross-stream motions generated by a particle are quickly disrupted by its neighbors.
 
Finally, we study the relation between particle rotation and secondary flows. 
We constrain a single particle to translate without rotation, and we observe that the intensity of the secondary vortices substantially 
decreases. We also notice that the focusing position (initially at the wall center) moves vertically closer to the duct core. We also inhibit 
particle rotation in the semi-dilute suspension with $\phi=5\%$, $Re_b=550$, $h/a=18$ and find that the mean intensity of the secondary flows is 
reduced by $55\%$. Therefore, these secondary flows strongly depend on particle rotation.

In the future, it will be interesting to study turbulent duct flows laden with finite-size spheres and observe the modification of secondary flows, 
particle statistics and turbulence modulation.

\begin{acknowledgments}

This work was supported by the European Research Council Grant No.\ ERC-2013-CoG-616186, TRITOS, from the Swedish Research Council (VR), through the
Outstanding Young Researcher Award. The authors gratefully acknowledge the COST Action MP1305: \emph{Flowing matter} and 
computer time provided by SNIC (Swedish National Infrastructure for Computing).

\end{acknowledgments}

\section*{Appendix:~Temporal evolution of the particle concentration.}

\begin{table}[h!]
  \caption{Time to reach the final steady-state particle distribution for the different cases considered.\\}
  \label{tab:timsim1}
  \begin{center}
\def~{\hphantom{0}}
  \begin{tabular}{ccccc}\hline\hline 
     $\phi ($\%$)$   & $Re_b$  & $\,$ $Re_p$     & $\,$ $\left(h/a\right)$   & $\,$ $\,$ $T^*$\\[3pt]\hline
     $0.4$           & $\,$ $144$   &  $\,$ $\,$ $1.7$      & $\,$ $\,$ $9$                  &  $\,$ $\,$ $ 56$\\
     $0.4$           & $\,$ $275$   &  $\,$ $\,$ $3.4$      & $\,$ $\,$ $9$                  &  $\,$ $\,$ $ 45$\\
     $0.4$           & $\,$ $550$   &  $\,$ $\,$ $6.8 $     & $\,$ $\,$ $9$                  &  $\,$ $\,$ $ 34$\\
     $0.4$           & $\,$ $300$   &  $\,$ $\,$ $1.7$      & $\,$ $\,$ $13$                 &  $\,$ $\,$ $ 64$\\
     $0.4$           & $\,$ $550$   &  $\,$ $\,$ $3.2$      & $\,$ $\,$ $13$                 &  $\,$ $\,$ $ 43$\\
     $0.4$           & $\,$ $550$   &  $\,$ $\,$ $1.7$      & $\,$ $\,$ $18$                 &  $\,$ $\,$ $ 79$\\           
     $5$             & $\,$ $144$   &  $\,$ $\,$ $1.7$      & $\,$ $\,$ $9$                  &  $\,$ $\,$ $ 60$\\
     $5$             & $\,$ $550$   &  $\,$ $\,$ $1.7$      & $\,$ $\,$ $18$                 &  $\,$ $\,$ $110$\\
     $10$            & $\,$ $550$   &  $\,$ $\,$ $1.7$      & $\,$ $\,$ $18$                 &  $\,$ $\,$ $ 80$\\
     $20$            & $\,$ $550$   &  $\,$ $\,$ $1.7$      & $\,$ $\,$ $18$                 &  $\,$ $\,$ $ 49$\\ \hline\hline 
\end{tabular}
  \end{center}
\end{table}

In this Appendix we briefly discuss the effect of $Re_b$, $Re_p$ and $\phi$ on the temporal evolution of the cases under the investigation. In table~\ref{tab:timsim1} we report the dimensionless time $T^*$ needed for the simulations to reach their final steady-state in terms of particle concentration distribution, $\Phi(y,z)$. We define this as the time needed by the local particle concentration around the focusing points to reach the final mean value. Here, time is non-dimensionalized by viscous units ($(2a)^2/\nu$).

For constant $h/a=9$ and $\phi=0.4\%$, the results show that the particles reach the equilibrium positions faster by increasing the bulk Reynolds number $Re_b$ form 144 to 550. For constant bulk Reynolds number $Re_b=550$ and increasing $h/a$, we notice that it takes longer for the particles to evolve and reach their equilibrium positions. Indeed, this is due to the fact that particle inertia, i.e. $Re_p$ is less significant at higher $h/a$. Overall, we see that for the dilute suspensions, $\phi=0.4\%$, the particle evolution time $T^*$ is reduced by increasing particle Reynolds number $Re_p$.

Finally, for semi-dilute cases, we show that $T^*$ decreases by increasing solid volume fraction $\phi$ from $5$ to $20\%$. At higher concentrations there is progressively less space available for particle migrations and the final average particle distribution is reached faster.

\section{References}

\bibliography{Paper}
\end{document}